\documentclass[aps,pra,twocolumn,amsmath,amssymb,superscriptaddress,floatfix]{revtex4-1}
     \usepackage{bbm}
    \usepackage{amsmath}
    \usepackage{verbatim}
    \usepackage{graphicx}
    \usepackage{times}
    \usepackage{amssymb}
    \usepackage{epsfig}
    \usepackage{graphicx}
    \usepackage{bm}
    \usepackage{times}
    \usepackage{txfonts}
    \usepackage{dsfont}
    \usepackage[bbgreekl]{mathbbol}
    \usepackage{physics}
    \usepackage{placeins}
    \usepackage{array}
    \usepackage{mathrsfs}
    \usepackage[utf8]{inputenc}
    \usepackage{wrapfig}

    \PassOptionsToPackage{hyphens}{url}
    \usepackage[colorlinks = true,
                linkcolor = blue,
                urlcolor  = blue,
                citecolor = blue,
                anchorcolor = blue]{hyperref}
    \usepackage{etoolbox}

    \newcolumntype{M}{&gt;{\centering}m{2cm}}
    \newcommand{\psigma}[1]{\hat{\sigma}_{#1}}
    
    \def\Lphase{\phi_\mathrm{L}}
    \def\SASphase{\theta}
    \def\dPhi{\Theta}
    \def\glP{\Phi}
    
    \def\tspread{\tau}
    \def\pspread{\Delta_p}
    \def\even{\mathrm{e}}
    \def\odd{\mathrm{o}}
    \def\g{+}
    \def\u{-}
    
    \def\diffLoss{\gamma}
    \def\sumLoss{\Gamma}
    \def\Fid{F}
    
    \newcommand{\asket}[3]{\ket{#3,#2,#1}}
    \newcommand{\asbra}[3]{\bra{#3,#2,#1}}
    \newcommand{\tasket}[4]{\ket{#2,#1;#4}}
    \newcommand{\tasbra}[4]{\bra{#2,#1;#4}}
    \newcommand{\pasket}[2]{\ket{#2,#1}}
    \newcommand{\pasbra}[2]{\bra{#2,#1}}
    \newcommand\difp{\mathop{}\!\mathrm{d}\kern-0.75pt p}

    \usepackage{multirow}

    \newcolumntype{L}[1]{>{\raggedright\arraybackslash}p{#1}}
    \newcolumntype{C}[1]{>{\centering\arraybackslash}p{#1}}
    \newcolumntype{R}[1]{>{\raggedleft\arraybackslash}p{#1}}
    
\begin{document}
    \title{Analytic theory for Bragg atom interferometry based on the adiabatic theorem}
    
    \author{Jan-Niclas Siem\ss}
    \email{jan-niclas.siemss@itp.uni-hannover.de}
    \affiliation{Institut f\"ur Theoretische Physik, Leibniz Universit\"at Hannover, Appelstra\ss e 2, D-30167, Hannover, Germany}
    \affiliation{Institut f\"ur Quantenoptik, Leibniz Universit\"at Hannover, Welfengarten 1, D-30167, Hannover, Germany}
    \author{Florian Fitzek}
    \affiliation{Institut f\"ur Quantenoptik, Leibniz Universit\"at Hannover, Welfengarten 1, D-30167, Hannover, Germany}
    \affiliation{Institut f\"ur Theoretische Physik, Leibniz Universit\"at Hannover, Appelstra\ss e 2, D-30167, Hannover, Germany}
    \author{Sven Abend}
    \affiliation{Institut f\"ur Quantenoptik, Leibniz Universit\"at Hannover, Welfengarten 1, D-30167, Hannover, Germany}
    \author{Ernst M. Rasel}
    \affiliation{Institut f\"ur Quantenoptik, Leibniz Universit\"at Hannover, Welfengarten 1, D-30167, Hannover, Germany}
    \author{Naceur Gaaloul}
    \affiliation{Institut f\"ur Quantenoptik, Leibniz Universit\"at Hannover, Welfengarten 1, D-30167, Hannover, Germany}
    \author{Klemens Hammerer}
    \affiliation{Institut f\"ur Theoretische Physik, Leibniz Universit\"at Hannover, Appelstra\ss e 2, D-30167, Hannover, Germany}
    
    \begin{abstract}
        High-fidelity Bragg pulses are an indispensable tool for state-of-the-art atom interferometry experiments. In this paper, we introduce an analytic theory for such pulses. Our theory is based on the pivotal insight that the physics of Bragg pulses can be accurately described by the adiabatic theorem. We show that efficient Bragg diffraction is possible with any smooth and adiabatic pulse shape and that high-fidelity Gaussian pulses are exclusively adiabatic. Our results give strong evidence that adiabaticity according to the adiabatic theorem is a necessary requirement for high-performance Bragg pulses. Our model provides an intuitive understanding of the Bragg condition, also referred to as the condition on the "pulse area". It includes corrections to the adiabatic evolution due to Landau-Zener processes as well as the effects of a finite atomic velocity distribution. We verify our model by comparing it to an exact numerical integration of the Schr\"odinger equation for Gaussian pulses diffracting four, six, eight and ten photon recoils. Our formalism provides an analytic framework to study systematic effects as well as limitations to the accuracy of atom interferometers employing Bragg optics that arise due to the diffraction process. 
    \end{abstract}
    \maketitle
    
    \section{Introduction}\label{sec:Introduction}
    Atomic optics and interferometry~\cite{TinoAI2014} is a powerful approach to test basic physical principles such as the equivalence principle~\cite{Bonnin2013,Zhou2015,Schlippert2014,Barret2016,Overstreet2018} and quantum electrodynamics~\cite{Arvanitaki2008,Bouchendira2011,Parker2018}. It also opens up important applications, for example in inertial sensor technology or for infrasound gravitational wave detection~\cite{Dimopoulos2009,Yu2011,Hogan2011,Canuel2018,Schubert2019,Canuel2019}. 
    The extreme sensitivity either requested or achieved in these tests relies on the extraordinary precision by which atom wave packets can be controlled using laser light to perform elementary atom optical elements such as beam splitters or mirrors. Accordingly, the imprecision of these operations makes a dominant contribution to the error budget in current experiments. A precise theoretical understanding of the atom optical elements is therefore essential for the evaluation and further improvement of the accuracy of atom interferometers.
    
    Bragg diffraction of atoms from optical lattice potentials \cite{Giltner1995,Martin1988} is a cornerstone of light-pulse atom interferometry~\cite{Hogan2009}. Often  paired with other methods, such as Bloch oscillations~\cite{Dahan1996,Wilkinson1996,Peik1997,Mueller2009PRL,Clade2009,Mcdonald2013,Gebbe2019}, it is at the heart of most elementary atom optical operations in modern atom interferometry aimed at transferring several photon recoils and precisely controlling the diffracted populations without changing the internal state of the atom~\cite{Mueller2008PRL,Chiow2011,Kovachy2012,Kovachy2015,Kueber2016,Ahlers2016,Plotkin-Swing2018,Gebbe2019}. 
    
    Theoretical models for Bragg diffraction of matter waves from light crystals have first covered the two limiting cases of short and intense or faint and long light pulses referred to as the Raman-Nath~\cite{Gould1986,Rasel1995,Dubetsky2001} or deep-Bragg regime~\cite{Martin1988,Giltner1995} respectively. An introduction to these regimes can be found in the textbook by Meystre~\cite{Meystre2001}, and an account of individual contributions towards a better understanding of the diffraction from optical lattices was given by Müller et al.~\cite{Mueller2008PRA}. Especially for rectangular pulses, the Raman-Nath and the deep-Bragg regime allow for simple analytic solutions of the Schr\"odinger equation and thus provide compact descriptions of elementary atom optical operations and interferometers composed of them.
    
    However, neither of these diffraction regimes allows for efficient large momentum transfer (LMT) operations~\cite{Mueller2008PRL,Mueller2009PRL,Chiow2011,Kovachy2012,Kovachy2015,Ahlers2016,Mcdonald2013,Plotkin-Swing2018,Gebbe2019} as desired for ultrasensitive atom interferometry. E.g. LMT pulses in the deep-Bragg regime require prohibitively long pulse durations rendering them extremely velocity selective due to Doppler shifts resulting from the finite velocity distribution of the atom.
    It has been found that efficient LMT operations can be achieved in between the two limiting cases of the Raman-Nath and deep-Bragg regime, in the so-called quasi-Bragg regime. Especially combined with smooth (e.g., Gaussian) temporal pulse envelopes~\cite{Keller1999,Mueller2008PRA}, quasi-Bragg pulses allow for weak couplings to off-resonant states comparable to the Bragg diffraction regime while relaxing its requirement of long interaction times which greatly improves diffraction efficiencies with ultracold atomic ensembles~\cite{Mueller2008PRA,Szigeti2012}. As a result, LMT Bragg pulses in state-of-the-art atom interferometer experiments predominantly operate in the quasi-Bragg regime~\cite{Mueller2008PRL,Mueller2009PRL,Chiow2011,Kovachy2012,Mcdonald2013,Kueber2016,Ahlers2016,Parker2018,Plotkin-Swing2018,Gebbe2019}.
    
    In this regime, the approximations that led to analytic solutions in the previous cases are not applicable, and no simple analytic description of the Schr\"odinger dynamics generated by quasi-Bragg pulses with time-dependent envelopes has been known so far.
    
    Existing approaches attempt to transfer the logic of deep-Bragg pulses to this intermediate regime by solving the effective dynamics of a two-level system after adiabatically eliminating all off-resonant couplings. In one of the most sophisticated descriptions along this line M\"uller et al.~\cite{Mueller2008PRA} showed how an effective two-level Hamiltonian can be systematically derived in a series expansion via eigenvalues of the Mathieu equation. Before that the Mathieu equation had already been used to describe the dynamics of an atom in a nonresonant standing light wave~ \cite{Fedorov1967,Horne1999}. The formalism developed by M\"uller et al. was supplemented with a description similar in spirit by Giese et al.~\cite{Giese2013}.
    
    For practically relevant parameters, the series expansion utilized in~\cite{Mueller2008PRA} results in rather cumbersome formulas. As a matter of fact, for lack of manageable analytic descriptions, measured data in experiments are typically compared against numerical solutions of the Schrödinger equation~\cite{Ahlers2016,Plotkin-Swing2018,Parker2018}.
    
    An alternative ansatz to describe the diffraction of an atom from a light crystal is to employ Bloch states providing analytic expressions for Bragg diffraction amplitudes~\cite{Champenois2001} and phases~\cite{Buechner2003} adequate in the weak lattice limit. Recently,  Gochnauer et al.~\cite{Gochnauer2019} have used the Bloch solutions to analyze quasi-Bragg pulses which agrees well with experimental results for the specific pulse shapes studied in Ref.~\cite{Gochnauer2019}. They show that the effective Rabi frequency is actually determined by the energy gaps in the Bloch spectrum of the optical lattice for a given potential depth, provided the Bragg pulse is sufficiently adiabatic. While this result does not directly provide analytic descriptions, it becomes clear that the Bloch state picture offers important insight into the functioning of quasi-Bragg diffraction.
    
    Here, we present a comprehensive and relatively simple analytic theory for Bragg atom interferometers. The key insight is that the dynamics in the quasi-Bragg regime can, in fact, be captured very accurately by a model based on the adiabatic theorem~\cite{Albash2018}. In our paper, we show that any smooth and adiabatic (in the sense of the adiabatic theorem) Bragg pulse can give rise to efficient atom optical operations. For the specific but widely used case of a Gaussian pulse we also show the reverse: 
    Efficient beam splitter or mirror operations are generated exclusively by adiabatic pulses. Whether this is generally true, i.e., whether nonadiabatic Bragg pulses generating diabatic dynamics can lead to high-performance atom-optical operations at all, is not clear, but seems doubtful to us on the basis of the description developed here. 
    
    We use the adiabatic theorem in combination with analytic methods from scattering theory to determine the transfer or scattering matrix for single quasi-Bragg pulses. We emphasize that this ansatz is conceptually different from the adiabatic elimination of off-resonant couplings mentioned previously~\cite{Mueller2008PRA,Giese2013}. The general form of the Bragg scattering matrix identified in this paper applies to adiabatic but otherwise arbitrary pulses of any order with constant laser phase. It depends on dynamic (energetic) phases and nonadiabatic first-order (in inverse pulse duration) corrections corresponding to Landau-Zener (LZ) losses and LZ phases. First-order Doppler shifts are included systematically in terms of perturbation theory to account for finite momentum widths of atom wave packets. This way, we obtain our central result, namely, an analytic formula for the Bragg scattering matrix as a function of the dynamic phases, LZ phases, LZ losses, and Doppler shifts, all of which are ultimately determined by the parameters of the Bragg pulse such as its peak intensity, duration and envelope. The only exception here is the formula for LZ losses, which as of yet applies only to second-order diffraction with Gaussian pulses. However, we show that LZ losses can be well understood by two-level dynamics, so a suitable adaptation of theoretical treatments as in~\cite{Vasilev2004,Vasilev2005,Davis1976,Dykhne1962} should give good analytic descriptions for more general cases. Our formulas for dynamic and LZ phases, in particular, provide in the quasi-Bragg regime simple expressions for the so-called Bragg condition on the pulse area (i.e., the combination of pulse duration and peak intensity) that must be met to implement beam splitter or mirror operations. In our formalism, this replaces the concept of the effective Rabi frequency, which has been used in earlier descriptions to formulate the Bragg condition~\cite{Martin1988,Giltner1995,Mueller2008PRA,Hogan2009}. We give a comprehensive comparison of the predictions of our analytic model with results from exact numerical solutions of the Schrödinger equation and find excellent agreement. The \textit{MATHEMATICA} code at the basis of this comparison is available~\cite{Siemss2020}.
    
    Finally, we show how scattering matrices for elementary Bragg operations can be combined to describe full atom interferometers and to extract analytic predictions about interferometer signals for given pulse parameters. Considering the accuracy of the description of all the atom optical components, we argue that our model provides a solid basis for a comprehensive evaluation of the error budget of Bragg atom interferometers as well as a powerful approach for a systematic optimization of interferometer sequences. For example, it can be used to determine the cause and magnitude of diffraction phases~\cite{Buechner2003,Estey2015}, i.e., interferometer phases generated in Bragg mirror and beam splitter operations, with high accuracy as we will report in future work. Whereas this paper covers exclusively Bragg atom interferometers it is quite clear that the general approach taken here can be employed to describe other mechanisms for atom optical operations, such as double Bragg diffraction~\cite{Giese2013,Kueber2016,Ahlers2016}, Bloch oscillations~\cite{Dahan1996,Wilkinson1996,Peik1997}, or Raman interferometers~\cite{Kasevich1991}.
    
    The paper is organized as follows. In Sec.~\ref{sec:BraggDiffraction} we formulate the problem of Bragg diffraction within the framework of scattering theory and explain how the scattering matrix can be used to quantify the quality of a Bragg pulse as a beam splitter or mirror in terms of a fidelity. We present exact numerical results for the fidelity achievable with Gaussian pulses, showing a rich phenomenology which is fully resolved in the following. Section~\ref{sec:BraggScatteringMatrix} contains the formal derivation of the scattering matrix. We exploit here the symmetries of the Bragg Hamiltonian, apply the adiabatic theorem, and calculate first-order corrections as explained above. Readers who are interested in the results rather than the technical aspects of the derivation are invited to skip this section, and read only its (almost) self-contained summary in Sec.~\ref{sec:summary}. The comparison of analytic and numerical results for Bragg beam splitters and mirrors is given in Sec.~\ref{sec:BSandMirror}, where we show that our analytic model accounts for the numerically observed phenomenology with high accuracy. Finally, we provide in Sec.~\ref{sec:Conclusions} a more detailed comparison to previous work, in particular to Refs.~\cite{Mueller2008PRA,Gochnauer2019}, and indicate in Sec.~\ref{sec:Outlook} possible extensions and generalizations of the results presented here.
    
    \section{Bragg diffraction}\label{sec:BraggDiffraction}
    \subsection{Bragg diffraction as a scattering problem}
    
        Consider an atom (mass $M$) in a state which is localized in momentum space at an average momentum $Mv_0$ with a cha\-rac\-teristic spread $\pspread\ll \hbar k$, where $k$ is the wave vector of the laser forming the optical lattice. A momentum $2N\hbar k$ shall be gained by the atom in $N$th-order Bragg diffraction, so that it is transferred to a final state with momentum $Mv_1=Mv_0+2N\hbar k$, or, just as well, into an arbitrary superposition state of momenta $Mv_0$ and $Mv_1$. If this is possible, the time-reversed process can be applied to any incoming superposition of momenta $Mv_0$ and $Mv_1$ too. Thus, for reasons of concreteness and without loss of generality, we can assume an incoming wave packet with average momentum $Mv_0$ as an initial condition.     
        To impart momentum, the atom is exposed to two counterpropagating light fields which are far detuned with respect to an atomic transition, and detuned with respect to each other by a detuning $\delta=k(v_0+N\hbar k/M)$. The atoms thus experience a time-dependent ac Stark potential, which gives rise to the following Hamiltonian in the laboratory frame
        
        \begin{align*}
           \mathcal{H}^\mathrm{LF}(t)&=\mathcal{K}+2\hbar\Omega(t)\cos\left(k\hat{z}-\delta t+\Lphase\right)^2, &
            \mathcal{K}&=\frac{\hat{p}^2}{2M}.
        \end{align*}
        Here we have also introduced a relative phase $\Lphase$ of the two laser fields. Both $\Omega(t)$ and $\Lphase$ may have a time dependence which is controlled through the intensities and phases of the two applied fields; the laser phase $\Lphase$, however, is assumed to be constant throughout our analysis and we will comment in the end of this paper on the case of a time-dependent tuning of the laser phase. Regarding the Rabi frequency, we assume a pulsed driving such that $\Omega(t)$ vanishes asymptotically, $\lim_{t\rightarrow\pm\infty}\Omega(t)=0$, and is nonzero only for a time interval on the order of $\tspread$ around $t=0$. In Sec.~\ref{sec:BSandMirror} as a concrete example we will consider the widely used Gaussian pulse
        \begin{align}\label{eq:GaussPulse}
            \Omega(t)=\Omega_0\;e^{-\frac{t^2}{2\tspread^2}},
        \end{align}
       as the choice of a smooth envelope reduced populations of unwanted states and parasitic phase shifts~\cite{Mueller2008PRA}.
        
        It is useful to change to a frame which is comoving with the lattice potential at a velocity $v_\mathrm{L}=\delta/k$, and to absorb a global phase $\Phi$ due to the average ac Stark potential and a shift in kinetic energy.  The corresponding transformation is effected by a unitary operator $\mathcal{G}(t)=\exp\left(-i(\hat{z}-\hat{p}t/M)Mv_\mathrm{L}/\hbar+i\Phi_{\mathcal{G}}(t)\right)$, where $\dot{\Phi}_{\mathcal{G}}(t)=\Omega(t) + Mv^2_\mathrm{L}/2\hbar$. The Hamiltonian  $\mathcal{H}^\mathrm{MF}=i\hbar\dot{\mathcal{G}}\mathcal{G}^\dagger+\mathcal{G}\mathcal{H}^\mathrm{LF}\mathcal{G}^\dagger$
        in the moving frame is
        \begin{align}\label{eq:Ham}
        \mathcal{H}^\mathrm{MF}(t)&=\mathcal{K}+\frac{\hbar\Omega(t)}{2}\left(e^{2i(k\hat{z}+\Lphase)}+e^{-2i(k\hat{z}+\Lphase)}\right).
        \end{align}
        In this frame, the incoming atomic wave packet is initially composed of momentum components around an average momentum $M(v_0-v_\mathrm{L})=-N\hbar k$, and the target momentum in $N$th-order Bragg diffraction is $M(v_1-v_\mathrm{L})=N\hbar k$. The Hamiltonian in Eq.~\eqref{eq:Ham} is the usual starting point to describe Bragg pulses \cite{Keller1999,Meystre2001,Mueller2008PRA,Giese2013,Gochnauer2019}. 
        
        We aim to understand and describe Bragg diffraction as a scattering process. For this purpose, it is suitable to assume an asymptotic initial condition for the state of the atom. This means that for $t\rightarrow -\infty$ we require that the incoming atomic wave packet $\ket{\psi(t)}$ satisfies
        \begin{align}\label{eq:asymp}
            \ket{\psi(t)}\overset{t\rightarrow-\infty}{\longrightarrow}e^{-i\mathcal{K}t/\hbar}\ket{\psi_\mathrm{in}},
        \end{align}    
        and $\ket{\psi_\mathrm{in}}$ is chosen to match the initial conditions discussed above. In an interaction picture with respect to the kinetic energy $\mathcal{K}$ the asymptotic initial condition assumes the simpler form  $\ket{\psi^{I}(t)}=\exp(i\mathcal{K}t/\hbar)\ket{\psi(t)}\overset{t\rightarrow-\infty}{\longrightarrow}\ket{\psi_\mathrm{in}}$. 
        
        The problem we are going to address in Sec.~\ref{sec:BraggScatteringMatrix} is to solve the Schr{\"o}dinger equation for the time evolution operator $\mathcal{U}(t,t_0)$ in the interaction picture,
        \begin{subequations}
        \begin{align}
            i\hbar\frac{\mathrm{d}}{\mathrm{d}t}\mathcal{U}(t,t_0)&=\mathcal{H}^\mathrm{I}(t)\mathcal{U}(t,t_0),\label{eq:EoMI}\\
            \mathcal{H}^\mathrm{I}(t)&=\frac{\hbar\Omega(t)}{2}e^{-i\mathcal{K}t/\hbar}\left(e^{2i(k\hat{z}+\Lphase)}+e^{-2i(k\hat{z}+\Lphase)}\right)e^{i\mathcal{K}t/\hbar},\label{eq:HamIntPic}
        \end{align}
        \end{subequations}
        from which we construct the scattering (or transfer) matrix corresponding to the Bragg pulse:
        \begin{align}\label{eq:scatmat}
            \mathcal{S}=\lim_{\substack{t\rightarrow\infty\\t_0\rightarrow-\infty}}\mathcal{U}(t,t_0).
        \end{align}
        In order for the limits in Eq.~\eqref{eq:scatmat} to be well defined, it is important to consider the time evolution in the interaction picture, where the Hamiltonian \eqref{eq:HamIntPic} vanishes asymptotically for $t\rightarrow\pm\infty$. 
        
        The Bragg scattering matrix maps asymptotic incoming onto asymptotic outgoing wave packets:
        \begin{align}\label{eq:InOut}
            \ket{\psi_\mathrm{out}}=\mathcal{S}\ket{\psi_\mathrm{in}}.
        \end{align}
        In Sec.~\ref{sec:BraggScatteringMatrix} we will derive the general structure of the Bragg scattering matrix without making any further assumption regarding the pulse shape $\Omega(t)$. We will then determine the specific shape that the scattering matrix takes on when the Rabi frequency is changed adiabatically. For Gaussian pulses as in Eq.~\eqref{eq:GaussPulse} we show that any Bragg pulse achieving a decent quality actually falls in this regime.
        
    \subsection{Quality of a Bragg pulse}\label{sec:BraggDiffQuality}
        
        For motivation and as a further reference we first explain here how we quantify the quality of a Bragg pulse. The form of the scattering matrix corresponding to an ideal $N$th-order Bragg pulse is well known (see, e.g.,~\cite{Hogan2009}). It can be written as
        \begin{align}\label{eq:BraggScatteringMatrixIdeal}
            \mathcal{S}^\mathrm{ideal}_{\dPhi}=\int_b \difp \sum_{s,s'=\pm} [B_{\dPhi}]_{ss'}
            \ket{\vphantom{'} s N\hbar k+p}\bra{s'N\hbar k+p}.
        \end{align}
        where the states $\ket{\pm N\hbar k+p}$ are momentum eigenstates, and $p$ denotes a (quasi)momentum relative to $\pm N\hbar k$. For narrow atomic wave packets the integration with respect to $p$ can be effectively constrained to a bandwidth on the order of the photon momentum, $b=[-\hbar k/2,\hbar k/2]$. For a beam splitter pulse ($\dPhi=\pi/2$), 
        \begin{subequations}\label{eq:idealScattM}
            \begin{align} \label{eq:idealScattMA}
                B_{\pi/2} &= \frac{1}{\sqrt{2}} \mqty(1&- i e^{- i 2 N \Lphase}   \\- i e^{+ i 2 N \Lphase} &1),
            \end{align}
        and for a mirror  pulse ($\dPhi=\pi$),
            \begin{align} \label{eq:idealScattMB} 
                B_{\pi} &=  \mqty(0&- i e^{- i 2 N \Lphase}   \\- i e^{+ i 2 N \Lphase} &0),
            \end{align}
        \end{subequations}
       the form  of the scattering matrix derived in Sec.~\ref{sec:BraggScatteringMatrix} will reproduce the expressions in Eqs.~\eqref{eq:idealScattM} in an ideal, hypothetical limit. 
         \begin{figure*}[t]
                \includegraphics[width=0.98\textwidth]{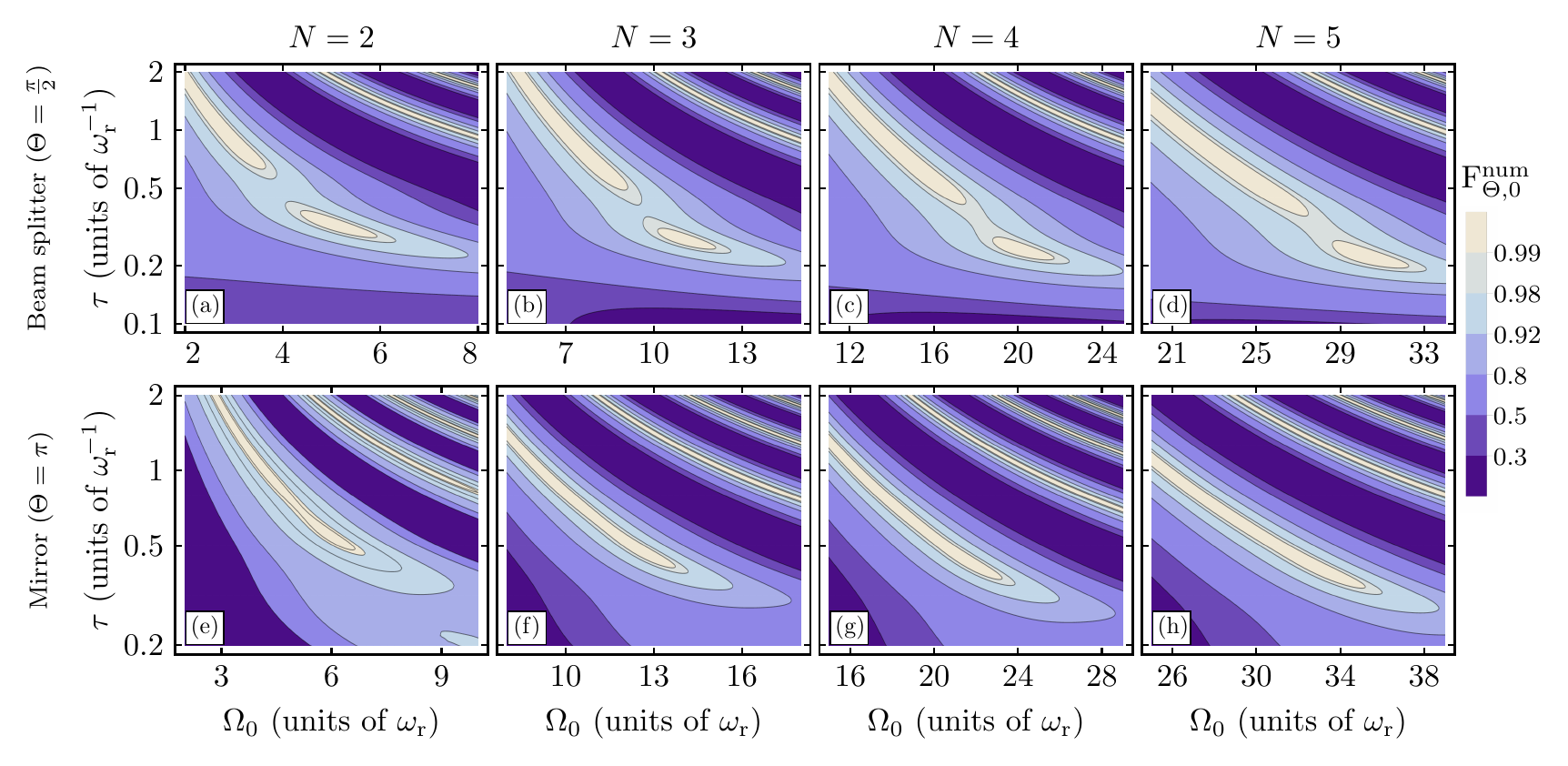}
                \caption{Numerically determined fidelities according to Eq.~\eqref{eq:fidelity2} of a single Gaussian quasi-Bragg pulse in case of a wave packet with vanishing momentum spread as a function of the peak Rabi frequency $\Omega_0$ and the temporal pulse width $\tspread$. Fidelities are depicted for beam splitters [top row, panels (a)-(d)] and mirrors [bottom row, panels (e)-(h)] of order $N=2,3,4,5$ (from left to right). Orders $N>5$ are not suitable for high-fidelity Bragg diffraction due to atom loss as pointed out in the main text. The parameters $(\Omega_0,\tspread)$ have been chosen to optimize the plot range for the pulse fidelities while maintaining experimentally relevant pulse durations for atomic clouds with finite momentum spread (see Sec.~\ref{sec:BSandMirror}). Quasi-Bragg beam splitting pulses feature a rich phenomenology that can be explained by LZ physics as we show in Sec. \ref{sec:BSandMirror}. For mirror pulses with longer pulse durations such features are less visible.}\label{fig:Bragg_Fid_NoNorm_Num}
            \end{figure*}  
        
        For the concrete case of an incoming atomic wave packet with average momentum $-N\hbar k$ and a narrow Gaussian envelope $g(p,\pspread)=(2\pi\pspread^2)^{-1/4}\exp(-p^2/4\pspread^2)$ of width $\pspread\ll\hbar k$ centered at $p=0$,
        \begin{align}
            |\psi_\mathrm{in}\rangle&=\int_b
            \difp\; g(p,\pspread)\ket{-N\hbar k+p},
        \end{align}
        the ideal outgoing state is accordingly $|\psi^\mathrm{ideal}_\mathrm{out,\dPhi}\rangle=\mathcal{S}^\mathrm{ideal}_{\dPhi}|\psi_\mathrm{in}\rangle$. For a Gaussian Bragg pulse with peak Rabi frequency $\Omega_0$ and temporal width $\tspread$ as in Eq.~\eqref{eq:GaussPulse} the true outgoing state is $|\psi_\mathrm{out}\rangle=\mathcal{S}(\Omega_0,\tspread)\ket{\psi_\mathrm{in}}$ where the scattering matrix $\mathcal{S}(\Omega_0,\tspread)$ denotes the limit in Eq.~\eqref{eq:scatmat} for the given pulse form.  
        
        We quantify the quality of a $\dPhi$-Bragg pulse by the fidelity between the ideal state and the true output state
        \begin{subequations}\label{eq:fidelities}
        \begin{align}\label{eq:fidelity1}
            \Fid_{\dPhi,\pspread}(\Omega_0,\tspread)&=\left|\langle\psi^\mathrm{ideal}_{\mathrm{out},\dPhi}
           |\psi_{\mathrm{out}}\rangle\right|^2.
        \end{align}
        In the limit of an infinitely narrow atomic wave packet,
        \begin{align}\label{eq:fidelity2}
        \Fid_{\dPhi,0}(\Omega_0,\tspread)&=\lim_{\pspread\rightarrow 0}\Fid_{\dPhi,\pspread}(\Omega_0,\tspread),
        \end{align}
        \end{subequations}
        we can infer the fidelity for the central momentum component $p=0$. Ultimately, the analytic approximation for the scattering matrix $\mathcal{S}(\Omega_0,\tspread)$ derived in Sec.~\ref{sec:BraggScatteringMatrix} will be gauged by comparing the corresponding analytic predictions for the fidelities \eqref{eq:fidelities} to the values $\Fid^\mathrm{num}_{\dPhi,\pspread}(\Omega_0,\tspread)$ and $\Fid^\mathrm{num}_{\dPhi,0}(\Omega_0,\tspread)$ inferred from numerical integrations of the Schr\"odinger equations. 
    
        In Fig.~\ref{fig:Bragg_Fid_NoNorm_Num} we display in anticipation the numerically determined exact fidelity $\Fid^\mathrm{num}_{\dPhi,0}(\Omega_0,\tspread)$ for beam splitter $(\dPhi=\pi/2)$ [Figs.~\ref{fig:Bragg_Fid_NoNorm_Num}(a)-(d)] and mirror $(\dPhi=\pi)$ [Figs.~\ref{fig:Bragg_Fid_NoNorm_Num}(e)-(h)] pulses for the Bragg orders $N=2,3,4,5$. 
        
        While our approach is applicable to arbitrary Bragg orders $N$, Fig.~\ref{fig:Bragg_Fid_NoNorm_Num} demonstrates the power requirements of increasing Bragg diffraction order. Szigeti et al.~\cite{Szigeti2012} show that Bragg pulses of orders $N>5$ suffer from substantial atom loss due to spontaneous emission rendering them unsuitable for state-of-the-art light pulse atom interferometers that rely on high-fidelity atom optics. Consequentially, LMT atom interferometry experiments have in the past either employed  single Bragg pulses of orders $N\leq 5$  \cite{Mueller2009PRL,Chiow2011,Mcdonald2013,Kovachy2012,Kovachy2015,Ahlers2016,Plotkin-Swing2018} or combined several ``low-order'' pulses sequentially \cite{Chiow2011,Kovachy2015,Ahlers2016,Plotkin-Swing2018,Gebbe2019}. For these reasons, we restrict our study to the experimentally relevant cases  $N\leq 5$. Details of the numerical treatment are given in Appendix~\ref{AppendixNumerics}. 
        
        We will show that the landscapes depicted in Fig.~\ref{fig:Bragg_Fid_NoNorm_Num} can be very well explained in terms of relatively simple formulas with concise physical interpretations. We emphasize that the fidelity is only used here as a figure of merit to demonstrate the quality of our approximation for the Bragg scattering matrix. Readers who are more interested in the physical results than in the technical details of our calculation can proceed directly to the summary of the next chapter of our paper in Sec.~\ref{sec:summary}.
    
    \section{Bragg Scattering Matrix}\label{sec:BraggScatteringMatrix}
    
        To solve the equation of motion \eqref{eq:EoMI} we will first use symmetries of the Hamiltonian \eqref{eq:Ham} to divide it into sub-blocks. This will greatly reduce the complexity of the problem and allow us to make quite general statements without assuming much about the specific shape $\Omega(t)$ of the Bragg pulse, and whether it operates in a diabatic or an adiabatic regime. To simplify the notation, we will suppress the explicit time dependence of the Rabi frequency $\Omega$ and the Hamiltonian $\mathcal{H}^\mathrm{MF}$ and restore the time argument only where necessary.
        
    \subsection{Hamiltonian in momentum basis}
    
        First, we exploit the well known property of the optical lattice potential to change the momentum of the atom only by a multiple of $2\hbar k$. If in the initial wave packet $N$ is an even (odd) number, then at a later point in time it will only consist of momentum components that are an even (odd) multiple of $\hbar k$.  This is formally reflected when the Hamiltonian is expanded in the momentum basis, and the momentum eigenstates $|n\hbar k+p\rangle$ are grouped into bins with $n\in\mathds{Z}$ and (quasi)momentum $p\in[-\hbar k/2,\hbar k/2]$. From now on, the momentum variable $p$ is always limited to this interval. Since efficient Bragg diffraction of atomic wave packets crucially depends on its narrow momentum width $\pspread \ll\hbar k $ \cite{Szigeti2012} we can further constrain the (quasi)momenta $p$ by assuming
        \begin{align}\label{eq:epsilon}
         \epsilon(p)=\frac{p}{\hbar k},
        \end{align} 
        to be a small parameter. This will allow us to solve the dynamics of the Bragg pulse for $p=0$ and take into account account first-order corrections in $p$ perturbatively. 
        
       The Hamiltonian in Eq.~\eqref{eq:Ham} then decomposes into blocks,
        \begin{align}\label{eq:splitHam}
            \mathcal{H}^\mathrm{MF}&=\int_{-\hbar k/2}^{\hbar k/2}\ \difp\left\{ \mathcal{H}^\mathrm{MF}_{\even}(p)+\mathcal{H}^\mathrm{MF}_{\odd}(p)\right\},
        \end{align}
        where the components $\mathcal{H}^\mathrm{MF}_{\alpha}(p)$ act on disjunct subspaces $\mathscr{H}_{p\alpha}=\mathrm{span}\{|n\hbar k+p\rangle\}_{n\in\mathds{Z}_\alpha}$  corresponding to even and odd momentum states for $\alpha=\even,\odd$, respectively. We denote the set of even and odd integers by  $\mathds{Z}_\even=2\mathds{Z}$ and $\mathds{Z}_\odd=2\mathds{Z}+1$. The total Hilbert space is $\mathscr{H}=\oplus_{p\alpha}\mathscr{H}_{p\alpha}$. Depending on whether $N$ of the initial mean momentum is even or odd, the dynamics of the $(n\hbar k+p)$-momentum components of the wave-packet are governed either by $\mathscr{H}^\mathrm{MF}_{\even}(p)$ or $\mathscr{H}^\mathrm{MF}_{\odd}(p)$. We will see that these Hamiltonians have a very similar structure, but still feature important differences. With the notation 
        \begin{align}\label{eq:sigma}
          \psigma{n,m}(p)\coloneqq|n\hbar k+p\rangle\langle m\hbar k+p|,
        \end{align}
        $n,m\in\mathds{Z}$, the components of the Hamiltonian in subspace $\mathscr{H}_{p\alpha}$ can be expressed as
        \begin{align}\label{eq:Halphap}
          \mathcal{H}^\mathrm{MF}_{\alpha}(p)&=\mathcal{K}_\alpha(p)+
          \sum_{n\in \mathds{Z}_\alpha}
          \frac{\hbar\Omega}{2}\left(e^{2i\Lphase}\psigma{n+2,n}(p)+\mathrm{H.c.}\right),\\
          \mathcal{K}_\alpha(p)&=\sum_{n\in \mathds{Z}_\alpha}\frac{(n\hbar k+p)^2}{2M}\psigma{n,n}(p),
        \end{align}
        for $\alpha=\even,\odd$. Appendix~\ref{App:Hamiltonian} provides some details of the derivation of this form of the Bragg Hamiltonian. We note that the symbols in Eq.~\eqref{eq:sigma} were already used by Shankar et al.~\cite{Shankar2019} in the context of atomic optics. We also would like to point out to the reader that (quasi)momentum variable $p$ is not identical, but closely related, to the quasimomentum in the sense of the Bloch theorem. Our choice is motivated by the fact that it naturally provides us with the decomposition of the Hilbert space into even and odd momentum bins in Eq.~\eqref{eq:splitHam} and allows us to describe the dynamics of even and odd diffraction orders equally. We further discuss the connection to the Bloch quasimomentum in Sec.~\ref{sec:Comparison}.
        
        It will be useful to expand the kinetic energy in two terms, $\mathcal{K}_\alpha(p)=\mathcal{L}_\alpha(p)+\mathcal{M}_\alpha(p)$, where the last term collects the components of the kinetic energy which are linear and quadratic in the (quasi)momentum variable $p$. That is,
        \begin{align}
          \mathcal{L}_\alpha(p)&=\sum_{n\in \mathds{Z}_\alpha}\hbar\omega_\mathrm{r} n^2 \psigma{n,n}(p),\label{eq:kineticL} \\
          \mathcal{M}_\alpha(p)&=\sum_{n\in \mathds{Z}_\alpha}\left(2\hbar\omega_\mathrm{r} n\epsilon(p)+ \frac{p^2}{2M}\right)\psigma{n,n}(p),
          \label{eq:kineticD}
        \end{align}
        where the recoil frequency is $\omega_\mathrm{r}=\hbar k^2/2M$ and we have used $\epsilon(p)$ introduced in Eq.~\eqref{eq:epsilon}.
        
        We now move to an interaction picture with respect to the term $\mathcal{M}_\alpha(p)$. In this picture the asymptotic initial condition in Eq.~\eqref{eq:asymp} becomes 
        \begin{align}\label{eq:inicond}
            \ket{\psi(t)}\overset{t\rightarrow-\infty}{\longrightarrow}e^{i\mathcal{M}_\alpha(p)t/\hbar}e^{-i\mathcal{K}t/\hbar}\ket{\psi_\mathrm{in}}=e^{-iN^2\omega_\mathrm{r} t}\ket{\psi_\mathrm{in}},
        \end{align}
        where we used that the initial state $\ket{\psi_\mathrm{in}}$ is localized in the momentum bin around $-N\hbar k$. The Hamiltonian in this interaction picture is
        \begin{align}\label{eq:Halphap1}
        \begin{split}
           \mathcal{H}_{\alpha}(p)&=\sum_{n\in \mathds{Z}_\alpha}\hbar\omega_\mathrm{r} n^2 \psigma{n,n}(p)\\
          &\quad+\frac{\hbar\Omega}{2}\left(e^{2i(\Lphase+2\epsilon(p)\omega_\mathrm{r}t)}\psigma{n+2,n}(p)+\mathrm{H.c.}\right). 
        \end{split}
        \end{align}
        The time dependence in the lattice potential reflects the Doppler shift of the two counterpropagating lattice beams seen by the components of the wave packet with (quasi)momentum $p$ in $\mathscr{H}_{p\alpha}$. 
        
        It is straightforward to check that if the unitary operator $\mathcal{V}_\alpha(p,t, t_0)$ on the subspace $\mathscr{H}_{p\alpha}$ is a solution of
        \begin{align}\label{eq:Vunitary}
            i\hbar\frac{\mathrm{d}}{\mathrm{d}t}\mathcal{V}(p,t,t_0)=\mathcal{H}_\alpha(p,t)\mathcal{V}(p,t,t_0),
        \end{align}
        then the time evolution operator solving Eq.~\eqref{eq:EoMI} on the same subspace is
        \begin{align}\label{eq:Usolve}
            \mathcal{U}_\alpha(p,t,t_0)=\exp\left(i\mathcal{L}_\alpha(p)(t-t_0)/\hbar\right)\mathcal{V}_\alpha(p,t, t_0).
        \end{align}
        Our strategy will be to solve Eq.~\eqref{eq:Vunitary}, and use this solution to construct the scattering matrix \eqref{eq:scatmat} using Eq.~\eqref{eq:Usolve}. So far, no approximation has been made.
        
        We now use that the initial state is a narrow wave packet with a momentum spread $\pspread\ll \hbar k$ which amounts to ${\abs{ \epsilon(p)}\ll1}$ for all (quasi)momentum components of the wave packet. Furthermore, we assume now that for the duration $\tspread$ of the Bragg pulse we have $N\omega_\mathrm{r}\tspread\pspread\ll \hbar k$ for $N$th-order Bragg scattering. With this assumption we can expand the time-dependent phase in Eq.~\eqref{eq:Halphap1} to first order in $\epsilon(p)$. Collecting the terms of zeroth and first order in $\epsilon(p)$ one finds
        \begin{align}\label{eq:fullHalpha}
         \mathcal{H}_{\alpha}(p)&=H_{\mathrm{\alpha}}(p)+\epsilon(p) V_{\mathrm{\alpha}}(p),
        \end{align}
        where
        \begin{subequations}\label{eq:HalphaValpha}
        \begin{align}
          H_{\alpha}(p)&=\sum_{n\in \mathds{Z}_\alpha} \left\{\hbar \omega_\mathrm{r} n^2 \psigma{n,n}(p)
          +\frac{\hbar\Omega}{2}\left(e^{2i\Lphase}\psigma{n+2,n}(p)+\mathrm{H.c.}\right)\right\},\label{eq:Halpha0}
          \\
          V_{\alpha}(p)&=i2\hbar\Omega\omega_\mathrm{r} t\sum_{n\in \mathds{Z}_\alpha} \left(e^{2i\Lphase}\psigma{n+2,n}(p)-\mathrm{H.c.}\right)\label{eq:H1}.
        \end{align}
        \end{subequations}
         We recall that the Hamiltonian $\mathcal{H}_{\alpha}(p)$ acts on the subspace $\mathscr{H}_{p\alpha}$. Its components $H_{\alpha}(p)$ and $V_{\alpha}(p)$ in Eqs.~\eqref{eq:HalphaValpha} are structurally identical for all (quasi)momentum $p$. It is just the strength $\epsilon(p)$ of the perturbation $V_{\alpha}(p)$ in Eq.~\eqref{eq:fullHalpha} due to the Doppler shift which has a nontrivial dependence on the (quasi)momentum $p$. In the next sections we will consider only the zeroth-order Hamiltonian \eqref{eq:Halpha0}. The perturbation \eqref{eq:H1} will be treated later on in Sec.~\ref{sec:Doppler}.
     
     \subsection{Hamiltonian in basis of symmetric and antisymmetric states}
    
        Within each subspace $\mathscr{H}_{p\alpha}$ we introduce a new basis which consists of (anti)symmetric states $\asket{\pm}{n}{p}$ defined by
        \begin{subequations}\label{eq:SymAsymEV}
        \begin{align}\label{eq:SymAsymEVa}
        \asket{\pm}{n}{p}\coloneqq\frac{1}{\sqrt{2}}\left(e^{in\Lphase}|n\hbar k+p\rangle\pm e^{-in\Lphase}|-n\hbar k+p\rangle\right),
        \end{align}
        for $n\in\mathds{N}/0$. We recall that $\Lphase$ is the laser phase. For $n=0$ there is a single state in $\mathscr{H}_{p\even}$:
        \begin{align}
            \asket{+}{0}{p}\coloneqq|p\rangle.
        \end{align}
        \end{subequations}
        The subspaces of symmetric and antisymmetric states in $\mathscr{H}_{p\alpha}$ are $\mathscr{H}_{p\alpha\pm}=\mathrm{span}\{\asket{\pm}{n}{p}\}_{n\in\mathds{Z}_\alpha}$, and the total Hilbert space is $\mathscr{H}=\oplus_{p\alpha\pm}\mathscr{H}_{p\alpha\pm}$. 
        When the Hamiltonian $H_{\alpha}$ in Eq.~\eqref{eq:Halpha0} is expressed in this new basis it decomposes further into a sum of two terms,
        \begin{align}\label{eq:Halpha}
            H_{\alpha}&=H_{\alpha+}+H_{\alpha-},
        \end{align}
        which act on the disjunct spaces $\mathscr{H}_{p\alpha\pm}$. 
        
        Before we explicitly construct the components $H_{\alpha\pm}$ we give an argument for why $H_{\alpha}$ has to be block diagonal in the basis of (anti)symmetric states. Consider the Hermitian operator   
        \begin{align}
            \Pi&\coloneqq\sum_{\alpha=\even,\odd}\int_{-\hbar k/2}^{\hbar k/2} \difp \Pi_{\alpha}(p) \\
            \Pi_{\alpha}(p)&=\sum_{n\in\mathds{Z}_\alpha} e^{2ni\Lphase}\psigma{n,-n}(p)
        \end{align}
        which fulfills $\Pi^2=\mathds{1}$. Its eigenvalues are $\pm1$, and the corresponding eigenvectors are the (anti)symmetric states, $\Pi\asket{\pm}{n}{p}=\pm\asket{\pm}{n}{p}$. It is straightforward to show that $H_{\alpha}$ in Eq.~\eqref{eq:Halpha0} is invariant under conjugation with $\Pi$, that is, $\Pi H_{\alpha}\Pi=H_{\alpha}$. Therefore, the commutator of these two operators vanishes, $[\Pi,H_{\alpha}]=0$, and $H_{\alpha}$ cannot couple states corresponding to different eigenvalues with respect to $\Pi$. In other words, $H_{\alpha}$ has to be block diagonal as in Eq.~\eqref{eq:Halpha}. We note that $\Pi$ is connected to reflections in momentum space, but is not equivalent to the parity operator. Setting the laser phase to zero, $\Lphase=0$, the operators $\Pi_{\alpha}(p)$ generate reflections in momentum space about (quasi)momentum $p$ in $\mathscr{H}_{p\alpha}$. The symmetry we are exploiting here will ultimately be broken by the Doppler detuning \eqref{eq:H1}. However, it will be almost conserved for sufficiently narrow initial wave packets and perturbation theory will be well suited to account for the effects of Doppler-induced breaking of this symmetry. We note that the basis of (anti)symmetric states in Eqs.~\eqref{eq:SymAsymEV} has been used recently also to analyze Bloch oscillations \cite{Pagel2019}.

        In order to identify the components $H_{\alpha\pm}$ of the Hamiltonian in Eq.~\eqref{eq:Halpha} we define, in correspondence to \eqref{eq:sigma},
        \begin{align}\label{eq:sigmabeta}
          \psigma{n,m}^{\pm}(p)\coloneqq\asket{\pm}{n}{p}\asbra{\pm}{m}{p}.
        \end{align}
        Here, $n$ and $m$ are non-negative integers and the operator $\psigma{n,m}^{-}(p)$ acting on the antisymmetric subspace is defined only for $n,m\neq 0$. The change to the basis of (anti)symmetric states is straightforward. A number of useful relations are given in the Appendix~\ref{App:Hamiltonian}. The result of this transformation is different for the Hamiltonian acting on the even and the odd subspace, that is, for Bragg scattering of even or odd order $N$. One finds for the Hamiltonians acting on the even subspaces $\mathscr{H}_{p\even\pm}$
        \begin{subequations}\label{eq:Hblocks}
        \begin{align}
        \begin{split}
          H_{\even-}&=\sum_{\substack{n\in \mathds{N}_\even\\ n\neq 0}}\left\{\hbar\omega_\mathrm{r} n^2 \psigma{n,n}^{-}+\frac{\hbar\Omega}{2}\left(\psigma{n+2,n}^{-}+\mathrm{H.c.}\right)\right\}\\
          H_{\even+}&=\sum_{\substack{n\in \mathds{N}_\even\\ n\neq 0}}\left\{\hbar\omega_\mathrm{r} n^2 \psigma{n,n}^{+}+\frac{\hbar\Omega}{2}\left(\psigma{n+2,n}^{+}+\mathrm{H.c.}\right)\right\}
          \\&\quad
          +\frac{\hbar\Omega}{\sqrt{2}}\big(\psigma{2,0}^{+}+\mathrm{H.c.}\big),
          \label{eq:Heven}
         \end{split}
        \end{align}
        and one finds for the Hamiltonians acting on the odd subspace $\mathscr{H}_{p\odd\pm}$
        \begin{align}
          H_{\odd\pm}&=\sum_{n\in \mathds{N}_\odd} \left\{\hbar\omega_\mathrm{r} n^2 \psigma{n,n}^{\pm}+\frac{\hbar\Omega}{2}\left(\psigma{n+2,n}^{\pm}+\mathrm{H.c.}\right)\right\}
          \pm\frac{\hbar\Omega}{2}\psigma{1,1}^{\pm}.\label{eq:Hodd}
        \end{align}
        \end{subequations}
        Thus in both even and odd subspaces $\mathscr{H}_{p\alpha}$ the symmetric and antisymmetric subspaces $\mathscr{H}_{p\alpha\pm}$ decouple in zeroth order of the Doppler detuning, as expected.
    
        In writing the Hamiltonians~\eqref{eq:Hblocks} we have suppressed the (quasi)momentum $p$ in all arguments. This can be done without loss of information, since all these Hamiltonians, just like the Hamiltonian of zeroth order in Eq.~\eqref{eq:Halpha0}, are structurally identical for all (quasi)momenta $p$. To simplify the notation, we therefore adhere to the following convention in this and all subsequent sections dealing exclusively with the zeroth-order Hamiltonian: The argument of $\psigma{n,m}^{\pm}$ and $\hat{\sigma}_{m,n}$ -- as well as of all operators composed thereof -- is $p$ everywhere, unless stated otherwise. We will also suppress the momentum $p$ in writing the basis vectors
        \begin{align*}
            \asket{\pm}{n}{p}\equiv \pasket{\pm}{n},
        \end{align*}
        and explicitly state the momentum $p$ as an argument again, when we treat Doppler detuning in Sec.~\ref{sec:Doppler}.
        
        In both even and odd subspaces the Hamiltonians in Eqs.~\eqref{eq:Heven} and \eqref{eq:Hodd} for the symmetric and antisymmetric subspace are very similar, but still show important differences: In the even subspace $\mathscr{H}_{p\even}$ the symmetric $(+)$ subspace  contains the state $\pasket{+}{0}$, while no such state exists for the antisymmetric $(-)$ subspace. As a consequence, the Rabi frequency of the coupling between the states $\pasket{+}{0}$ and $\pasket{+}{2}$ [see last term in Eq.~\eqref{eq:Heven}] is larger by a factor of $\sqrt{2}$ than the Rabi frequency in the coupling of other levels $\pasket{\pm}{2n}\leftrightarrow\pasket{\pm}{2n+2}$ for $n>0$. In order to make this more transparent, and for later reference, we give here a truncated representation of the Hamiltonians in the basis  $(\pasket{-}{6},\pasket{-}{4},\pasket{-}{2})$ for $H_{\even-}$, and $(\pasket{+}{6},\pasket{+}{4},\pasket{+}{2},\pasket{+}{0})$ for $H_{\even+}$:
         \begin{subequations}\label{eq:Htrunc}
         \begin{align}\label{eq:Heventrunc}
            \begin{split}
             H_{\even-}&=\hbar\omega_\mathrm{r}
             \begin{pmatrix}
             36 & w & 0 \\
             w & 16 & w \\
             0 & w & 4 
             \end{pmatrix},\\
             H_{\even+}&=\hbar\omega_\mathrm{r}
             \begin{pmatrix}
             36 & w & 0 & 0 \\
             w & 16 & w & 0 \\
             0 & w & 4 & \sqrt{2} w \\
             0 & 0 & \sqrt{2} w & 0
             \end{pmatrix},
             \end{split}
         \end{align}
         where we used $w=\Omega/2\omega_\mathrm{r}$.
        
        In the odd subspace $\mathscr{H}_{p\odd}$ the levels $\pasket{\pm}{1}$ have energies $\hbar\omega_\mathrm{r}\pm \frac{\hbar\Omega}{2}$ shifted proportionally to the Rabi frequency in opposite directions for the symmetric and the antisymmetric subspace [see last term in Eq.~\eqref{eq:Hodd}]. The energies of higher lying levels $\pasket{\pm}{2n+1}$ for $n>0$ are independent of the Rabi frequency. In a truncated basis $(\pasket{\pm}{7},\pasket{\pm}{5},\pasket{\pm}{3},\pasket{\pm}{1})$ one finds,
         \begin{align}\label{eq:Hoddtrunc}
             H_{\odd\pm}&=\hbar\omega_\mathrm{r}
             \begin{pmatrix}
             49 & w & 0 & 0 \\
             w & 25 & w & 0 \\
             0 & w & 9 &  w \\
             0 & 0 & w & 1\pm w
             \end{pmatrix}.
         \end{align}
         \end{subequations}
    
        After transforming the Hamiltonian to the basis of (anti)symmetric states we also have to consider, how the initial condition in Eq.~\eqref{eq:inicond} reads in this basis. An initial wave packet $\ket{\psi_\mathrm{in}}$ composed of momentum states around an average momentum $-N\hbar k$ corresponds to an odd super\-position of states in the symmetric and the antisymmetric subspace,  $\ket{-N\hbar k+p}=\exp(iN\Lphase)(\asket{+}{N}{p}-\asket{-}{N}{p})/\sqrt{2}$. If we were to perform, e.g. a mirror pulse transferring a momentum $2N\hbar k$ to the atom, the challenge is to change this state into the even superposition $\exp(-iN\Lphase)(\asket{+}{N}{p}+\asket{-}{N}{p})/\sqrt{2}=\ket{N\hbar k+p}$. This intuition is expressed more formally in terms of the scattering matrix. 
    
    \subsection{General structure of the Bragg scattering matrix}    
    
        Based on the decomposition of the Bragg Hamiltonian into its sub-blocks \eqref{eq:Hblocks} we will now determine the scattering matrix $\eqref{eq:scatmat}$ for a Bragg pulse. To zeroth order in the Doppler detuning, the dynamics in the subspace $\mathscr{H}_{p\alpha}$ is governed by the Hamiltonian $H_{\alpha}(t)$ in Eq.~\eqref{eq:Halpha} which is block-diagonal in the subspaces $\mathscr{H}_{p\alpha\pm}$. Therefore, the unitary evolution operator will be of the form
        \begin{align}\label{eq:BraggScatteringMatrixFull}
             U_{\alpha}(t,t_0)&=U_{\alpha+}(t,t_0)+U_{\alpha-}(t,t_0),
        \end{align}
        where $U_{\alpha\pm}(t,t_0)$ acts on $\mathscr{H}_{p\alpha\pm}$ only, and fulfills the Schr{\"o}\-dinger equation
        \begin{align}\label{eq:dynamics}
            i\hbar\frac{\mathrm{d}}{\mathrm{d}t}{U}_{\alpha\pm}(t,t_0)&=H_{\alpha\pm}(t)U_{\alpha\pm}(t,t_0).
        \end{align}
        In zeroth order of Doppler detuning, that is, in zeroth order of $\epsilon(p)$, the formal solution \eqref{eq:BraggScatteringMatrixFull} provides already the solution to Eq.~\eqref{eq:Vunitary}. Using $\mathcal{V}_\alpha(t,t_0)=U_{\alpha}(t,t_0)$ in Eq.~\eqref{eq:Usolve}, we find that the Bragg scattering matrix from Eq.~\eqref{eq:scatmat} on the subspace $\mathscr{H}_{p\alpha}$ is
        \begin{align}
            \mathcal{S}_\alpha&=\lim_{\substack{t\rightarrow\infty\\ t_0\rightarrow-\infty}}\exp\left(i\mathcal{L}_\alpha(t-t_0)/\hbar\right){U}_\alpha(t, t_0)
            =\mathcal{S}_{\alpha+}+\mathcal{S}_{\alpha-}.\label{eq:ScattMatLimit}
        \end{align}
        The block diagonal structure of the formal solution \eqref{eq:BraggScatteringMatrixFull} and the diagonal form of $\mathcal{L}_\alpha$ [cf. Eq.~\eqref{eq:kineticL}] imply that the scattering matrix is also block diagonal in the (anti)symmetric basis. 
        
        Single $N$th-order Bragg diffraction pulses are supposed to couple the momentum eigenstates in the incoming wave packet $\ket{- N \hbar k + p} \longleftrightarrow \ket{ N \hbar k + p}$ (for $N>0$), and ideally execute $\pi/2$ or $\pi$ pulses in this two-dimensional subspace. What ultimately enters in an interferometer sequence is not the full Bragg scattering matrix of Eq.~\eqref{eq:ScattMatLimit}, but rather its projection into this two-dimensional subspace. In terms of the basis of (anti)symmetric states this subspace is spanned by the states $\pasket{\pm}{N}$, see Eq.~\eqref{eq:SymAsymEVa}. Due to the block-diagonal structure, the projection of the scattering matrix in \eqref{eq:ScattMatLimit} yields a diagonal matrix in the basis $(\pasket{+}{N},\pasket{-}{N})$:
        \begin{align}
            \mathcal{S}_\alpha=
            \sum_{s,s'=\pm} S_{ss'}
            \pasket{ s }{N}\pasbra{s'}{N},\nonumber\\
            S  =\begin{pmatrix} e^{-i\SASphase_{N+}-\gamma_{N+}} & 0 \\ 0 & e^{-i\SASphase_{N-}-\gamma_{N-}} \end{pmatrix},\label{eq:ProjectedBragg}
        \end{align}
        where 
        \begin{align}\label{eq:gammaphi}
        e^{-i\SASphase_{N\pm}-\gamma_{N\pm}}&=\pasbra{\pm}{N}\mathcal{S}_{\alpha\pm}\pasket{\pm}{N}\nonumber\\
        &=\lim_{\substack{t\rightarrow\infty\\ t_0\rightarrow-\infty}}
        e^{iN^2\omega_\mathrm{r}(t-t_0)}\pasbra{\pm}{N}U_{\alpha\pm}(t,t_0)\pasket{\pm}{N}.
        \end{align}
        Parameters $\SASphase_{N\pm}$ and $\gamma_{N\pm}$ describe scattering phases and population loss from the states $\pasket{\pm}{N}$. Since the scattering matrices $\mathcal{S}_{\alpha\pm}$ are unitary, we have $\gamma_{N\pm}\geq0$. It is important to note that the general form of the scattering matrix $S$ applies regardless of the exact shape $\Omega(t)$ of the Bragg pulse. Moreover, it is instructive to write the projected Bragg scat\-tering matrix \eqref{eq:ProjectedBragg} in the basis of momentum states $(\ket{+N\hbar k+p},\ket{-N\hbar k+p})$. The transformation from the (anti)symmetric states $\pasket{\pm}{N}$ to momentum states can be read off from Eqs.~\eqref{eq:SymAsymEV}: 
        \begin{align}\label{eq:Transform}
            T=\frac{1}{\sqrt{2}}\begin{pmatrix} e^{iN\Lphase} & e^{-iN\Lphase} \\ e^{iN\Lphase} & -e^{-iN\Lphase} \end{pmatrix}.
        \end{align}
        With Eq.~\eqref{eq:ProjectedBragg} one finds the projected Bragg scattering matrix in the momentum basis, $B\coloneqq T^\dagger S T$, which evaluates to
        \begin{multline}\label{eq:GeneralScattMatrix}
            B(\glP-i\sumLoss,\dPhi-i\diffLoss)\\
            = e^{-i \frac{\glP-i\sumLoss}{2}}\mqty(\cos{\left(\frac{\dPhi-i\diffLoss}{2}\right)}&- ie^{-i2N \Lphase} \sin{\left(\frac{\dPhi-i\diffLoss}{2}\right)}\\- ie^{+i2N \Lphase} \sin{\left(\frac{\dPhi-i\diffLoss}{2}\right)}&\cos{\left(\frac{\dPhi-i\diffLoss}{2}\right)}).
        \end{multline}
        We define the differential phase between the symmetric and the antisymmetric state $\pasket{\pm}{N}$ and the global phase imprinted on this subspace, 
        \begin{align}\label{eq:phases}
            \dPhi &= \SASphase_{N+}-\SASphase_{N-}, &
            \glP &= \SASphase_{N+}+\SASphase_{N-},    
        \end{align}
        and the corresponding parameters characterizing differential and total loss:
        \begin{align}\label{eq:losses}
                \diffLoss  &= \gamma_{N+} -\gamma_{N- }, &
                \sumLoss  &= \gamma_{N+} +\gamma_{N- }.
        \end{align}
        We remind the reader that $\Lphase$ denotes the relative laser phase between the two light fields generating the optical lattice. We also note that the global phase $\Phi$ should not be confused with the global phase $\Phi_{\mathcal{G}}$ which includes the average ac Stark shift and has been gauged out in the picture of the fundamental Hamiltonian \eqref{eq:Ham}. 
     
        Comparing the scattering matrix in Eq.~\eqref{eq:GeneralScattMatrix} to the ones of an ideal beam splitter or mirror pulse, as given in Eqs.~\eqref{eq:idealScattM}, we can identify conditions to achieve high-quality pulse operations: First of all, the differential phase collected between symmetric and asymmetric subspace needs to be tuned to $\dPhi = \pi/2$ for a beam splitter and to $\dPhi = \pi$ for a mirror pulse. We thus see that the differential phase $\dPhi$ turns out to be identical to what is usually referred to as the pulse area. The global phase $\glP$ does not necessarily have to be nulled in order to achieve a good pulse quality, but it must be controlled and included in the phase budget of an interferometer. Finally, to maintain the population in the subspace $\pasket{\pm}{N}$ and avoid losses to other momentum states, ideally the condition $\gamma_{N\pm}=0$ should be fulfilled. In view of Eq.~\eqref{eq:gammaphi} this is tantamount to
        \begin{align}\label{eq:unitariescond}
        \lim_{\substack{t\rightarrow\infty\\ t_0\rightarrow-\infty}}|\pasbra{\pm}{N}U_{\alpha\pm}(t,t_0)\pasket{\pm}{N}|=1,
        \end{align}
        where the unitaries $U_{\alpha\pm}(t,t_0)$ are the solutions to the Schr{\"o}\-dinger Eqs.~\eqref{eq:dynamics}. Thus, in both the symmetric and the antisymmetric subspace an initial population of $\pasket{\pm}{N}$ ultimately has to return to this state. This presents a highly nontrivial constraint in view of the fact that the  Hamiltonians $H_{\alpha\pm}(t)$ in these two subspaces differ structurally but are controlled through the same Rabi frequency $\Omega(t)$. 
        
        The challenge is to identify a pulse $\Omega(t)$ that meets all of these requirements. As we will establish in the next section, a sufficient condition on $\Omega(t)$ for achieving this is that the Rabi frequency is tuned adiabatically in the sense of the adiabatic theorem: thereby the initial population of $\pasket{\pm}{N}$ is maintained at all times in a corresponding instantaneous energy eigenstate of $H_{\alpha\pm}(t)$, and is thus perfectly restored to $\pasket{\pm}{N}$ at the end of the pulse, satisfying Eq.~\eqref{eq:unitariescond}. As an ideal adiabatic tuning requires infinitely long pulse durations, it is important to consider also effects of nonadiabaticity, and to determine the impact of a finite pulse duration on the Bragg pulses. We do so in Secs.~\ref{sec:LandauZener} and \ref{sec:LandauZenerLosses}. In Sec.~\ref{sec:BSandMirror} we will show for the specific but most relevant case of a Gaussian pulse [see Eq.~\eqref{eq:GaussPulse}] that each pair of peak Rabi frequencies $\Omega_0$ and pulse durations $\tspread$ leading to a high-quality Bragg $\pi/2$ or $\pi$ pulse with losses at an acceptable level does indeed correspond to adiabatic dynamics with first-order nonadiabatic corrections. Thus, for Gaussian pulses adiabaticity in the sense of the adiabatic theorem is a necessary and sufficient condition. It is interesting, but outside the scope of this paper, to ponder whether a nonadiabatic Bragg pulse $\Omega(t)$ , i.e., a pulse that produces real transitions among the instantaneous energy eigenstates of $H_{\alpha\pm}(t)$, can at all give rise to high-quality atom optics operations. 
        
    \subsection{Scattering matrix for adiabatic Bragg pulse}
       
\begin{figure}[ht]
            \includegraphics[width=0.4\textwidth]{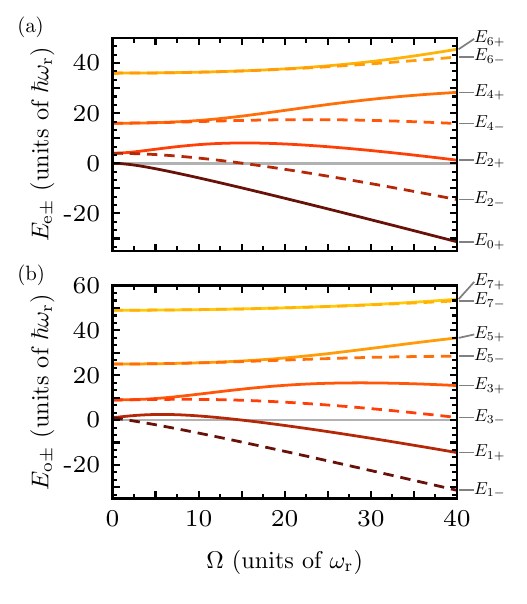}
            \caption{Spectra of the lowest-energy eigenstates of the Hamiltonians in the (anti)symmetric subspaces vs Rabi frequency $\Omega$ in (a) the even subspace, $H_{\even\pm}$ in Eq.~\eqref{eq:Heven}, and (b) the odd subspace, $H_{\odd\pm}$ in Eq.~\eqref{eq:Hodd}, with truncations $n_\mathrm{max,e}=8$ and $n_\mathrm{max,o}=11$, respectively. The range for $\Omega$ includes Rabi frequencies required for high-fidelity quasi-Bragg pulses up to order $N=5$ as depicted in Fig. \ref{fig:Bragg_Fid_NoNorm_Num}.  }\label{fig:SpectraOmegaScan}
        \end{figure}

        We consider now the important special case of an adiabatic tuning of the Rabi frequency $\Omega(t)$. As shown in Fig.~\ref{fig:SpectraOmegaScan}, the energy spectrum of the Hamiltonians $H_{\alpha\pm}$ is nondegenerate for any value of $\Omega$, and no level crossing occurs. This means that the quantum numbers labeling the eigenstates $\pasket{\pm}{n}$ corresponding to eigenenergies $n^2\hbar\omega_\mathrm{r}$ for vanishing Rabi frequency, $\Omega=0$, remain good quantum numbers also for $\Omega\neq0$. We note here that this is only the case because we are working in an interaction picture with respect to the Doppler shift term in Eq.~\eqref{eq:kineticD}.
        
        For a time-dependent Rabi frequency $\Omega(t)$ we denote the instantaneous eigenstates and eigenenergies by
        \begin{subequations}\label{eq:eigenvalueproblem}
        \begin{align}
            H_{\alpha\pm}(t)\tasket{\pm}{n}{p}{t}&=E_{n\pm}(t)\tasket{\pm}{n}{p}{t}.
        \end{align}
        with $n\in\mathds{N}_\alpha$ and $n>0$. For $\alpha=\even$ and $n=0$ there is only one eigenstate, 
        \begin{align}
            H_{\even+}(t)\tasket{+}{0}{p}{t}&=E_{0+}(t)\tasket{+}{0}{p}{t}.
        \end{align}
        \end{subequations}
        In the asymptotic limits, where $\lim_{t\rightarrow\pm\infty}\Omega(t)=0$, we have
        \begin{align}\label{eq:energystateslimit}
            \lim_{t\rightarrow\pm\infty}\tasket{\pm}{n}{p}{t}=\pasket{\pm}{n}.    
        \end{align}
        The instantaneous eigenstates and eigenenergies can be calculated from Eqs.~\eqref{eq:Heven} and \eqref{eq:Hodd} for a given Rabi frequency $\Omega$ with a suitable truncation of the Hilbert space. Due to the block decomposition of the Hamiltonian excellent results can be achieved for a relatively low order of truncation, as will be seen in Sec.~\ref{sec:BSandMirror}. In the following, we will write all results in a form which only requires the numerical calculation of instantaneous energy eigenvalues, which is an efficient subroutine. The much more laborious calculation of energy eigenstates can be avoided by suitable approximations. 
      
        The adiabatic theorem states that for an infinitely slow tuning, that is, for an infinitely long pulse $\tspread\rightarrow\infty$, no transitions among the energy eigenstates of $H_{\alpha}(t)$ occur. Thus, the ideal adiabatic solution to Eq.~\eqref{eq:dynamics} is
        \begin{subequations}\label{eq:unitary}
        \begin{align}
             U_{\alpha\pm}(t,t_0)&=\sum_{n\in\mathds{N}_\alpha}
             e^{-i\SASphase_{n\pm}(t,t_0)}\tasket{\pm}{n}{p}{t}\tasbra{\pm}{n}{p}{t_0},\label{eq:Usol}
        \end{align}
        with dynamic phases
        \begin{align}
             \SASphase_{n\pm}(t,t_0)&=\frac{1}{\hbar}\int_{t_0}^{t}\mathrm{d}t_1 E_{n\pm}(t_1).
        \end{align}
        \end{subequations}
        In the present case, in which the Hamiltonian depends on time only via a single parameter $\Omega(t)$, no geometric phase can occur. 
        
        In the ideal adiabatic regime and to zeroth order in Doppler detuning, Bragg diffraction simply imprints phases on the (anti)symmetric states $\pasket{\pm}{n}$. With Eqs.~\eqref{eq:unitary} and \eqref{eq:energystateslimit} the limit in Eq.~\eqref{eq:gammaphi} yields a unitary scattering matrix $S$ with $\gamma_{N\pm}=0$ and dynamic scattering phases $\SASphase_{N\pm}=\SASphase^\mathrm{dyn}_{N\pm}$ with
        \begin{align}\label{eq:BraggPhases}
            \SASphase^\mathrm{dyn}_{N\pm}&=\frac{1}{\hbar}\int_{-\infty}^{\infty}\mathrm{d}t\left( E_{N\pm}(t)-\hbar\omega_\mathrm{r} N^2\right).
        \end{align}
        From the structure of the Hamiltonians $\eqref{eq:Hblocks}$ it is clear that any differential phase between symmetric and antisymmetric subspace can only arise by coupling the incoming momentum states $\pasket{\pm}{N}$ to the lowest states in the spectrum of $H_{\alpha\pm}$, since these Hamiltonians differ only there. From the spectrum shown in Fig.~\ref{fig:SpectraOmegaScan} it is also evident that a suitable energy splitting between the states $\pasket{\pm}{N}$ for practical peak Rabi frequencies $\Omega_0$ is only possible for low orders of Bragg diffraction. E.g., one can expect that, for $\Omega_0\lesssim 40\omega_\mathrm{r}$, Bragg diffraction will be efficient only for $N\leq 5$, which can already be inferred from Fig.~\ref{fig:Bragg_Fid_NoNorm_Num}. As explained in Sec.\ref{sec:BraggDiffQuality} the loss of atoms due to spontaneous emission is setting an effective limitation for the Rabi frequency given a certain threshold above which losses can no longer be tolerated~\cite{Szigeti2012}.
    
        For later reference it will be useful to rewrite the dynamic phase as
        \begin{align}
            \SASphase^\mathrm{dyn}_{N\pm}&=\tspread\omega_\mathrm{r} x_{N\pm}(\Omega_0), \label{eq:BraggPhasesNum}\\ x_{N\pm}(\Omega_0)&=\int_{-\infty}^{\infty}\mathrm{d}\zeta\left( \frac{E_{N\pm}(\zeta\tspread)}{\hbar\omega_\mathrm{r}} - N^2\right) \label{eq:BraggPhasesW}
        \end{align}
        where $x_{N\pm}(\Omega_0)$ is a dimensionless quantity which depends on the exact pulse form and in particular on the peak Rabi frequency and which we display in the top Figs.~\ref{fig:xyz_Parameters}(a)-(d). We also introduced a dimensionless time $\zeta=t/\tspread$ for a characteristic pulse duration $\tspread$. We will see that the dynamic phases \eqref{eq:BraggPhasesNum} largely capture the physics of the Bragg pulses, but not with the precision we want to achieve here. In the next section we will therefore treat corrections beyond the ideal adiabatic limit.
        
         Corrections beyond the ideal adiabatic limit come in two ways. First, nonadiabatic transitions from $\pasket{\pm}{N}$ to other states in the respective subspace $\mathscr{H}_{p\alpha\pm}$ result in losses of population, $\gamma_{N\pm}\neq0$. These losses can be described by LZ theory, as done in the next section. Second, by nonadiabatic off-resonant coupling of the states $\pasket{\pm}{N}$ to other states within $\mathscr{H}_{p\alpha\pm}$ a further phase is generated, which in addition to the dynamic phase contributes in order $\tspread^{-1}$ to the net scattering phase of the states $\pasket{\pm}{N}$. We refer to this contribution as LZ phases $\SASphase_{N\pm}^\mathrm{LZ}$.  We will now illustrate how the LZ phases and loss parameters can be calculated, at least approximately, from the Hamiltonians in Eqs.~\eqref{eq:Hblocks} and their  eigenenergies in Eq.~\eqref{eq:eigenvalueproblem}. As motivated above, we will focus on Bragg diffraction of order $N\leq 5$ (corresponding to a momentum transfer of at most $10\hbar k$), and demonstrate that both LZ phases and losses can be understood largely in terms of two-level physics.

        \subsection{LZ phases}\label{sec:LandauZener}
        
       Regarding LZ phases, we show in Appendix~\ref{app:LandauZenerPhase} that the phase acquired by the states $\pasket{\pm}{N}$ due to their off-resonant nonadiabatic coupling to other states in $\mathscr{H}_{p\alpha\pm}$ is given by the quite intuitive expression
         \begin{align}\label{eq:LZphase}
            \SASphase_{N\pm}^\mathrm{LZ}&=\hbar\int_{-\infty}^\infty\mathrm{d}t \sum_{\overset{n\in\mathds{N}_\alpha}{n\neq N}}
            \frac{\left|\tasbra{\pm}{n}{p}{t}\partial_t\tasket{\pm}{N}{p}{t}\right|^2}{E_{N\pm}(t)-E_{n\pm}(t)}.
        \end{align}
        This formula follows from a straightforward application of perturbation theory beyond the adiabatic approximation, and is similar in spirit to corrections derived in Ref.~\cite{Rigolin2008}. The result is correct to first order in the adiabaticity parameter $\big|\tasbra{\pm}{n}{p}{t}\partial_t\tasket{\pm}{N}{p}{t}/(E_{N\pm}(t)-E_{n\pm})\big|$. As it is, the expression for the LZ phase is not very useful for making quantitative statements. This is because the sum runs over all states in $\mathscr{H}_{p\alpha\pm}$ different from $\pasket{\pm}{N}$ and, moreover, because it is cumbersome to calculate the matrix elements in the numerator of the integrand. 
        
        Both of these difficulties can be remedied by invoking an appropriate two-level approximation. The idea is to restrict the sum to its dominant term, which describes the coupling of the state $\pasket{\pm}{N}$ to the energetically closest state, that is $\pasket{\pm}{N-2}$. 
        While the state $\pasket{\pm}{N+2}$ also is energetically close, the coupling to it becomes relevant only for Rabi frequencies $\Omega_0$ that turn out to be prohibitive for high-fidelity Bragg diffraction. For these parameters the Bragg condition cannot be fulfilled perfectly anymore due to nonadiabatic phase contributions (see numerical results for a Gaussian pulse in Fig.~\ref{fig:BraggCondition}) and nonadiabatic losses increase drastically (see Sec.~\ref{sec:LandauZenerLosses} as well as Figs.~\ref{fig:Bragg_Fid_NoNorm_Num},\ref{fig:BraggBeamSplitter},\ref{fig:BraggMirror} for a Gaussian pulse).
        
        If in addition a suitable truncation of the Hamiltonian \eqref{eq:Hblocks} to the two-dimensional subspace composed of $\pasket{\pm}{N}$ and $\pasket{\pm}{N-2}$ is used, the matrix element in the numerator  on the right-hand side of Eq.~\eqref{eq:LZphase} can be evaluated exactly. In Sec.~\ref{sec:BSandMirror} we will show that this approximation indeed gives excellent agreement when compared to exact numerical results.
        
        In order to explain the idea in more detail, we consider Bragg diffraction of order $N=2$ as a concrete example. In this case, in the symmetric subspace the level closest to $\pasket{+}{2}$ is $\pasket{+}{0}$ [see  Hamiltonian~\eqref{eq:Heventrunc}]. The coupling of $\pasket{+}{2}$ to the higher-lying state $\pasket{+}{4}$ is discarded. The Hamiltonian \eqref{eq:Heventrunc} restricted to the two-level subspace $(\pasket{+}{2},\pasket{+}{0})$ is
        \begin{align}\label{eq:H2e+}
            H^{(2)}_{\even,+}(t)&=\hbar\omega_\mathrm{r}\begin{pmatrix}4&\sqrt{2}w(t)\\\sqrt{2}w(t)&0\end{pmatrix},
        \end{align}
        where $w(t)=\Omega(t)/2\omega_\mathrm{r}$. In the antisymmetric subspace the state $\pasket{-}{2}$ has no lower-lying partner, and its coupling to the higher-lying state $\pasket{-}{4}$ is of the same order as the coupling already discarded in the symmetric subspace. Thus, $\pasket{-}{2}$ will not acquire a LZ phase in the order considered here. For the truncated Hamiltonian~\eqref{eq:H2e+} eigenenergies and state overlaps in Eq.~\eqref{eq:LZphase} can be evaluated analytically. The corresponding LZ phase can then be expressed as a simple time integral which has to be evaluated numerically for a given pulse form $\Omega(t)$. 
        
\begin{figure*}[ht]
                \includegraphics[width=0.98\textwidth]{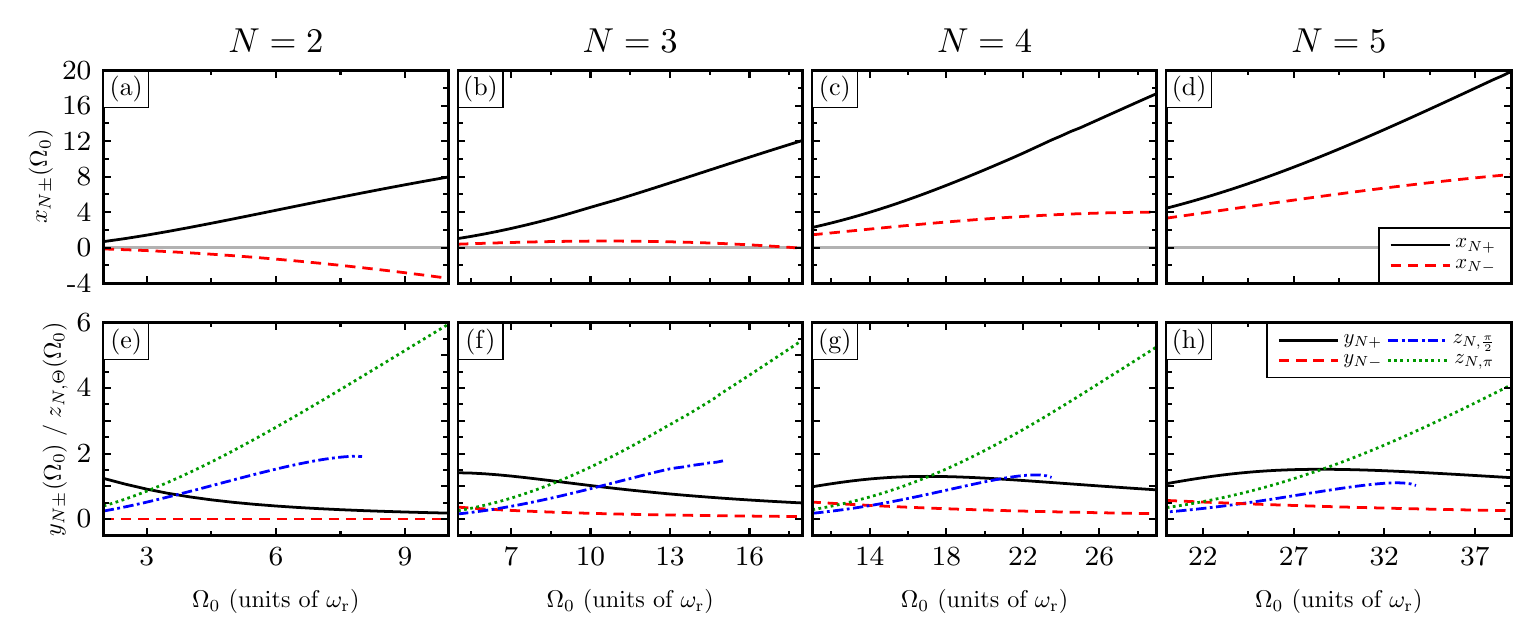}
                \caption{Values of the dimensionless parameters introduced in this paper that are linked to the dynamics phase $x_{N,\pm}$~\eqref{eq:BraggPhasesW} [top row, panels (a)-(d)], the LZ phase $y_{N,\pm}$~\eqref{eq:yparams}, and the Doppler detuning $z_{N,\dPhi}$~\eqref{eq:zint} [both bottom row, panels (e)-(h)]  for Bragg orders $N=2,3,4,5$ (left to right). They are plotted as functions of the peak Rabi frequency $\Omega_0$. The temporal pulse width is fixed by Eq.~\eqref{eq:pulseduration}. Note that for large $\Omega_0$ the blue dash-dotted lines representing $z_{N,\frac{\pi}{2}}$ (bottom row) break off as solutions of Eq.~\eqref{eq:pulseduration} become imaginary. Such short pulse durations require higher-order corrections in $\tspread^{-1}$ than included in Eq.~\eqref{eq:LZphaseapprox1}, which we do not consider here. The numerical results visible in Figs.~\ref{fig:Bragg_Fid_NoNorm_Num}(a)-(e) demonstrate, however, that this regime is not relevant for high-fidelity beam splitter pulses with Gaussian envelopes.}\label{fig:xyz_Parameters}
            \end{figure*}

        Bragg diffraction of higher order can be treated in a similar way with minor complications due to ac Stark shifts, as explained in Appendix~\ref{app:LZTruncation}. In all cases $N\leq 5$ considered here, the resulting approximation for the LZ phase can be expressed as
        \begin{align}\label{eq:LZphaseapprox1}
            \SASphase_{N\pm}^\mathrm{LZ}&=\frac{y_{N\pm}(\Omega_0)}{256(N-1)^3}\frac{\Omega^2_0}{\tspread\omega^3_r} ,
        \end{align}
        where $\Omega_0$ is the peak Rabi frequency and $\tspread$ is the (effective) pulse duration. The dimensionless parameter $y_{N\pm}(\Omega_0)$ is of order unity and absorbs the time integral in Eq.~\eqref{eq:LZphase}. The explicit form of $y_{N\pm}(\Omega_0)$ is given in Eq.~\eqref{eq:yparams} in Appendix~\ref{app:LZTruncation}. For the particular example of a Gaussian pulse, we present the dependence on $N,\,\pm$ and $\Omega_0$ in Fig.~\ref{fig:xyz_Parameters}(e)-(h). Eq.~\eqref{eq:LZphaseapprox1} clearly depicts that the LZ phase is a first-order correction in $\tspread^{-1}$ the weight of which relative to the dynamic phase will become more important for short pulses. As we will see, this approximation gives excellent results for all relevant orders of Bragg diffraction with Gaussian pulses. 
        
        In summary, the net scattering phase of the state $\pasket{\pm}{N}$ entering Eq.~\eqref{eq:phases} is 
        \begin{align}\label{eq:totalscattphase}
            \SASphase_{N\pm}=\SASphase_{N\pm}^\mathrm{dyn}+\SASphase_{N\pm}^\mathrm{LZ}.
        \end{align}
        The dynamic phase is given by Eq.~\eqref{eq:BraggPhasesNum} and the correction due to the LZ phase is given by Eq.~\eqref{eq:LZphaseapprox1}. Both can be evaluated numerically for a given pulse form $\Omega(t)$ by means of the time integrals in Eqs.~\eqref{eq:BraggPhasesW} and \eqref{eq:yparams} for $x_{N\pm}(\Omega_0)$ and $y_{N\pm}(\Omega_0)$, respectively.
        
        It is important to note that this provides a (quasi)analytic expression for the Bragg condition linking  the pulse duration $\tspread$ and the peak Rabi frequency $\Omega_0$: With the help of the now known dependence of the dynamic and LZ phases on the peak Rabi frequency and pulse duration we can determine for a given $\Omega_0$ the pulse duration $\tspread$ necessary to attain a desired differential phase $\dPhi$ (such as $\dPhi=\pi/2$ or $\pi$). Computing the total scattering phase in Eq.~\eqref{eq:totalscattphase} by means of the dynamic phase in Eq.~\eqref{eq:BraggPhasesNum} as well as the LZ phase in Eq.~\eqref{eq:LZphaseapprox1}, and inserting the result into the first of Eqs.~\eqref{eq:phases}, yields a quadratic equation for $\tspread$. The physically relevant solution is the one corresponding to longer pulse duration, and is given by
       \begin{align}\label{eq:pulseduration}
                   \tspread(\dPhi,\Omega_0)=\frac{\dPhi}{2x_{N}(\Omega_0)\omega_\mathrm{r}}
            \left(\vphantom{ \frac{x_N(\Omega_0)y_N(\Omega_0)\Omega_0^2}{64(N-1)^3\dPhi^2\omega_\mathrm{r}^2}}  1 +\sqrt{1
            -\frac{x_N(\Omega_0)y_N(\Omega_0)\Omega_0^2}{64(N-1)^3\dPhi^2\omega_\mathrm{r}^2}}\right),
       \end{align}
        where $x_N(\Omega_0)=x_{N+}(\Omega_0)-x_{N-}(\Omega_0)$ and $y_N(\Omega_0)=y_{N+}(\Omega_0)-y_{N-}(\Omega_0)$. In this solution the dynamic phase makes the dominant contribution, while the LZ phase is a correction which becomes relevant only for large peak Rabi frequency and, correspondingly, short pulses. In the other (formal) solution for $\tspread$ this relation is inverted and the LZ phase makes the dominant contribution. In this regime, however, higher-order corrections to Eq.~\eqref{eq:LZphase} as well as LZ losses become significant and impede high-quality Bragg pulses.
        
        \subsection{LZ losses}\label{sec:LandauZenerLosses}
        
        Next, we consider LZ losses from the states $\pasket{\pm}{N}$ to other states in their respective subspace $\mathscr{H}_{p\alpha\pm}$. As with the LZ phase, it is to be expected that the dominant loss can again be attributed to the energetically closest-lying state. With the same logic and approximations as used before the problem is thus reduced to the determination of LZ losses in a two-level system. 
        
        For the simplest case of  $N=2$ the coupling in the symmetric subspace of $\pasket{+}{2}$ to $\pasket{+}{0}$ is still given by the truncated Hamiltonian in  Eq.~\eqref{eq:H2e+}. Now, in principle LZ theory can be used to determine for a certain pulse form $\Omega(t)$ the population loss from level $\pasket{+}{2}$ to $\pasket{+}{0}$. For the particular Hamiltonian \eqref{eq:H2e+} and Gaussian pulses as in Eq.~\eqref{eq:GaussPulse} Vasilev and Vitanov \cite{Vasilev2004} derived an approximate analytic formula for the LZ loss, which reads in our notation
        \begin{align}\label{eq:gamma2+}
            \gamma_{2+}&=-\frac{1}{2}\ln{
            \left(1-2\frac{\sin(a_\dPhi(\Omega_0,\tspread))^2}{\cosh(b_\dPhi(\Omega_0,\tspread))^2}\right)}.
        \end{align}
        Here, $a_\dPhi(\Omega_0,\tspread)$ and $b_\dPhi(\Omega_0,\tspread)$ are functions of the peak Rabi frequency and pulse duration, the explicit form of which is rather cumbersome and therefore included in Appendix~\ref{app:LandauZenerLosses} [see Eqs.~\eqref{eq:abcoeffs}]. In Appendix ~\ref{app:LandauZenerLosses} the $\dPhi$ dependence of these functions is explained as well, which we drop for $\gamma_{2+}$~\eqref{eq:gamma2+} in the interest of readability. For the same reason as given above, the corresponding LZ loss in the antisymmetric subspace can be neglected to within the order considered here, $\gamma_{2-}=0$. We will see in Sec.~\ref{sec:BSandMirror} that these expressions match very well with exact numerical results. Most notable, the harmonic modulation of the LZ losses due to the sine function in the numerator on the right-hand side of Eq.~\eqref{eq:gamma2+} will be clearly visible.
        
        For higher orders of Bragg diffraction $N=3,4,5$ the problem of LZ losses can still be reduced to two-level physics. However, the relevant truncated Hamiltonians given in Appendix~\ref{app:LZTruncation} involve time-dependent ac Stark shifts which are not covered by the result of Vasilev and Vitanov. The same authors reported an extension of their work to account for a linear sweep in time of energy levels \cite{Vasilev2005}, but this is still very different from the present case, where the relevant ac Stark shift is proportional to $\Omega(t)^2$. An extension of LZ theory to this case would be very desirable, but is beyond the scope of this paper. From the numerical results presented in Sec.~\ref{sec:BSandMirror} for the cases $N=3,4,5$ it will become clear that the relevant physics still corresponds to LZ dynamics in a two-level system, and that one can expect a formula very similar to Eq.~\eqref{eq:gamma2+} to hold also for loss parameters $\gamma_{N\pm}$ in higher-order Bragg diffraction. 
    
    \subsection{Doppler detuning}\label{sec:Doppler}
    
        As a last step, we will further generalize the shape of the scattering matrix \eqref{eq:GeneralScattMatrix} by also taking into account the effect of first-order Doppler detuning. In order to do so, we have to consider the Hamiltonian $\mathcal{H}_{\alpha}(p)$ in Eq.~\eqref{eq:fullHalpha} which includes the Doppler shift term $V_\alpha$ from Eq.~\eqref{eq:H1}. Instead of Eq.~\eqref{eq:dynamics} we now have to construct a solution of Eq.~\eqref{eq:Vunitary} on the subspace $\mathscr{H}_{p\alpha}$ valid to first order in the Doppler detuning. That is, we aim to solve
        \begin{align}
            i\hbar\frac{\mathrm{d}}{\mathrm{d}t}\mathcal{V}_\alpha(p,t,t_0)=\Big({H}_\alpha(t)+\epsilon(p){V}_\alpha(t)\Big)\mathcal{V}_\alpha(p,t,t_0),
        \end{align}
        to first order in $\epsilon(p)$. Using the fact that $U_\alpha(t,t_0)$ in Eq.~\eqref{eq:Usol} solves Eq.~\eqref{eq:dynamics} one finds
        \begin{subequations}
        \begin{align}
            \mathcal{V}_\alpha(p,t,t_0)&=U_\alpha(t,t_0)\Big(1-i\epsilon(p)Z_\alpha(t,t_0)\Big),\\
            Z_\alpha(t,t_0)&=\frac{1}{\hbar}
            \int_{t_0}^t\mathrm{d}t_1 U_\alpha^\dagger(t_1,t_0){V}_\alpha(t_1)U_\alpha(t_1,t_0).\label{eq:Z}
        \end{align}
        \end{subequations}
        We can now take the limit 
        \begin{align}\label{eq:ScattMatp}
            \lim_{\substack{t\rightarrow\infty\\ t_0\rightarrow-\infty}}\exp\left(i\mathcal{L}_\alpha(t-t_0)/\hbar\right)\mathcal{V}_\alpha(t, t_0)
            =\mathcal{S}_{\alpha}\Big(1-i\epsilon(p)Z_\alpha\Big),
        \end{align}
        where $\mathcal{S}_{\alpha}$ from Eq.~\eqref{eq:ScattMatLimit} is the zeroth-order scattering ma\-trix. Regarding the first-order correction $Z_\alpha$, it is simplest to consider directly the relevant matrix elements in the (anti)symmetric basis $\pasket{\pm}{N}$ from Eq.~\eqref{eq:Z}. We show in Appendix~\ref{app:Doppler} that the diagonal elements vanish, $\pasbra{\pm}{N}Z_\alpha\pasket{\pm}{N}=0$, and the off-diagonal elements  $\pasbra{-}{N}Z_\alpha\pasket{+}{N}=\pasbra{+}{N}Z_\alpha\pasket{-}{N}^*$ are nonzero. This reflects the fact that Doppler detuning breaks the decoupling of symmetric and antisymmetric subspace. One finds
        \begin{align}\label{eq:approxmatelemZ}
            \pasbra{+}{N}Z_\alpha\pasket{-}{N}&= 2N \tspread^2\omega_\mathrm{r}^2  e^{i\dPhi/2}z_{N,\dPhi}(\Omega_0),
        \end{align}
        where $\dPhi$ is the differential phase from Eq.~\eqref{eq:phases} and $z_{N,\dPhi}(\Omega_0)$ is a positive real parameter of order unity given in Eq.~\eqref{eq:zint}. It absorbs a time integral of overlaps of instantaneous energy eigenstates and is shown in Fig.~\ref{fig:xyz_Parameters}(e)-(h) up to order $N\leq 5$. 
        
        Overall, we find that the scattering matrix, projected into the subspace $\ket{\pm N\hbar k+p}$ and written in the basis of (anti)symmetric states $(\asket{+}{N}{p},\asket{-}{N}{p})$ is
        \begin{align}\label{eq:GeneralScatMat}
            S(p)=\begin{pmatrix} e^{-i\SASphase_{N+}-\gamma_{N+}} & 0 \\ 0 & e^{-i\SASphase_{N-}-\gamma_{N-}} \end{pmatrix}\begin{pmatrix} 1 & i\eta(p)e^{i\dPhi/2} \\ i\eta(p)e^{-i\dPhi/2} & 1 \end{pmatrix}.
        \end{align}
        From here on we explicitly write again the dependence on the (quasi)momentum $p$ and have introduced in Eq.~\eqref{eq:GeneralScatMat} the dimensionless Doppler parameter:
        \begin{align}\label{eq:eta}
            \eta(p)=- 2 N \tspread^2 \omega_\mathrm{r}^2 z_{N,\dPhi}(\Omega_0) \frac{p}{\hbar k}.
        \end{align}
        Eq.~\eqref{eq:GeneralScatMat} generalizes Eq.~\eqref{eq:ProjectedBragg} and includes Doppler detuning to first order. Thus, Doppler detuning causes a mixing of the (anti)symmetric states $\asket{\pm}{N}{p}$, but no real loss out of this subspace, like the LZ losses do. As it stands, the projected scattering matrix is nonunitary due to both effects, Doppler detuning and LZ losses, as $\mathrm{tr}\left(S^\dagger(p) S(p)\right)=(1+\eta(p)^2)(e^{-2\gamma_{N+}}+e^{-2\gamma_{N-}})$. The nonunitarity due to the Doppler effect is, however, an artifact of the perturbation series expansion adopted here. In contrast, the nonunitarity due to LZ losses is due to actual losses out of the relevant subspace. It is important to account for this difference by renormalizing the scattering matrix in order to remove the artificial nonunitarity due to the Doppler shift. This requires us to replace $S(p)\rightarrow S(p)/\sqrt{1+\eta(p)^2}$. Finally, the transformation of the scattering matrix \eqref{eq:GeneralScatMat} from the (anti)symmetric basis to the basis momentum eigenstates $\ket{\pm N\hbar k+p}$ is again achieved by means of $T$ in Eq.~\eqref{eq:Transform} and $B(p)=T^\dagger S(p) T$. The result is given in the next section in Eq.~\eqref{eq:BraggScatteringMatrixFinal}.
    
        \subsection{Summary}\label{sec:summary}
        In the following we will give an -- as far as possible -- self-contained summary of the results of Sec.~\ref{sec:BraggScatteringMatrix}. We have successfully applied the adiabatic theorem to describe single Bragg diffraction of any order $N$ with smooth temporal pulse shapes and arrive at intuitive analytical expressions linking the products of this elastic-scattering process and the experimental parameters of the Bragg pulse. More specifically, our results are based on the realization that instead of driving an effective two-level system diabatically it is far more useful to understand the physics of a single Bragg pulse by conceiving the evolution of the atom interacting with the optical lattice to be adiabatic. This enables the formulation of a comprehensive and relatively simple analytic model accurately capturing the dynamics of single Bragg diffraction with smooth temporal pulse profiles for any order $N$. In the following, we briefly summarize our findings to stress why applying the adiabatic theorem is natural and provides an intuitive picture of the physics of an atom being elastically scattered from a pulsed optical lattice.
        
\begin{figure}[ht]
            \includegraphics[width=0.48\textwidth,height=5cm]{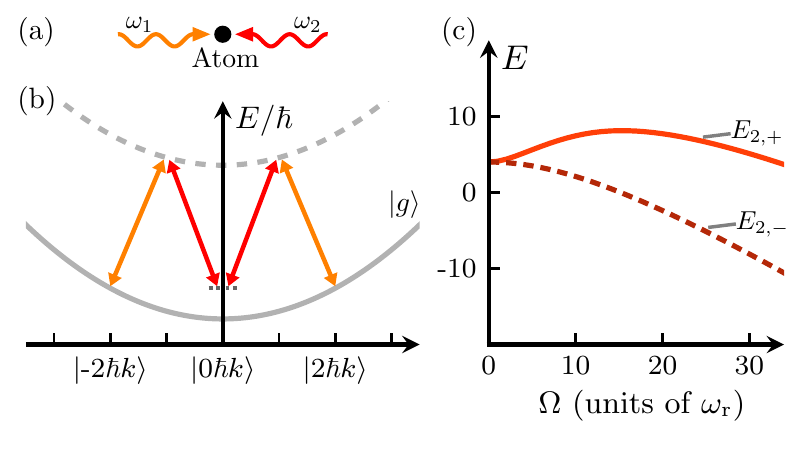}
                \caption{(a) Schematic of an atom interacting with a pulsed optical lattice realized by two counterpropagating light fields $\omega_1$ and $\omega_2$. Depending on the relative momentum between the atom and the optical lattice, the atom can undergo multiple $2N$-photon transitions via stimulated absorption and emission imparting $2N$-photon recoils $\hbar k$ on the atom. The dispersion relation in the inertial frame co-moving with the optical lattice (b) shows the resonant coupling between momentum states $\ket{\pm 2\hbar k}$ via a four-photon transition satisfying energy and momentum conservation. As a single Bragg diffraction imprints a differential phase between the symmetric and antisymmetric components of the initial and final state [see Eq.~\eqref{eq:PlaneWaves}] it is natural to describe the dynamics of such pulses in terms of the adiabatic theorem. The eigenenergies of the symmetric (+) and antisymmetric (-) components for a $4\hbar k$ Bragg pulse ($N=2$) are plotted in (c) as a function of the Rabi frequency $\Omega$. The difference of these eigenenergies integrated over the duration of the single Bragg pulse determines the differential phase [see, e.g., Eq.~\eqref{eq:BraggDiffPhaseSummary}].}\label{fig:AtomLightCouplingScheme}
       \end{figure}

        To impart momentum an atom is exposed to two counterpropagating light fields as shown in Fig.~\ref{fig:AtomLightCouplingScheme}(a) which are far detuned with respect to an atomic transition. The average relative velocity between the resulting optical lattice and the atom is controlled by the detuning $\omega_1 - \omega_2$ and can be chosen such that it is a multiple $N$ of the laser photon recoil velocity $\hbar k/M$.
        
        Assuming that both the initial and the final state of the atom, being Bragg diffracted, can be idealized as plane waves,
        \begin{align} \label{eq:PlaneWaves}
            \begin{split}
                e^{- i N k x} &= \cos{(Nkx)} - i \sin{(Nkx)} \\
                e^{+ i N k x} &= \cos{(Nkx)} + i \sin{(Nkx)}, \\
            \end{split}
        \end{align}
        they are degenerate in kinetic energy when choosing an inertial frame comoving with the lattice potential. The resonant coupling of these states via $2N$-photon transitions is therefore allowed in terms of energy and momentum conservation. Fig.~\ref{fig:AtomLightCouplingScheme}(b) illustrates this coupling for $N=2$ meaning a $4\hbar k$ momentum transfer via Bragg diffraction. Expanding the initial and final momentum eigenstates in the basis of their symmetric ($\cos{(Nkx)}$) and antisymmetric ($\sin{(Nkx)}$) components according to Eq.~\eqref{eq:PlaneWaves} reveals that a Bragg pulse ideally imprints a differential phase between these basis states. This differential phase is exactly $\pi$ for a mirror pulse and $\pi/2$ for a beam splitting pulse. Through the expansion in symmetric(+) and antisymmetric(-) components the evolution of this differential phase over the course of the pulse can be well understood by calculating the evolution of the associated eigenenergies according to the adiabatic theorem. As an example in Fig.~\ref{fig:AtomLightCouplingScheme}(c) the relevant energies $E_{2\pm}$ for a $4\hbar k$ Bragg pulse are plotted.
        The integral over the difference in eigenenergies [see Eqs.~\eqref{eq:phases} and \eqref{eq:BraggPhases}],
        \begin{align}\label{eq:BraggDiffPhaseSummary}
                   \dPhi^{\mathrm{dyn}}=\frac{1}{\hbar}\int_{-\infty}^{\infty}\mathrm{d}t\left( E_{N+}(t)- E_{N-}(t)\right),
        \end{align}
        basically is the equivalent to what is typically (in the context of the diabatic description of Bragg diffraction) referred to as the ``pulse area''. It can readily be calculated by diagonalizing finite dimensional Hamiltonians. In our paper, we also include corrections to the adiabatic evolution via perturbation theory and find that LZ phases and LZ losses play important roles for the dynamics of the Bragg pulse. Together these quantities determine the scattering matrix  for $N$th-order Bragg diffraction, that we have derived in the previous section:
        \begin{widetext}
        \begin{subequations}\label{eq:BraggScatteringMatrixComplete}
        \begin{align}
            S(\Omega_0,\tspread)=\int\displaylimits_{-\hbar k/2}^{\hbar k/2}\!\!\difp\sum_{s,s'=\mp} \left(B(p,\Omega_0,\tspread)\right)_{ss'}
            \ket{\vphantom{'} s N\hbar k+p}\bra{s'N\hbar k+p},
        \end{align}
        where
        \begin{align}\label{eq:BraggScatteringMatrixFinal}
            B(p,\Omega_0,\tspread)=
            \frac{\exp(-i \frac{\glP-i\sumLoss}{2})}{\sqrt{1+\eta(p)^2}}\begin{pmatrix}\cos{\left(\frac{\dPhi-i\diffLoss}{2}\right)}&- ie^{-i2N \Lphase} \sin{\left(\frac{\dPhi-i\diffLoss}{2}\right)}\\[3pt]
            - ie^{i2N \Lphase} \sin{\left(\frac{\dPhi-i\diffLoss}{2}\right)}&\cos{\left(\frac{\dPhi-i\diffLoss}{2}\right)}\end{pmatrix}
            \begin{pmatrix} 1+i\eta(p) \cos{\left(\frac{\dPhi}{2}\right)}& e^{-i2N \Lphase} \eta(p)\sin{\left(\frac{\dPhi}{2}\right)}\\[3pt]- e^{i2N \Lphase} \eta(p)\sin{\left(\frac{\dPhi}{2}\right)}&
            1-i\eta(p) \cos{\left(\frac{\dPhi}{2}\right)} \end{pmatrix}
            \coloneqq \begin{pmatrix} B_{--} & B_{-+} \\ B_{+-} & B_{++} \end{pmatrix}.
        \end{align}
        \end{subequations}
        \end{widetext}
        
Table~\ref{tab:params} summarizes all parameters entering the scattering matrix. In addition we provide references to their respective de\-fi\-ni\-tions which link them to the Rabi frequency $\Omega(t)$, and for the special case of a Gaussian pulse to the peak Rabi frequency $\Omega_0$ and the pulse width $\tspread$. The general structure of the scattering matrix \eqref{eq:BraggScatteringMatrixFinal} holds for arbitrary pulse forms $\Omega(t)$ and accounts for Doppler detuning (to first order in $|p|/\hbar k\ll 1$) as well as for population loss out of the subspace $\left(\ket{N\hbar k+p},\ket{-N\hbar k+p}\right)$. The formulas presented in Table~\ref{tab:params} assume an adiabatic tuning of $\Omega(t)$, and include the dominant nonadiabatic corrections due to LZ processes. For Gaussian pulses we will see in the next section that high-quality quasi-Bragg pulses do indeed always fall within this regime.
        \begin{table}[t]
            \caption{\label{tab:params}%
            Parameters determining the Bragg scattering matrix~\eqref{eq:BraggScatteringMatrixComplete}.}
        {\renewcommand{\arraystretch}{1.2} 
          \begin{ruledtabular}
            \begin{tabular}{l c c c }
                \textrm{Parameter} & \textrm{Symbol} & \textrm{defined by} & \textrm{Equation}  \\
                \colrule
                Global phase    &   $\Phi$      &   $\Phi=\SASphase_{N+}+\SASphase_{N-}$  &   \eqref{eq:phases}\\
                Global LZ loss  &   $\Gamma$    &   $\Gamma=\gamma_{N+}+\gamma_{N-}$    &   \eqref{eq:losses}\\
                Differential phase  &   $\dPhi$ &   $\dPhi=\SASphase_{N+}-\SASphase_{N-}$ &   \eqref{eq:phases}\\
                Differential LZ loss    &   $\gamma$    &   $\gamma=\gamma_{N+}-\gamma_{N-}$    &   \eqref{eq:losses}\\
                Doppler shift   &   $\eta(p)$   &   &   \eqref{eq:eta}  \\
                Laser phase &   $\Lphase$  &   &   \eqref{eq:Ham} \\
                Total phase of $\asket{\pm}{N}{p}$  &   $\SASphase_{N\pm}$   &   $\SASphase_{N\pm}=\SASphase_{N\pm}^\mathrm{dyn}+\SASphase_{N\pm}^\mathrm{LZ}$  &    \eqref{eq:totalscattphase}\\[2pt]
                Dynamic phase   &   $\SASphase_{N\pm}^\mathrm{dyn}$  &   &   \eqref{eq:BraggPhasesNum}\\[2pt]
                LZ phase    &   $\SASphase_{N\pm}^\mathrm{LZ}$   &   &   \eqref{eq:LZphaseapprox1}\\[2pt]
                LZ loss from $\asket{\pm}{N}{p}$    &   $\gamma_{N\pm}$ &  &   \eqref{eq:gamma2+},\eqref{eq:gammaNum+-}\\
          \end{tabular}
          \end{ruledtabular}
          }
        \end{table}
        
       In the hypothetical case of vanishing LZ losses (${\Gamma=\gamma=0}$), no Doppler detuning ($\eta(p)=0$), and zero global phase ($\glP =0$) the scattering matrix in Eq.~\eqref{eq:BraggScatteringMatrixFinal} assumes familiar forms if the pulse $\Omega(t)$ is tuned such that the differential phase $\dPhi$ takes on specific values: $\dPhi=\pi/2$ provides a beam splitter operation, and $\dPhi=\pi$ yields a mirror pulse as given in Eqs.~\eqref{eq:idealScattM}. The scattering matrix~\eqref{eq:BraggScatteringMatrixFinal} provides a systematic generalization to account for nonideal phases $\dPhi$ as well as unavoidable global phases, population losses, and Doppler shifts. Our model gives a microscopic explanation and analytic characterization (except for LZ losses in Bragg diffraction of higher order $N>2$) for all of these effects in leading order. The approach taken here provides also a systematic framework for deriving higher-order corrections.
       
       An important insight that can be gained from our analytic characterization of the differential phase concerns the so-called Bragg condition: For a Gaussian pulse the requirement to achieve a desired phase $\dPhi$ links $\Omega_0$ to $\tspread$, such that the pulse duration $\tspread(\dPhi,\Omega_0)$ can be expressed as a function of the peak Rabi frequency for a given differential phase [see Eq.~\eqref{eq:pulseduration}]. For a desired operation, such as a beam splitter ($\dPhi=\pi/2$) or a mirror ($\dPhi=\pi$) pulse, this leaves a single free parameter, $\Omega_0$, which fully determines the scattering matrix \eqref{eq:BraggScatteringMatrixComplete}. What is left is to choose the peak Rabi frequency to balance the dominant imperfections: LZ losses will become large for short pulses, that is, for a large Rabi frequency. The effects of Doppler detuning will be stronger for long, spectrally narrow pulses with correspondingly small Rabi frequencies. The trade-off implied by this is very well covered by our analytic model, as will be demonstrated in the Sec.~\ref{sec:BSandMirror}. To improve accessibility of the results presented in this paper and to facilitate further research all of the codes used to generate these results are available \cite{Siemss2020}. 
    \section{Comparison of analytic model with numerics}\label{sec:BSandMirror}
    
    In this section, we are going to compare our analytical model to numerical solutions of the Schr\"odinger equation corresponding to the exact Bragg Hamiltonian in Eq.~\eqref{eq:Ham} for a Gaussian Bragg pulse. We remind the reader of the fidelities introduced in Sec.~\ref{sec:BraggDiffQuality} in order to quantify the quality of Bragg operations. With the analytic form of the scattering matrix in Eq.~\eqref{eq:BraggScatteringMatrixComplete} we can re-express the fidelity from Eq.~\eqref{eq:fidelity1} as
    \begin{subequations}\label{eq:fidelitiesAna}
    \begin{align}\label{eq:FidAverageGeneral}
    \begin{split}
        \Fid_{\dPhi,\pspread}(\Omega_0,\tspread)&=\left|\langle\psi^\mathrm{ideal}_{\mathrm{out},\dPhi}|\psi_{\mathrm{out}}(\Omega_0,\tspread)\rangle\right|^2\\
        &=\int_{-\hbar k/2}^{\hbar k/2}\!\!\difp |g(p)|^2 \left|\left[B^\dagger_\dPhi B(p,\Omega_0,\tspread)\right]_{11}\right|^2.
    \end{split}
    \end{align}
    The last term denotes the squared modulus of the top-left element of the matrix $B^\dagger_\dPhi B(p,\Omega_0,\tspread)$ which we express explicitly in Appendix $\ref{app:Fidelities}$.
    It will be useful to consider also the fidelity~\eqref{eq:fidelity2} for the hypothetical situation of an infinitely narrow atomic wave packet which does not experience a Doppler effect: 
    \begin{multline}\label{eq:Fidp0}
            \Fid_{\dPhi,0}(\Omega_0,\tspread)=\lim_{\pspread\rightarrow 0}\Fid_{\dPhi,\pspread}(\Omega_0,\tspread)\\
            =\left|\left[B^\dagger_\dPhi B(0,\Omega_0,\tspread)\right]_{11}\right|^2 =\frac{e^{-\Gamma}}{2}(1+\cosh\left(\gamma\right)).
    \end{multline}
    \end{subequations}
    This can also be looked at as the exact fidelity achieved for the center component with momentum $p=0$ of a finite atomic wave packet, or equivalently as the fidelity attained within each subspace $\left(\ket{-N\hbar k+p},\ket{N\hbar k+p}\right)$ in zeroth order of Doppler detuning.
    
    These approximate analytic expressions for the fidelities can be compared to the fidelities inferred from the exact numerical solution of the Schr\"odinger equation $|\psi^\mathrm{num}_{\mathrm{out}}(\Omega_0,\tspread)\rangle$ for given pulse parameters. We denote the numerically inferred fidelities corresponding to Eqs.~\eqref{eq:fidelitiesAna} by
    \begin{subequations} \label{eq:FidNum}
    \begin{align}
        \Fid^\mathrm{num}_{\dPhi,\pspread}(\Omega_0,\tspread)&= \left|\langle\psi^\mathrm{ideal}_{\mathrm{out},\dPhi}|\psi^\mathrm{num}_{\mathrm{out}}(\Omega_0,\tspread)\rangle\right|^2, \label{eq:FidNumpAvg}\\   
        \Fid^\mathrm{num}_{\dPhi,0}(\Omega_0,\tspread)&=\lim_{\pspread\rightarrow 0}\Fid^\mathrm{num}_{\dPhi,\pspread}(\Omega_0,\tspread). \label{eq:FidNump0}
    \end{align}
    \end{subequations}
    The fidelity~\eqref{eq:FidNump0} is shown in Fig.~\ref{fig:Bragg_Fid_NoNorm_Num} for beam splitter $(\dPhi=\pi/2)$ [Figs.~\ref{fig:Bragg_Fid_NoNorm_Num}(a)-(d)] and mirror $(\dPhi=\pi)$ [Figs.~\ref{fig:Bragg_Fid_NoNorm_Num})(e)-(h)] pulses of Bragg diffraction orders $N=2,3,4,5$, corresponding to momentum transfers of $4\hbar k, 6\hbar k, 8\hbar k, 10\hbar k $, respectively.
    
    One last figure of merit which will be useful in the following discussion is a fidelity, where both the Doppler effect and the LZ losses are masked out. This can be achieved by considering the fidelity for the central $p=0$ momentum component of the wave packet from Eq.~\eqref{eq:Fidp0} but calculating it with respect to the normalized state $|\psi^\mathrm{num}_{\mathrm{out}}(\Omega_0,\tspread)\rangle/\norm{|\psi^\mathrm{num}_{\mathrm{out}}(\Omega_0,\tspread)\rangle}$. This vector describes the state of atoms conditioned on the fact, that they actually remain in the correct $\left(\ket{\pm N\hbar k}\right)$-subspace. When Doppler effect and LZ losses are ignored in this way, the conditional fidelity for the conditional, normalized state is 
    \begin{align}\label{eq:fidelityNormalized}
        \mathcal{F}^\mathrm{num}_{\dPhi,0}(\Omega_0,\tspread)=\frac{\Fid^\mathrm{num}_{\dPhi,0}(\Omega_0,\tspread)}{\norm{|\psi^\mathrm{num}_{\mathrm{out}}(\Omega_0,\tspread)\rangle}}.
    \end{align} 
    It will be reduced below 1 only when the pulse parameters $(\Omega_0,\tspread)$ fail to generate the desired differential phase $\dPhi$. Thus, $\mathcal{F}^\mathrm{num}_{\dPhi,0}(\Omega_0,\tspread)$ is a suitable figure of merit to benchmark the analytic formula for the prediction of the pulse duration~\eqref{eq:pulseduration}  
    $\tspread(\dPhi,\Omega_0)$  necessary to achieve a desired differential phase $\dPhi$. 
    \subsection{Bragg condition}
    
    
         \begin{figure*}[t]
                \includegraphics[width=0.98\textwidth]{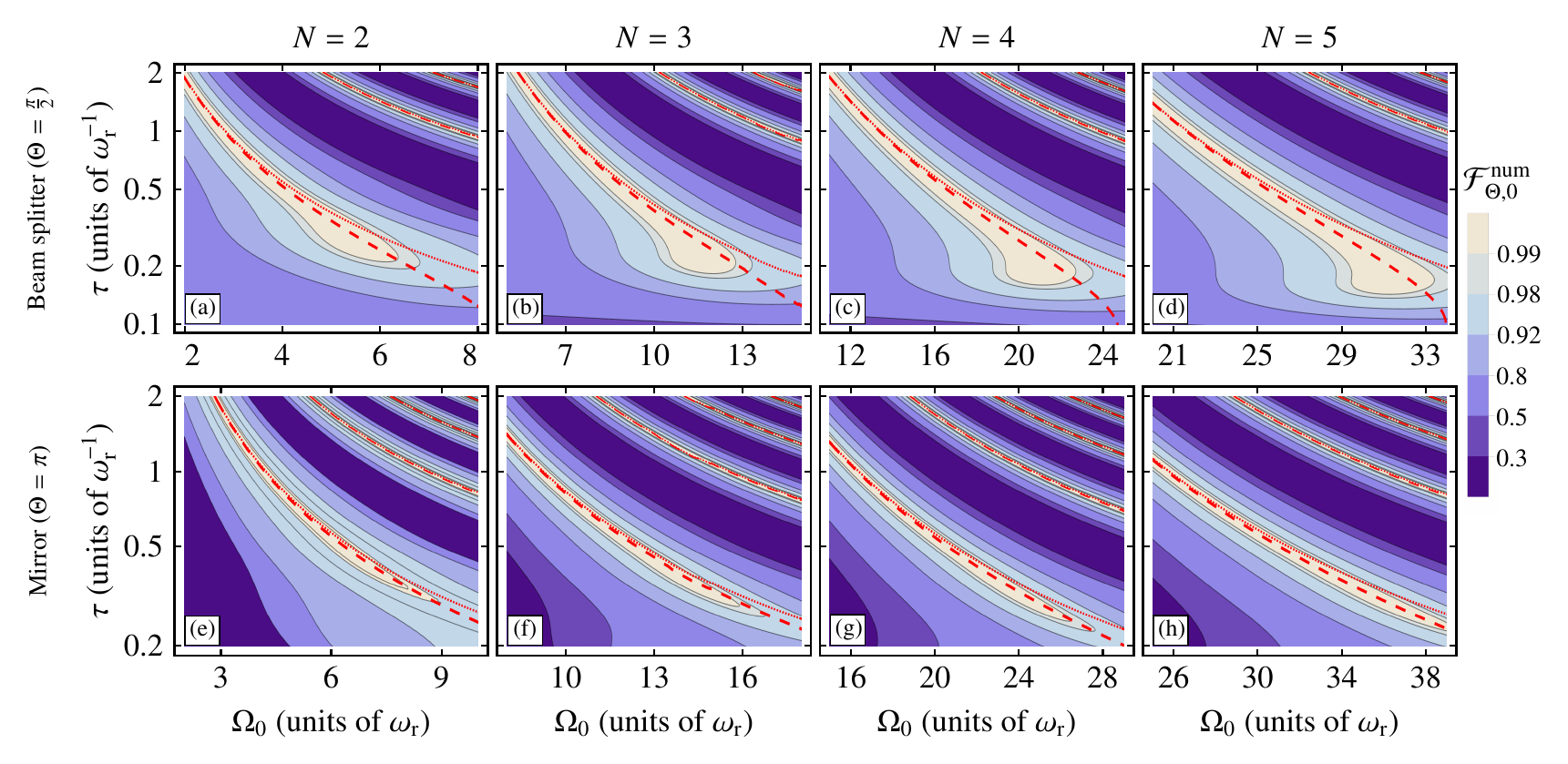}
                \caption{Similar to Fig.~\ref{fig:Bragg_Fid_NoNorm_Num}. Here, the normalized beam splitter [top row, panels (a)-(d)] and mirror [bottom row, panels (e)-(h)] fidelities~\eqref{eq:fidelityNormalized} of a single Gaussian quasi-Bragg pulse in case of a wave packet with vanishing momentum spread as a function of the peak Rabi frequency $\Omega_0$ and the temporal pulse width $\tspread$ are plotted for diffraction orders $N=2,3,4,5$ (from left to right). The red lines represent the calculated temporal pulse width  $\tspread(\dPhi,\Omega_0)$ in Eq.~\eqref{eq:pulseduration} with (dashed) and without (dotted) the phase contribution from LZ physics. In contrast to Figs.~\ref{fig:Bragg_Fid_NoNorm_Num}(a)-(d), beam splitter fidelities possess a simplified structure as LZ losses from the subspace $\ket{\pm N\hbar k}$ are blanked out in the conditional fidelity $\mathcal{F}^\mathrm{num}_{\dPhi,0}$. }\label{fig:BraggCondition}
            \end{figure*}
    
    Figure~\ref{fig:BraggCondition} shows the conditional fidelity $\mathcal{F}^\mathrm{num}_{\dPhi,0}(\Omega_0,\tspread)$ introduced in Eq.~\eqref{eq:fidelityNormalized} for Bragg beam splitters and mirrors.  These plots are similar to the ones shown already in Fig.~\ref{fig:Bragg_Fid_NoNorm_Num} but blank out the effects of LZ losses. We immediately observe that the rich fidelity landscapes showcased in Fig.~\ref{fig:Bragg_Fid_NoNorm_Num} simplify considerably when evaluating this fidelity instead of the unconditional fidelity $F^\mathrm{num}_{\dPhi,0}(\Omega_0,\tspread)$~\eqref{eq:FidNump0}.
    
    Considering first the numerical data represented by the shaded regions in Fig.~\ref{fig:BraggCondition}, one clearly recognizes the Bragg condition: The shorter the temporal width $\tspread$  of the pulse, the stronger its coupling must be to achieve Bragg operations of decent quality. It is also visible, that for sufficiently large parameters $(\Omega_0,\tspread)$ one can re\-a\-lize an efficient beam splitter (mirror) with a differential phase of $\dPhi = \pi/2 + m\;2\pi$ $(\dPhi = \pi + m\;2\pi)$ with $m \in \mathds{N}$.
    More importantly, the numerical data highlight the fact that for all Bragg orders depicted, even when disregarding LZ losses, there exists a minimal temporal pulse width beyond which fidelities degrade quickly. Rising nonadiabatic couplings such as the LZ phase introduced in Sec.~\ref{sec:LandauZener} and higher-order corrections to the adiabatic theorem make it impossible to perfectly match the Bragg condition with Gaussian pulses featuring pulse widths shorter than that. 
    
    The Bragg condition visible in Fig.~\ref{fig:BraggCondition} can now be compared to the predictions our analytic model provides regarding the pulse timings $\tspread(\dPhi,\Omega_0)$~\eqref{eq:pulseduration}. We show the pulse timings including (red dashed line) and excluding (red dotted line) the contribution of the LZ phase to the differential phase, that is, with and without the second term under the square root in Eq.~\eqref{eq:pulseduration}, respectively. Clearly, Eq.~\eqref{eq:pulseduration} provides an excellent approximation for the necessary pulse duration in all regimes, where it is even possible to perform a high-quality operation. Thus, for Gaussian pulses adiabaticity is indeed a necessary and sufficient condition for performing efficient Bragg diffraction.
    
    From Fig.~\ref{fig:BraggCondition} it is also evident that the  LZ phase has a significant contribution to the Bragg condition even when the LZ losses themselves have been re-normalized. Naturally, these corrections are less important when operating with small peak Rabi frequencies and accordingly long pulse durations, i.e., for more adiabatic pulses. Following the same logic, it is straightforward to understand that in the case of a mirror pulse corrections to the differential phase due to LZ physics are suppressed compared to a beam splitting pulse: For the same value of $\Omega_0$ the latter operates with half the temporal width of the former. That is why for Bragg mirror operations the mismatch in differential phase when considering only the dynamic phase is reduced, which is visible in Figs.~\ref{fig:BraggCondition}(e)-(h). Nonetheless, the figure makes it fairly obvious that using the full Eq.~\eqref{eq:pulseduration} instead of just the dynamic phase contribution considerably improves predictions for both operations and especially in the case of a beam splitter. 
    
    In fact, from the weight of the LZ-phase term in Eq.~\eqref{eq:pulseduration} we can deduce the adiabaticity of the Bragg diffraction process for a given Rabi frequency. This is considerably more precise than the estimation done in Sec.~V~A of \cite{Mueller2008PRA} or the usual adiabaticity criterion derived from the separation of the $N$-th and $(N-1)$-th energy levels~\cite{Mueller2008PRA} which in the case of Gaussian pulse profile transforms into~\cite{Gochnauer2019} $\displaystyle \tspread \omega_\mathrm{r} \gg [4 (N-1)]^{-1}$.
    
    The discrepancy between the numerically determined pulse parameters $(\Omega_0,\tspread)$ maximizing the fidelity and the results of the full Eq.~\eqref{eq:pulseduration} for values $\tspread\omega_\mathrm{r} < 0.2$ is a consequence of the rising nonadiabaticity and the limitation of perturbation theory developed in Appendix $\ref{app:LandauZenerPhase}$. Since Bragg pulses implemented in state-of-the-art atom interferometry experiments typically aim at efficiencies approaching unity~\cite{Chiow2011,Plotkin-Swing2018,Gebbe2019}, this regime can be, however, considered unsuitable for high-performance quasi-Bragg beam splitters and mirrors as fidelities quickly degrade.
    
    In the following two subsections we return to the unconditional fidelities in Eqs.~\eqref{eq:FidNum} which include losses from the subspace $\ket{\pm N\hbar k +p}$, and identify them as product of LZ processes that give rise to the features observed in Fig.~\ref{fig:Bragg_Fid_NoNorm_Num}.  
    
    \subsection{Bragg beam splitters and mirrors}
    \subsubsection*{Gaussian beam splitter pulses}
\begin{figure*}[t]
                \includegraphics[width=0.9\textwidth]{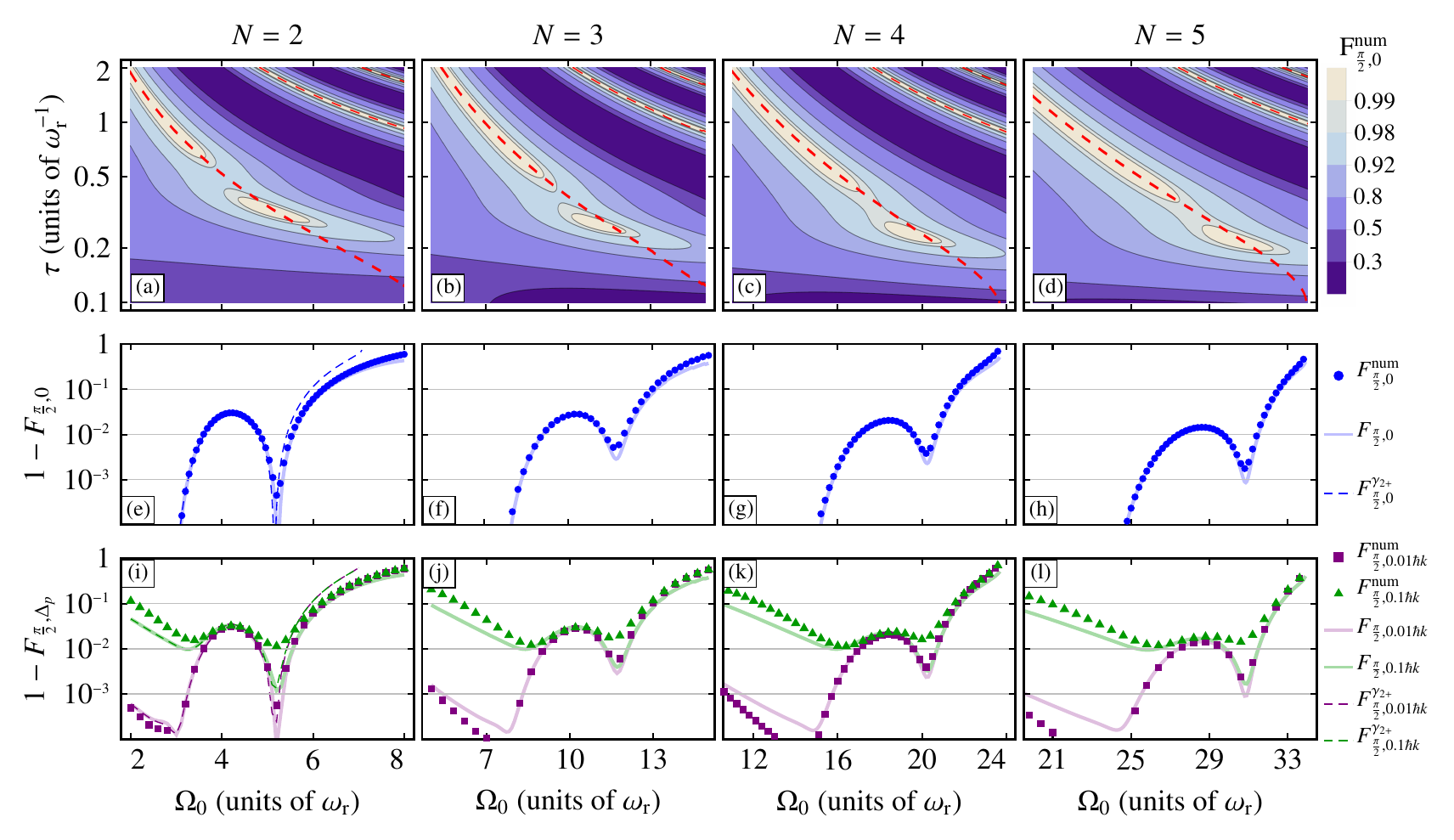}
                \caption{\textbf{Bragg beam splitter}. Top row: Pulse duration~\eqref{eq:pulseduration} (red dashed line) on top of the numerically determined beam splitter fidelities~\eqref{eq:FidNump0} introduced in Figs.~\ref{fig:Bragg_Fid_NoNorm_Num}(a)-(d). Middle row: Fidelity loss as a function of peak Rabi frequency $\Omega_0$ and pulse durations computed via Eq.~\eqref{eq:pulseduration}. Beam splitter fidelities have been determined numerically (blue disks), via Eq.~{\eqref{eq:Fidp0}} with values~\eqref{eq:gammaNum+-} (solid line). The dashed line in panel (e) for $N=2$  is obtained by inserting $\gamma_{2,+}$~\eqref{eq:gamma2+} into Eq.~{\eqref{eq:Fidp0}} and denoted $F^{\gamma_{2,+}}_{\frac{\pi}{2},0}$ in the legend on the right.   Bottom row: Fidelity loss extracted from exact numerics (squares \& triangles) is again compared to analytic results (solid \& dashed lines) similar to middle row. Fidelities~\eqref{eq:FidNumpAvg} are now averaged over a wave packet with finite momentum width $\pspread = 0.01 \hbar k$ (purple squares and lines) and $\pspread = 0.1 \hbar k$ (green triangles and lines). $\tilde{\gamma}_{N,\pm}$ values have been obtained via Eq.~\eqref{eq:gammaNum+-} (solid lines) or for $\gamma_{2,+}$ using Eq.~\eqref{eq:gamma2+} (dashed lines) in panel (i).  }\label{fig:BraggBeamSplitter}
        \end{figure*}

    We start by discussing the Bragg beam splitter pulse $(\dPhi=\pi/2)$ in Fig.~\ref{fig:BraggBeamSplitter} for diffraction orders $N=2,3,4,5$. We consider first the lowest order $N=2$. In Fig.~\ref{fig:BraggBeamSplitter}(a) we present the result of our approximate analytic formula for $\tspread(\dPhi,\Omega_0)$ in Eq.~\eqref{eq:pulseduration} on top of the numerically inferred fidelities Eq.~\eqref{eq:FidNump0} in the $(\Omega_0,\tspread)$ plane assuming a vanishing momentum width. The figure shows good agreement between our model and the peak fidelities over the relevant range of peak Rabi frequencies. In addition, it illustrates that pulse parameters complying with the Bragg condition in the subspace $\ket{\pm N\hbar k}$ (red dashed line, cf. Fig.~\ref{fig:BraggCondition}) are subject to losses with an intricate dependency on the peak Rabi frequency $\Omega_0$. 
    
    To demonstrate that this dependency can be understood applying LZ theory, Fig.~\ref{fig:BraggBeamSplitter}(e) depicts the fidelity loss for pulse parameters $\{\Omega_0,\tspread(\pi/2,\Omega_0)\}$ highlighted by the dashed red line in Fig.~\ref{fig:BraggBeamSplitter}(a). The blue circles are obtained evaluating Eq.~\eqref{eq:FidNump0} numerically. The corresponding analytic fidelity \eqref{eq:Fidp0} is dependent on the LZ loss parameters ${\Gamma}$ and ${\gamma}$ that have been derived in Sec.~\ref{sec:LandauZenerLosses}. Within our approximation both are entirely determined by $\gamma_{2,+}$, the loss of amplitude from the symmetric state $\ket{0,2,+}$, hence the superscript $  \Fid^{\gamma_{2,+}}_{\dPhi,0}$.  We insert $\gamma_{2,+}$~\eqref{eq:gamma2+} into Eq.~\eqref{eq:Fidp0} and plot the dashed blue line in Fig.~\ref{fig:BraggBeamSplitter}(e).
    
    The analytic results exactly mirror the functional dependence on $\Omega_0$ of the numerical data. Both analytics and numerics show an exponential increase of losses towards large values of $\Omega_0$ that is harmonically modulated as showcased in Eq.~\eqref{eq:gamma2+}. On top of that, we find good quantitative agreement up to Rabi frequencies of $\Omega_0 < 5 \omega_\mathrm{r}$. Towards larger $\Omega_0$, time-dependent ac Stark shifts proportional to $\Omega^2(t)$ become increasingly significant. These are not taken into account in the LZ rates defined in Appendix~\ref{app:LandauZenerLosses}. Still, for relevant values of the Rabi frequency, formula \eqref{eq:gamma2+} gives remarkably good results in light of the fact that it is based on a simple two-level approximation accounting for losses to the energetically closest-lying level only.
    
    Unfortunately, we cannot simply apply Eq.~\eqref{eq:gamma2+} to higher orders of Bragg diffraction, as it turns out that in these cases ac Stark shifts are relevant for all values of the Rabi frequencies. Thus, we currently do not have an analytic expression for LZ losses applicable to higher Bragg orders. However, we can show that also for these orders LZ losses are simply a result of two-level dynamics.  
    
\begin{figure*}[t]
                 \includegraphics[width=0.9\textwidth]{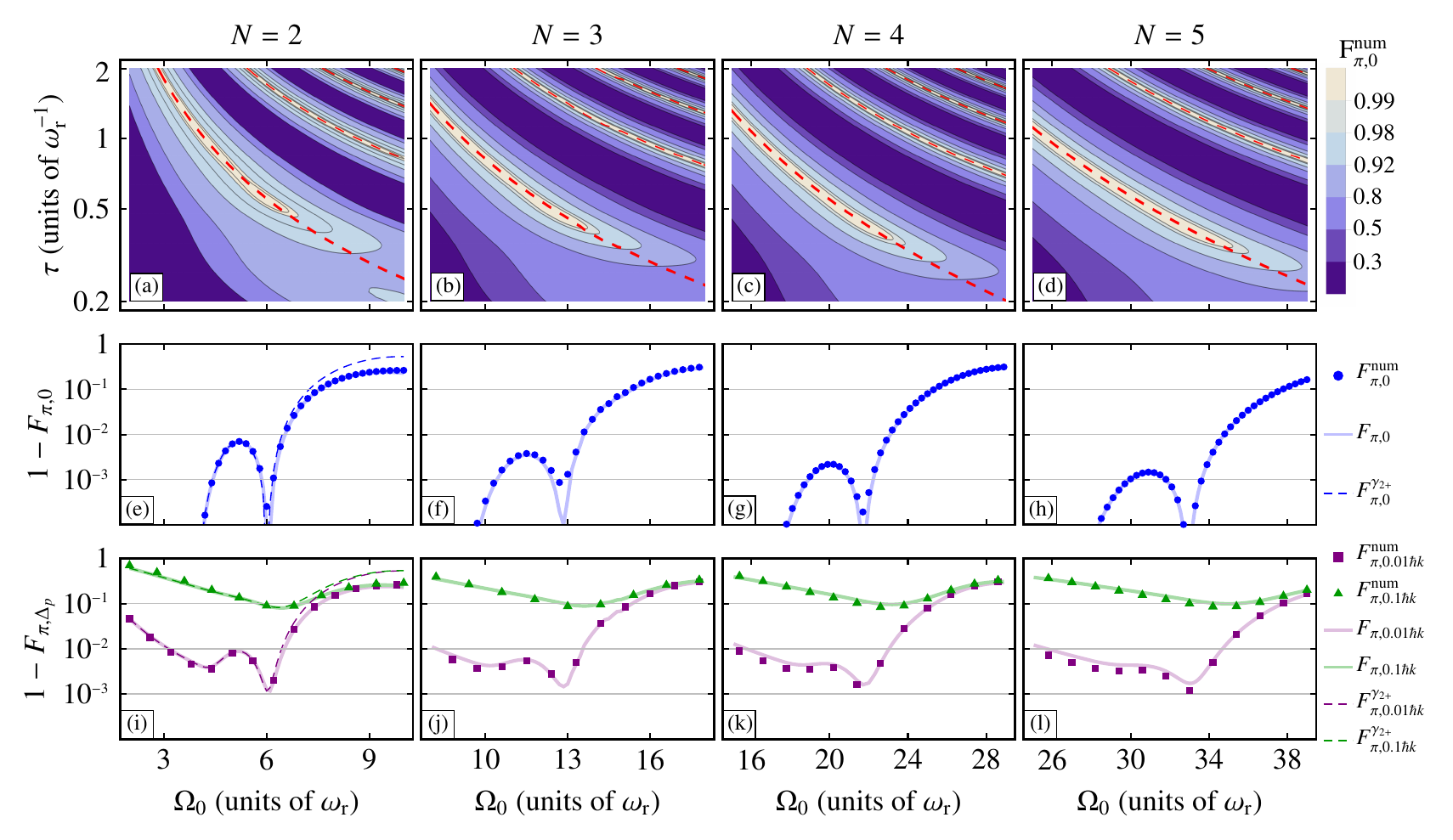}
                 \caption{\textbf{Bragg mirror}. Top row: Pulse duration~\eqref{eq:pulseduration} (red dashed line) on top of the numerically determined mirror fidelities~\eqref{eq:FidNump0} introduced in Figs.~\ref{fig:Bragg_Fid_NoNorm_Num}(e)-(h). Middle row: Fidelity loss as a function of peak Rabi frequency $\Omega_0$ and pulse durations computed via Eq.~\eqref{eq:pulseduration}. Mirror fidelity loss has been determined numerically (blue disks), via Eq.~{\eqref{eq:Fidp0}} with values~\eqref{eq:gammaNum+-} (solid line). The dashed line in panel (e) for $N=2$  is obtained by inserting $\gamma_{2,+}$~\eqref{eq:gamma2+} into Eq.~{\eqref{eq:Fidp0}} and denoted $F^{\gamma_{2,+}}_{\pi,0}$ in the legend on the right.   Bottom row: Fidelity loss extracted from exact numerics (squares \& triangles) is again compared to analytic results (solid \& dashed lines) similar to middle row. Fidelities~\eqref{eq:FidNumpAvg} are now averaged over a wave packet with finite momentum width $\pspread = 0.01 \hbar k$ (purple squares and lines) and $\pspread = 0.1 \hbar k$ (green triangles and lines). $\tilde{\gamma}_{N,\pm}$ values have been obtained again via Eq.~\eqref{eq:gammaNum+-} (solid lines) or for $\gamma_{2,+}$ using Eq.~\eqref{eq:gamma2+} (dashed lines) in panel (i).}\label{fig:BraggMirror}
        \end{figure*}

    To see this we infer values for $\tilde{\gamma}_{2,+}$ and $\tilde{\gamma}_{N,\pm}$, for orders $N=3,4,5$ from the exact numerical solution of the Schr\"odinger equation in an equivalent two-level approximation. We numerically calculate the populations $P_{N-2,\pm}$ of the states $\ket{N-2,\pm}$ which are energetically closest to the states $\ket{N,\pm}$ (for $N=2$, only  $\ket{0,+}$ with $P_{0,+}$ is relevant). In accordance with the two-level approximation we assume the amplitude loss parameters to be defined entirely by these populations:
    \begin{subequations}\label{eq:gammaNum+-}
    \begin{align}
    \tilde{\gamma}_{2,+} &= -\frac{1}{2} \ln{\left(1-2 P_{0,+}\right)},\\
    \tilde{\gamma}_{3,\pm} &= -\frac{1}{2} \ln{\left(1-2 P_{1,\pm}\right)}.
    \end{align}
    We remark that, as pointed out in Appendix~\ref{app:LZTruncation} in the context of the LZ phase, the spectra of the Hamiltonians make it necessary in the cases of $N=4$ and $5$ to include the coupling to the states $\ket{N-4,\pm}$
    \begin{align}
    \tilde{\gamma}_{N,\pm} &= -\frac{1}{2} \ln{\left(1-2 (P_{N-2,\pm}+P_{N-4,\pm})\right)},
    \end{align}
    \end{subequations}
    which can be still achieved in the spirit of a two-level description when performing the appropriate hybridization of the states $\ket{N+2,\pm}$ and $\ket{N+4,\pm}$ also discussed in Appendix~\ref{app:LZTruncation}. The fidelities $\Fid_{\dPhi,0}$~\eqref{eq:fidelitiesAna} in Fig.~\ref{fig:BraggBeamSplitter} and Fig.~\ref{fig:BraggMirror} using values $\tilde{\gamma}_{2,+}$ as well as $\tilde{\gamma}_{N,\pm}$~\eqref{eq:gammaNum+-}, $N=3,4,5$, have no superscript to differentiate them from $\Fid^{\gamma_{2,+}}_{\dPhi,0}$ using $\gamma_{2,+}$~\eqref{eq:gamma2+}.
    
    As the solid blue line in Fig.~\ref{fig:BraggBeamSplitter}(e) splendidly matches the exact numerics, it is clear that losses from the subspace $\ket{\pm N\hbar k}$ can be linked to two-level dynamics for all Bragg orders treated here. An accurate description requires the adaptation of the LZ theory in Ref.~\cite{Vasilev2004} that provides the LZ coefficients~\eqref{eq:abcoeffs} used for $N=2$.
    
    In Fig.~\ref{fig:BraggBeamSplitter}(i), we extend our discussion of the pulse parameters indicated  by the dashed red line in Fig.~\ref{fig:BraggBeamSplitter}(a) to the experimentally more relevant case of an atomic wave packet with finite momentum width. Here, we contrast the results of Eq.~\eqref{eq:FidNumpAvg} (squares and triangles) with our analytic fidelity~\eqref{eq:FidAverageGeneral} assuming a Gaussian momentum distribution with widths $\pspread = 0.01\hbar k$ and $0.1\hbar k$. Again, we find good agreement between numerics and analytics for both the numerically extracted LZ loss (solid lines) and the fully analytic expression Eq.~\eqref{eq:gamma2+} (dashed lines). The Doppler detuning mixes the (anti)symmetric states $\asket{\pm}{N}{p}$ and consequently changes the differential phase and therefore the Bragg condition reducing the fidelities in Fig.~\ref{fig:BraggBeamSplitter}(j) compared to Fig.~\ref{fig:BraggBeamSplitter}(e). In the limit of small $\Omega_0$ and long pulse durations fidelities are decreased as a result of the velocity filter effect caused by the Doppler detuning that is determined by the parameter $\eta(p)\propto \tspread^2$ [see Eq.~\eqref{eq:eta}]. Hence, towards shorter pulse widths the velocity filter effect quickly diminishes, especially for the $\pspread = 0.01\hbar k$ wave packet. While in case of such narrow momentum widths, for values  $\Omega_0 \gtrapprox 3 \omega_\mathrm{r}$ LZ physics  already discussed for $p=0$ quickly dominates the pulse fidelities, averaging over a larger uncertainty in momentum $\pspread = 0.1 \hbar k$ considerably washes out these features.
    
    The remaining panels of Fig.~\ref{fig:BraggBeamSplitter} confirm that our analytic model equally applies to Gaussian Bragg pulses of orders $N=3,4,5$. As before, we select the pulse parameters represented by the red dashed lines in Figs.~\ref{fig:BraggBeamSplitter}(b)-(d) for pulses investigated in Figs.~\ref{fig:BraggBeamSplitter}(f)-(h) and \ref{fig:BraggBeamSplitter}(j)-(l). Figures~\ref{fig:BraggBeamSplitter}(f)-(h) prove that as well for these higher orders the fidelity is reduced due to transitions to the closest state in energy $\ket{N-2,\pm}$ (hybridized level of $\ket{N+2,\pm}$ and $\ket{N+4,\pm}$ for $N=4,5$). Despite the fact that our perturbative model for the Doppler detuning underestimates the magnitude of the velocity filter for $\pspread = 0.1\hbar k$ and overestimates it for $\pspread = 0.01\hbar k$ in case of a finite momentum width, we still find good qualitative agreement in Figs.~\ref{fig:BraggBeamSplitter}(j)-(l) with regards to the exact numerics.
    
    \subsubsection*{Gaussian mirror pulses}
    In the case of Bragg mirrors $(\dPhi=\pi)$ for diffraction orders $N=2,3,4,5$, corresponding to momentum transfers of $4\hbar k,6\hbar k,8\hbar k,10\hbar k $, respectively, Fig.~\ref{fig:BraggMirror} paints a picture very similar to the discussion of the beam splitter. Nonetheless,
    Figs.~\ref{fig:BraggMirror}(e)-(h) show that in contrast to the beam splitter the longer pulse widths (in case of a given value of $\Omega_0$) suppress nonadiabatic losses. For the very same reason fidelity loss is visibly reduced in Figs.~\ref{fig:BraggMirror}(e)-(h) when directly comparing it to Figs.~\ref{fig:BraggBeamSplitter}(e)-(h). Following the same logic as before, however, the results confirm that LZ transitions to the closest state in energy are responsible for losses in amplitude during the Bragg mirror process. Moreover, it can be seen looking at Figs.~\ref{fig:BraggMirror}(i)-(l) that our perturbative treatment of the Doppler detuning accurately models the velocity filtering properties of a Bragg mirror on a quantitative level for all orders considered here. Owing to the fact that temporal mirror pulse widths are about twice the ones for beam splitters for the same $\Omega_0$, there is reduced acceptance for off-resonant momentum classes visible in Figs.~\ref{fig:BraggMirror}(i)-(l) which feature fidelity losses in the limit of small peak Rabi frequencies approaching unity. 
    
    In fact, when performing a Gaussian quasi-Bragg pulse of order $N=2,3,4,5$ for wave packets with finite momentum widths the results demonstrate that there exist optimal combinations of parameters $\{\Omega_0,\tspread\}$ for Bragg beam splitting pulses [see Figs.~\ref{fig:BraggBeamSplitter}(i)-(l)] and mirror pulses alike [see Figs.~\ref{fig:BraggMirror}(i)-(l)]  that minimize nonadiabatic losses as well as the impact of the velocity filter.
    
    \section{Conclusions}\label{sec:Conclusions}
    The comprehensive comparison with exact results in this paper, obtained by numerically integrating the Schr\"odinger equation, exemplifies that our scattering matrix precisely describes the dynamics of Gaussian Bragg pulses in the so-called quasi-Bragg regime. At the same time, we have provided simple formulas highlighting the analytic dependence of the dynamic phase, the LZ phases, as well as the LZ losses on the Bragg pulse parameters. Although, with regards to the LZ losses we only give such a formula for the diffraction order $N=2$, our analysis leaves no doubt that the logic of nonadiabatic losses within a two-level system can be extended to diffraction orders $N>2$. Nonetheless, the LZ formula needs to be adapted to these cases. 
    
    At this point, we want to underline once more that the fidelities introduced in Sec.~\ref{sec:BraggDiffQuality} of this paper to benchmark our analytic theory against the exact integration of the Schr\"odinger equation are of limited value for experiments. The effects of spontaneous emission have not been part of our analysis although they pose significant constraints on the Rabi frequencies available to perform high-fidelity Bragg pulses~\cite{Szigeti2012}. In their study, Szigeti et al. conclude that for the example of $^{87}$Rb atoms due to the effects of spontaneous emission viable Bragg orders are restricted to $N\leq 5$, which is the same range of orders discussed in this article. Yet, the logic developed in this paper equally applies to quasi-Bragg pulses of higher orders $N>5$.
    
    The Bragg scattering theory developed in the previous sections allows for perspectives on Bragg pulses that serve as a basis for analytic models of complete interferometry sequences. This foundation promises significantly increased insight and precision when studying systematic effects related to the imperfections of the diffraction process. Before we elaborate on how to proceed to the discussions of the signal of atom interferometers, we put our model in clear context regarding the preceding theory that we seek to complement.
    
    \subsection{Comparison to existing theory}\label{sec:Comparison}
    As pointed out earlier, the majority of existing descriptions aim at transferring the concept of a two-level system being diabatically coupled by a Rabi frequency which is valid in the limit of asymptotically long pulse durations (deep-Bragg) to quasi-Bragg pulses. They do so by averaging over the non-negligible off-resonant transitions arising in this regime and introducing an effective Rabi frequency that couples the two resonant momentum states. A more quantitative discussion of the relationship between our paper and the methods and results obtained while relying on the adiabatic elimination, in particular presented by M\"uller et al.~\cite{Mueller2008PRA} and Giese et al.~\cite{Giese2013}, is certainly fruitful but would undoubtedly exceed the scope of this paper.
   
    Instead, we would like to emphasize once again the conceptual differences between the application of the adiabatic theorem in our model and the adiabatic elimination of off-resonant states. We stress foremost the efficiency of our approach when increasing the Bragg order $N$. Since our formalism only requires us to calculate eigenenergies of finite-dimensional Hamiltonians, taking into account more states does not significantly increase the complexity of computing the quantities in Tab.~\ref{tab:params}, whereas the adiabatic elimination of additional states that become relevant when increasing the Bragg order $N$ is more complicated. M\"uller et al. arrive at an expression for the effective Rabi frequency as a power series expansion in $\Omega(t)$ and $\dot{\Omega}(t)$ in Eq.~(48) of~\cite{Mueller2008PRA}. This result requires in particular to either numerically calculate the eigenvalues $a_n$ and $b_n$ of the Mathieu equation and to ensure their convergence for the desired orders in $\Omega(t)$ or to find closed expressions for these parameters (see Appendix B in Ref.~\cite{Mueller2008PRA}).
    
    Our solution for the differential dynamic phase \eqref{eq:BraggPhases} can be also expanded in orders of $\Omega(t)$ and $\dot{\Omega}(t)$ and we expect to be able to reproduce Eq.~(48) in~\cite{Mueller2008PRA} with similar or even higher accuracy, the reason being that our formalism allows us to obtain the necessary eigenvalues by diagonalizing finite-dimensional matrices which can be achieved with efficient and accurate numerical routines even for high truncation orders. Gochnauer et al.~\cite{Gochnauer2019} successfully calculate effective Rabi frequencies and diffraction phases in a similar fashion and find good agreement with their experimental data taking into account nine states for a second-order Bragg mirror pulse provides sufficient convergence (see Ref.~[34] in~\cite{Gochnauer2019}). While our formalism is closely related to the Bloch band picture (see below for more details on the relation of both approaches), instead of solving the entire Bloch band structure for arbitrary Rabi frequencies and all possible (quasi)momenta p we reduce the problem to the diagonalization of low dimensional Hamiltonians only for p=0.
    
    This procedure results in expressions like~\eqref{eq:BraggPhases} that are instructive as well as efficient and provide us excellent agreement with exact numerical calculations in the experimentally most relevant regime of quasi-Bragg pulses with Gaussian envelopes~\cite{Keller1999,Mueller2008PRA}. Furthermore, our model applies to atoms with velocity distributions that are narrow on the scale of the photon recoil of the Bragg lattice as we include linear Doppler shifts via perturbation theory up to first order. In our formalism a nonvanishing (quasi)momentum $p$ couples the (in zeroth order of $p$) disjoint (anti)symmetric Hilbert spaces $\mathscr{H}_{p\alpha\pm}$ [see Eq.~\eqref{eq:Halpha}]. Ac\-cor\-ding to the theory presented here, this coupling influences the differential phase~\eqref{eq:phases}, equivalent to an effectively reduced Rabi frequency,  and results in a velocity filter that depends on the (quasi)momentum $p$. The reduction of the transfer efficiencies of Bragg pulses due to a Doppler detuning has previously been modelled numerically in the work published by Szigeti et al. \cite{Szigeti2012} and perturbatively in the case of rectangular double Bragg pulses by Giese et al.~\cite{Giese2013}.
    
    Finally, we discuss in more detail how the model developed here connects to the picture of Bloch bands in optical lattices. The latter is the natural framework to treat matter wave diffraction via Bloch oscillations \cite{Dahan1996,Wilkinson1996,Peik1997}. This ansatz has previously been shown to provide analytic insight into the process of Bragg diffraction in the limit of weak lattice coupling ($\Omega \lesssim 2 \omega_\mathrm{rec}$) \cite{Champenois2001,Buechner2003} while Gochnauer et al.~\cite{Gochnauer2019} numerically extract the Bloch energy bands to calculate effective Rabi frequencies without restricting the potential depth.
    
\begin{figure}[!htb]
                \includegraphics[width=0.48\textwidth]{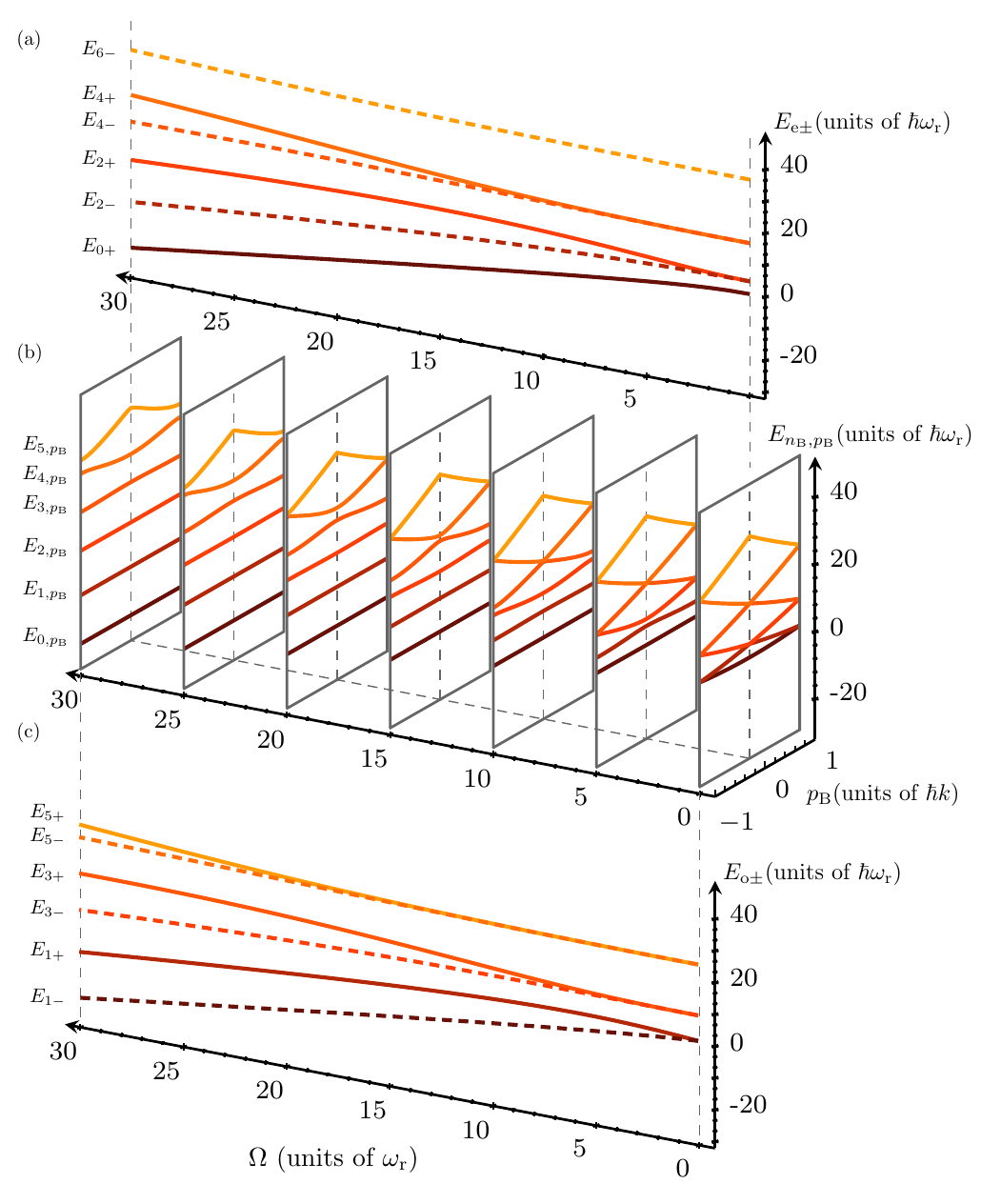}
                \caption{(b) The lowest six energy bands $E_{n_\mathrm{B},p_\mathrm{B}} (\Omega)$ $(n_\mathrm{B}=0,1,\dots,5)$  in the first Brillouin zone for Bloch band quasimomenta $p_\mathrm{B}\in [-\hbar k, \hbar k]$  and different values of the Rabi frequency $\Omega$. They are obtained by diagonalizing the Hamiltonian in Eq.~\eqref{eq:splitHam} after truncating orders $n>5$. In case of $\Omega=0$, always two energy bands are degenerate either at the edges ($p_\mathrm{B}=\pm 1 \hbar k$) or the center ($p_\mathrm{B}=0 \hbar k$) of the Brillouin zone. For nonzero values of $\Omega$ the degeneracy is lifted and the bands separate. (a,c) show cuts along the $\Omega$ axis at the edge ($p_\mathrm{B}=- 1 \hbar k$)  and the center ($p_\mathrm{B}=0\hbar k $) of the Brillouin zone respectively. The energies of the Bloch spectrum in (a) [(c)] are identical to the spectra of the Bragg Hamiltonians in Eqs.~\eqref{eq:Hblocks} displayed in Fig.~\ref{fig:SpectraOmegaScan}(a) [Fig.~\ref{fig:SpectraOmegaScan}(b)].
                }\label{fig:BlochSpectra}
        \end{figure}

    Fig.~\ref{fig:BlochSpectra}(b) displays the lowest-energy bands $E_{n_\mathrm{B},p} (\Omega)$ $(n_\mathrm{B}=0,1,\dots    ,5)$ of the fundamental Hamiltonian in Eq.~\eqref{eq:splitHam} in the first Brillouin zone for quasimomenta $p_B\in [-\hbar k, \hbar k]$ and for different Rabi frequencies increasing from $\Omega=0$ to $30\omega_\mathrm{r}$. We included a subscript in $p_\mathrm{B}$ at this point to differentiate it from the Bragg (quasi)momentum variable $p$ we introduced earlier. For a free atom, that is, for $\Omega=0$, a narrow wave packet with mean momentum $-N\hbar k$ (in the rest frame of the lattice) consists of a superposition of Bloch states around the points of degeneracy of the $N$-th and the $(N-1)$-th band in the Bloch spectrum. For odd $N$ this degeneracy occurs at a quasimomentum $p_\mathrm{B}=\pm\hbar k$, and for even $N$ at $p_\mathrm{B}=0$ [see rightmost panel in Fig.~\ref{fig:BlochSpectra}(b)]. When the optical lattice is ramped up adiabatically, the atom remains in the superposition of states in the $N$-th and the $(N-1)$-th band the degeneracy of which will now be lifted [see panels in Fig.~\ref{fig:BlochSpectra}(b) for $\Omega>0$]. 
    
    Based on this picture Gochnauer et al.~\cite{Gochnauer2019} explain that the band gap is equivalent to the effective Rabi frequency for oscillations between the momentum eigenstates $\ket{\pm N \hbar k}$ coupled by the Bragg pulse. This explanation was confirmed in~\cite{Gochnauer2019} by a comparison of numerically calculated band gaps and experimentally determined Rabi frequencies measured at constant potential depth. Gochnauer et al. also show that the Bragg diffraction is accompanied by a global phase (diffraction phase) which corresponds to the energetic shift of the center of the band gap with respect to the position of the degeneracy point at vanishing potential. 
    
    All of these important observations are fully confirmed and complemented with further insights by our analytic model. The spectra of the Hamiltonians derived here for Bragg scattering of even and odd order, as shown in Fig.~\ref{fig:SpectraOmegaScan}, correspond exactly to cuts through the Bloch spectra at constant quasimomentum, $p_\mathrm{B}=\pm\hbar k$ and $=0$ [see Figs.~\ref{fig:BlochSpectra}(a) and~\ref{fig:BlochSpectra}(c) and their connection to Fig.~\ref{fig:BlochSpectra}(b) indicated by the vertical, dashed, grey lines]. The decomposition of the Hamiltonians into their symmetric and antisymmetric components allows us to determine the energy gap very accurately already for low truncation orders, and avoids the need to numerically determine Bloch spectra for variable potential depths. The net differential dynamic phase $\dPhi^{\mathrm{dyn}}=\SASphase^\mathrm{dyn}_{N+} - \SASphase^\mathrm{dyn}_{N-}$, with $\SASphase^\mathrm{dyn}_{N\pm}$ given in Eq.~\eqref{eq:BraggPhases}, is of course nothing else than what Gochnauer et al. refer to as the integrated effective Rabi frequency. However, we emphasize that the very concept of an effective Rabi frequency alludes to the concept of diabatic dynamics described by an effective Hamiltonian. We hope that the present paper has made it sufficiently clear that it is much more economic and appropriate to consider this phase as a differential dynamic phase in the sense of the adiabatic theorem. After all, it is this interpretation of the phase which allows us to systematically determine corrections beyond the ideal adiabatic limit. The application of the adiabatic theorem to Bragg diffraction, which was clearly anticipated in Ref.~\cite{Gochnauer2019}, together with the first-order corrections regarding LZ phases, LZ losses, and Doppler shifts indeed give an exhaustive analytic description of all high-quality Bragg pulses.
    
    \subsection{Introduction to interferometry sequences}\label{sec:Interferometer}
    In this section we briefly outline how to apply the description we have developed to systematically analyze the phases of atom interferometers employing Bragg optics. At this point our goal is to convey the idea of the calculation only, details will be given in a future publication. The aim of this endeavour is to arrive at an analytic expression for the signal of the interferometer that explicitly depends on the pulse parameters, as well as on properties of the atomic wave packets such as its momentum width.
    
\begin{figure}[t]
                \includegraphics[width=0.48\textwidth]{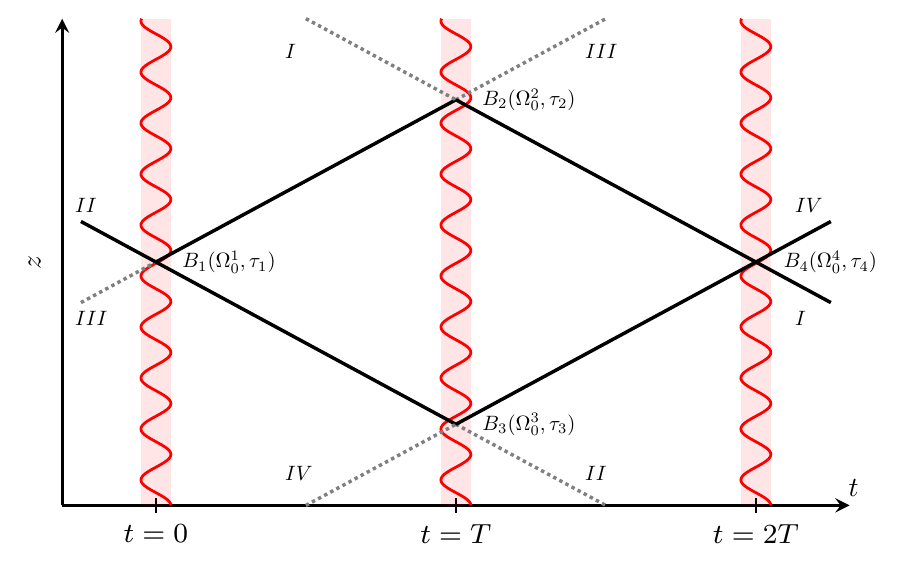}
                \caption{$(z,t)$ representation of a Mach-Zehnder interferometer as a four-port device possessing four input and four output ports $I,\,II,\,III,\,IV$. The beam splitters $B_1$ ($B_4$) at time $t=0$ $(t=2T)$ and mirrors $B_2$ ($B_3$) at time $t=T$ are individual scattering matrices \eqref{eq:BraggScatteringMatrixFinal}. The trajectories marked by the gray dotted lines are either not populated or loss channels. Given a sufficiently small expansion rate of the atomic ensemble on the order of the photon recoil $\hbar k$, these channels are spatially well separated from the detection ports and can thus be assumed not to contribute to the signal.}\label{fig:AIScheme}
        \end{figure}

    Figure~\ref{fig:AIScheme} depicts an arrangement of four of these atom optic elements building up a Mach-Zehnder interferometer in a space-time diagram. Assuming the beam splitters and mirrors are based on $N$th-order Bragg scattering, we define a momentum basis in the four input and output channels $I$ to $IV$, as shown in Fig.~\ref{fig:AIScheme}, by 
    \begin{alignat*}{4}
        &\ket{p}_{\mathrm{in},I}&&=\ket{N\hbar k+p}_I\quad
        &&\ket{p}_{\mathrm{out},I}&&=\ket{N\hbar k+p}_I \\
        &\ket{p}_{\mathrm{in},{II}}&&=\ket{-N\hbar k+p}_{II}\quad &&\ket{p}_{\mathrm{out},{II}}&&=\ket{-N\hbar k+p}_{II} \\
        &\ket{p}_{\mathrm{in},{III}}&&=\ket{N\hbar k+p}_{III}\quad &&\ket{p}_{\mathrm{out},{III}}&&=\ket{N\hbar k+p}_{III} \\
        &\ket{p}_{\mathrm{in},{IV}}&&=\ket{-N\hbar k+p}_{IV}\quad &&\ket{p}_{\mathrm{out},{IV}}&&=\ket{-N\hbar k+p}_{IV}. 
    \end{alignat*}
    In analogy to the notation of the scattering matrix of a single pulse~\eqref{eq:BraggScatteringMatrixIdeal} we can define the scattering matrix of the complete interferometer:
    \begin{align}\label{eq:MZScatteringMatrixOperator}
            \mathcal{S}_{\mathrm{MZ}}=\int^{\hbar k /2}_{- \hbar k /2}\difp\sum_{l,m=I}^{IV} [I(p)]_{lm}\hspace{3pt}
            \smash{\ket{p}}_{\mathrm{out},l} \hspace{1.5pt}{}_{\mathrm{in},m\!\!}\bra{p}.
    \end{align}
    The matrix $I(p)$ represents the combination of the individual scattering matrices~\eqref{eq:BraggScatteringMatrixFinal} that make up the interferometer  including two beam splitters, $B_1$ and $B_4$, and a pair of mirrors, $B_2$ and $B_3$. They depend on the individual pulse parameters, i.e., $B_j = B_j (p,\Omega^{j}_{0},\tspread^{j})$ in case of Gaussian pulses with peak Rabi frequency $\Omega^{j}_{0}$ and temporal pulse width $\tspread^{j}$. Every scattering matrix $B_j$ is a two-port device with two input and two output ports. The matrices ideally map the momentum eigenstates $\ket{\pm N \hbar k + p}$ onto a state vector in the Hilbert space spanned by $\left(\ket{- N \hbar k + p},\ket{N \hbar k + p}\right)$. The Mach-Zehnder interferometer is consequently a four-port device the scattering matrix of which can be constructed from a simple multiplication of the individual scattering matrices $B_j$. 
    
    Specifically, we find
    \begin{align}\label{eq:MZMatrix}
        I(p) = 
        \left(\begin{array}{@{}c|c|c|c@{}}
            [B_4]_{11} &\multicolumn{2}{c|}{0} & [B_4]_{12} \\ \hline
            \multirow{2}{*}{0}& \multicolumn{2}{c|}{\multirow{2}{*}{\scalebox{1.3}
        {$\mathds{1}$}}}& \multirow{2}{*}{0}\\
            &\multicolumn{2}{c|}{}&\\\hline
            [B_4]_{21}&\multicolumn{2}{c|}{0} & [B_4]_{22} \\
        \end{array}\right) 
        \cdot 
        \mathcal{C}
        \cdot
        \left(\begin{array}{@{}c|c|c|c@{}}
            1 &\multicolumn{2}{c|}{0} & 0 \\ \hline
            \multirow{2}{*}{0}&\multicolumn{2}{c|}{\multirow{2}{*}{\scalebox{1.3}
        {$B_1$}}}& \multirow{2}{*}{0}\\
            &\multicolumn{2}{c|}{\multirow{2}{*}{}}&\\\hline
            0&\multicolumn{2}{c|}{0} & 1 \\
        \end{array}\right)\quad ,
    \end{align}
    where 
    \begin{align*}
      \mathcal{C} = \left(\begin{array}{@{}c|c|c|c@{}}
         [B_2]_{11} & 0 & [B_2]_{12} & 0 \\ \hline
         0 & [B_3]_{11} & 0 & [B_3]_{12} \\ \hline
         [B_2]_{21} & 0 & [B_2]_{22} & 0 \\ \hline
         0 & [B_3]_{21} & 0 & [B_3]_{22} \\
        \end{array}\right).
    \end{align*}
   In a standard Mach-Zehnder interferometer, three of the input ports $I,\,III,\,IV$  are empty and two output ports $II,\,III$ are normally considered loss channels from the interferometer (represented by the dotted gray lines). Assuming these paths do not contribute to the signal, e.g., due to spatial separation from the detection zone, we can trace over them and only take into account the main interferometric paths marked by the black solid lines in Fig.~\ref{fig:AIScheme}. This is justified, as ultracold sources for atom interferometry like Bose-Einstein condensates feature effective temperatures on the order of $50-100~\mathrm{pK}$~\cite{Kovachy2015PRL,RudolphPHDthesis}. Adopting this assumption Eq.~\eqref{eq:MZScatteringMatrixOperator} simplifies to
    \begin{align}\label{eq:ReducedMZScatteringMatrixOperator}
            \tilde{\mathcal{S}}_{\mathrm{MZ}}&=\int^{\hbar k /2 }_{-\hbar k  /2}\difp\sum_{l=I,IV}\sum_{m=II,III} [I(p)]_{lm}\hspace{3pt}
            \smash{\ket{p}}_{\mathrm{out},l}\hspace{1.5pt} {}_{\mathrm{in},m\!\!}\bra{p}\\
            &=\int^{\hbar k /2}_{-\hbar k /2}\difp\sum_{s,s'=\mp} [\tilde{I}(p)]_{ss'}
            \ket{\vphantom{'} s N \hbar k + p}\bra{s' N\hbar k +p},
    \end{align}
    where in the second line we regained a familiar basis notation in momentum states $\ket{\pm N\hbar k +p}$ by introducing the $(2x2)$-matrix
    \begin{align}
             \tilde{I}(p) &= B^T_4 \cdot \tilde{C}\cdot B_1 .
    \end{align}
    Here, the mirror matrix $\tilde{C}$ combines the transfer coefficients of both mirror scattering matrices, $\left[B_2\right]_{12}$ and $\left[B_3\right]_{21}$, as diagonal entries
    \begin{align*}
        \tilde{C} &= 
             \left(\begin{array}{@{}cc@{}}
            0 &\left[B_2\right]_{12}\\
             \left[B_3\right]_{21}& 0\\
            \end{array}\right).
    \end{align*}
    After adequate multiplication of matrices $B_j$ we arrive at analytic expressions for the output port populations that crucially depend on the pulse parameters $\{\Omega^{j}_{0},\tspread^{j}\}$ with $j=1,2,3,4$. Using this signal, we can study quantities like amplitude, contrast and phase offset of the interferometer and analyze their dependence of the pulse parameters. 
    \section{Outlook}\label{sec:Outlook}
    Our analytic model of Bragg diffraction based on the adiabatic theorem is the stepping stone towards numerous subsequent investigations into the optic elements in atom interferometry with great utility and precision. It will be curious to see in the future how the logic behind our approach can be adapted to Raman diffraction~\cite{Kasevich1991}, the other dominant atom optics operation in atom interferometry, with its differentiating features of internal state labeling and reduced velocity selectivity.
    In a more immediate next step, it will be worthwhile to seek the extension of our formalism to double Bragg diffraction~\cite{Giese2013,Kueber2016,Ahlers2016}, a technique that allows for symmetric interferometer configurations with state-of-the-art momentum separations when paired with Bloch oscillations~\cite{Gebbe2019}.
    
    In Sec.~\ref{sec:Conclusions} we have pointed out that the theory developed in this paper is closely related to the Bloch band picture. It is therefore natural to suspect that we can extend our findings to an analytic description of the dynamics of Bloch oscillations with the theory presented here by introducing a time-dependent laser phase, $\Lphase(t)$. 
    
    The avenue of a comprehensive theoretical framework unifying the regimes of Bragg pulses and Bloch oscillations holds exciting new opportunities. With analytic insight into how to seamlessly traverse from one to the other it is not difficult to imagine the existence of new pulse shapes specifically tailored to the requirements of modern atom interferometry devices, similar to the numerically devised proposal by Kovachy et al.~\cite{Kovachy2012}. Such research efforts can undoubtedly be facilitated through the application of optimal control algorithms which will be able to leverage the findings provided in this paper. More to the point, we have outlined in Sec.~\ref{sec:Conclusions} how to model complete interferometers based on our description enabling optimization routines that target interferometric quantities rather than the features of a single element of the interferometer. 
    
    In future work we will illuminate the impact the diffraction processes have on the signal of atom interferometers. A better understanding of diffraction phases~\cite{Buechner2003} is paramount to facilitate the development of new and improvement of existing mitigation strategies~\cite{Estey2015,Plotkin-Swing2018,Gochnauer2019,McAlpine2019}. A comprehensive study of these phenomena requires the inclusion of realistic three-dimensional light pulses considering as well the effects of the profile of the laser beam. Even though we have restricted ourselves to the case of one-dimensional scattering in this paper, the introduction of a position dependence into the amplitude and phase of the laser will allow for a systematic discussion of diffraction processes with realistic optical lattices on a microscopic level.
    
    The codes used to generate these results are available~\cite{Siemss2020}.
    
    \begin{acknowledgements}
        We thank E. Giese and S. Loriani for carefully reading the manuscript and C. Schubert for fruitful discussions. This work was funded by the Deutsche Forschungsgemeinschaft (German Research Foundation) under Germany’s Excellence Strategy (EXC-2123 QuantumFrontiers Grant No. 390837967) and through CRC 1227 (DQ-mat) within Projects No. A05 and No. B07, the Verein Deutscher Ingenieure (VDI) with funds provided by the German Federal Ministry of Education and Research(BMBF) under Grant No. VDI 13N14838 (TAIOL), and the Ger\-man Space Agency (DLR) with funds provided by the German Federal Ministry of Economic Affairs and Energy (BMWi) due to an enactment of the German Bundestag under Grant No. DLR 50WM1952 (QUANTUS-V-Fallturm), 50WP1700 (BECCAL), 50WM1861 (CAL), 50WM2060 (CARIOQA) as well as 50RK1957 (QGYRO). We furthermore acknowledge financial support from "Niedersächsisches Vorab" through "Förderung von Wissenschaft und Technik in Forschung und Lehre" for the initial funding of research in the new DLR-SI Institute and  the “Quantum- and Nano Metrology (QUANOMET)” initiative within the project QT3.
    \end{acknowledgements}
    
    \appendix
    \section{HAMILTONIAN}\label{App:Hamiltonian}
    Here we provide some details on the decomposition of the Bragg Hamiltonian. The initial Hamiltonian \eqref{eq:Ham} can be rewritten as
    \begin{align*}
     \mathcal{H}^{\mathrm{MF}}&=\frac{\hat{p}^2}{2M}+\frac{\hbar\Omega(t)}{2}\left(e^{2i(k\hat{z}+\Lphase)}+e^{-2i(k\hat{z}+\Lphase)}\right)\\
    &=\int_{-\infty}^\infty\difp\left\{ \frac{{p}^2}{2M}|p\rangle\langle p|+\frac{\hbar\Omega}{2}\left(e^{2i\Lphase}|p+2\hbar k\rangle\langle p|+\mathrm{H.c.}\right)\right\}\\
    &=\int_{-\hbar k/2}^{\hbar k/2}\difp\sum_{n=-\infty}^\infty \left\{ \frac{(n\hbar k+p)^2}{2M}|n\hbar k+p\rangle\langle n\hbar k+p|\right.\\
    &\qquad\qquad\left.+\frac{\hbar\Omega}{2}\left(e^{2i\Lphase}|(n+2)\hbar k+p\rangle\langle n\hbar k+p|+\mathrm{H.c.}\right)\right\}\\
    &=\int_{-\hbar k/2}^{\hbar k/2}\difp\left\{  \mathcal{H}^{\mathrm{MF}}_{\mathrm{e}}(p)+ \mathcal{H}^{\mathrm{MF}}_{\mathrm{o}}(p)\right\}
    \end{align*}
    where
    
    \begin{align}
       \begin{split}
        \mathcal{H}^{\mathrm{MF}}_\mathrm{\alpha}(p)=\sum_{n\in \mathds{Z}_\alpha}\left\{\vphantom{\frac{(n\hbar k+p)^2}{2M} }\right.& \frac{(n\hbar k+p)^2}{2M}  \psigma{n,n}(p)\\
        &+ \frac{\hbar\Omega}{2}\left(e^{2i\Lphase}\psigma{n+2,n}(p)+\mathrm{H.c.}\left. \vphantom{\frac{(n\hbar k+p)^2}{2M} }\right)\right\}
       \end{split}
    \end{align}
    for $\alpha\in\{\mathrm{\mathrm{e},\,\mathrm{o}}\}$ and the summation is over even or odd numbers $\mathds{Z}_\mathrm{e}=2\mathds{Z}$ or $\mathds{Z}_\mathrm{o}=2\mathds{Z}+1$, respectively. This is Eq.~\eqref{eq:Halphap} of the main text.
    
    For the transformation of the Hamiltonian in the form \eqref{eq:fullHalpha} to the basis of (anti)symmetric states the following identities are useful,
        \begin{align*}
          \psigma{n,n}+\psigma{-n,-n}&=\psigma{n,n}^{\g}+\psigma{n,n}^{\u},\qquad(n>0),\\
          \psigma{0,0}&=\psigma{0,0}^{\g},
        \end{align*}
    and for the case $n\geq 2$
        \begin{align*}
          e^{2i\Lphase}(\psigma{n+2,n}+\psigma{-n,-(n+2)})+\mathrm{H.c.}&=\psigma{n+2,n}^{\g}+\psigma{n+2,n}^{\u}+\mathrm{H.c.},
        \end{align*}
    as well as 
        \begin{align*}
          e^{2i\Lphase}(\psigma{2,0}+\psigma{0,-2})+\mathrm{H.c.}&=\sqrt{2}\big(\psigma{2,0}^{\g}+\mathrm{H.c.}\big),\\
          e^{2i\Lphase}\psigma{1,-1}+\mathrm{H.c.}&=\psigma{1,1}^{\g}-\psigma{1,1}^{\u}.
        \end{align*}
    \section{LZ PHASES AND LOSSES}\label{app:LandauZener}
        \subsection{Derivation of Eq.~\eqref{eq:LZphase} for the LZ phase}\label{app:LandauZenerPhase}
        
        In order to cover LZ phases, we look for a solution of Eq.~\eqref{eq:dynamics} which is nondiagonal in the basis of instantaneous energy eigenstates, that is,
        \begin{align}\label{eq:UAnsatz}
             U_{\alpha\beta}(t,t_0)&=\sum_{n,m\in\mathds{N}_\alpha}
             e^{-i\SASphase_{n\beta}(t,t_0)}c^{\,\beta}_{nm}(t)|\beta,n,p;t\rangle\langle\beta,m,p;t_0|.
        \end{align}
        In the ideal adiabatic limit we have $c^{\,\beta}_{nm}(t)=\delta_{nm}$. Beyond the adiabatic limit we are particularly interested in the corrections to the coefficients $c_{NN}^{\,\beta}(t)$ as these enter the scattering matrix \eqref{eq:ProjectedBragg}. Inserting the ansatz for $U_{\alpha\beta}(t,t_0)$ in Eq.~\eqref{eq:UAnsatz} in the equation of motion \eqref{eq:dynamics}, using $\partial_t\SASphase_{k\beta}(t,t_0)=E_{k\beta}(t)/\hbar$, and taking the matrix element $\bra{\beta,n,p;t}\ldots\ket{\beta,m,p;t_0}$ one finds
        \begin{align}\label{eq:ccoeffs}
            \dot{c}^{\,\beta}_{nm}(t)=-\sum_{k\in\mathds{N}_\alpha}e^{-i[\SASphase_{k\beta}(t,t_0)-\SASphase_{n\beta}(t,t_0)]}
            G_{nk}^{\,\beta}(t){c}^{\,\beta}_{km}(t)
        \end{align}
        where $G_{nk}^{\,\beta}(t)=\bra{\beta,n,p;t}\partial_t\ket{\beta,k,p;t}$. As usual in the analysis of LZ dynamics, it is convenient to impose the gauge condition of parallel transport where $G_{nn}^{\,\beta}(t)=0$ (see~\cite{Xiao2010}). The set of equations \eqref{eq:ccoeffs} should be solved with initial condition $c^{\,\beta}_{nm}(0)=\delta_{nm}$. For the relevant coefficient $c_{NN}^{\,\beta}(t)$ one finds
        \begin{align}
            \dot{c}^{\,\beta}_{NN}(t)&=-\sum_{\overset{k\in\mathds{N}_\alpha}{k\neq N}}
            e^{-i[\SASphase_{k\beta}(t,t_0)-\SASphase_{N\beta}(t,t_0)]}
            G_{Nk}^{\,\beta}(t){c}^{\,\beta}_{kN}(t),\label{eq:cNN}\\
            \dot{c}^{\,\beta}_{kN}(t)&=-e^{-i[\SASphase_{N\beta}(t,t_0)-\SASphase_{k\beta}(t,t_0)]}
            G_{kN}^{\,\beta}(t){c}^{\,\beta}_{NN}(t).\nonumber
        \end{align}
        In the last equation we kept only the leading term in the sum. The adiabatic solution to the last equation is
        \begin{align*}
            c^{\,\beta}_{kN}(t)=-i\hbar e^{-i[\SASphase_{N\beta}(t,t_0)-\SASphase_{k\beta}(t,t_0)]}
            \frac{ G_{kN}^{\,\beta}(t)}{E_{N\beta}(t)-E_{k\beta}(t)}{c}^{\,\beta}_{NN}(t).
        \end{align*}
        Inserting this into \eqref{eq:cNN} yields
        \begin{align*}
            \dot{c}^{\,\beta}_{NN}(t)&=i\hbar\sum_{\overset{k\in\mathds{N}_\alpha}{k\neq N}}
            \frac{\left|G_{kN}^{\,\beta}(t)\right|^2}{E_{N\beta}(t)-E_{k\beta}(t)}{c}^{\,\beta}_{NN}(t).
        \end{align*}
        Solving this equation and taking the limit for final or initial times to $\pm\infty$, respectively, gives ${c}^{\,\beta}_{NN}=\exp(i\SASphase_{N\beta}^\mathrm{LZ})$ where the LZ phase is given by
        \begin{align}
            \SASphase_{N\beta}^\mathrm{LZ}&=\hbar\int_{-\infty}^\infty\mathrm{d}t \sum_{\overset{k\in\mathds{N}_\alpha}{k\neq N}}
            \frac{\left|G_{kN}^{\,\beta}(t)\right|^2}{E_{N\beta}(t)-E_{k\beta}(t)}
        \end{align}
        which is Eq.~\eqref{eq:LZphase} of the main text.
        
        \subsection{Truncation of the Hamiltonian for $N=2,3,4,5$}\label{app:LZTruncation}
        
        As stated in Eq.~\eqref{eq:H2e+}, for $N=2$ the truncated Hamiltonian in the symmetric subspace is
        \begin{align*}
            H^{(2)}_{\even,+}&=\hbar\omega_\mathrm{r}\begin{pmatrix}4&\sqrt{2}w\\\sqrt{2}w&0\end{pmatrix},
        \end{align*}
        where $w=\Omega(t)/2\omega_\mathrm{r}$. The corresponding truncated Hamiltonian in the antisymmetric subspace is trivial, $H^{(2)}_{\even,-}=4\hbar\omega_\mathrm{r}$ in the same approximation, and does not contribute a LZ phase. For $N=3$ one gets from Eq.~\eqref{eq:Hoddtrunc}
        \begin{align*}
            H^{(3)}_{\odd,\pm}&=\hbar\omega_\mathrm{r}\begin{pmatrix}9&w\\w&1\pm w\end{pmatrix}.
        \end{align*}
        For the other relevant cases $N=4$ and $5$ a similar truncation can be performed, but will produce worse results, since now also lower-lying states ($\ket{+,0,p}$ for $N=4$, and $\ket{\pm,1,p}$ for $N=5$) are neglected, which are of great importance for the spectrum of the Hamiltonian. This is clearly visible in the avoided crossings in Fig.~\ref{fig:SpectraOmegaScan}. In order to reduce the truncation error it is appropriate to perform a pre-diagonalization of the lower two levels, and discard the lowest lying dressed state. This procedure yields for $N=4$ and $5$
        \begin{align*}
            H^{(4)}_{\even,\pm}&=\hbar\omega_\mathrm{r}\begin{pmatrix}16&q_{4\pm}w\\q_{4\pm}w& e_{4\pm}(w)\end{pmatrix}, &
            H^{(5)}_{\odd,\pm}&=\hbar\omega_\mathrm{r}\begin{pmatrix}25&q_{5\pm}w\\q_{5\pm}w& e_{5\pm}(w)\end{pmatrix}.
        \end{align*}
        where $e_{4-}=4$, $e_{4+}(w)=2+\sqrt{4+2w^2}$, and $e_{5\pm}(w)=5\pm w/2+\sqrt{16\mp 4w+5 w^2/4}$. The last three expressions correspond to the larger eigenvalues of $H^{(2)}_{\even,+}$ and $H^{(3)}_{\odd,\pm}$, respectively. The off-diagonal elements in the last two Hamiltonians are as well affected by the prediagonalization and in principle have a more complicated $w$ dependence. In effect, the coupling will be somewhat smaller than $w$ on average. We cover this by including a parameter $q_{N\beta}$ which we fit to numerical data.
        \begin{table}[ht]
            \caption{\label{tab:LZvalues}%
            Parameters for LZ phases.}
            {\renewcommand{\arraystretch}{1.5} 
            \begin{ruledtabular}
            \begin{tabular}{c c c c }
                $N$ &   $\beta$ &   $q_{N\beta}$    &   $e_{N\beta}(w)$   \\[0.5ex] 
                \colrule
                2   &   +   &   $\sqrt{2}$  &   0\\
                2   &   -   &   0   &   0\\
                3   &   +   &   1   &   $1+w$\\
                3   &   -   &   1   &   $1-w$\\
                4   &   +   &   0.58&   $2+\sqrt{4+2w^2}$\\
                4   &   -   &   1   &   $4$ \\
                5   &   +   &   0.45&   $5 + w/2+\sqrt{16 - 4w+5 w^2/4}$\\
                5   &   -   &   1   &   $5 - w/2+\sqrt{16 + 4w+5 w^2/4}$\\
          \end{tabular}
          \end{ruledtabular}
          }
        \end{table}
        Thus, truncated Hamiltonians in all cases $N=2,3,4,5$ are of the form
        \begin{align*}
            H^{N}_{\alpha,\beta}=\hbar\omega_\mathrm{r}\begin{pmatrix}N^2&q_{N\beta}w\\q_{N\beta}w&e_{N\beta}(w)\end{pmatrix},
        \end{align*}
        where the parameters $q_{N\beta}$ and functions $e_{N\beta}(w)$ are summarized in Table~\ref{tab:LZvalues}. Let the normalized eigenvectors and eigenvalues of this matrix be $\ket{v_i}$ and $E_i$ for $i=1,2$. One can check that ($E_1>E_2$)
        \begin{align}
            \hbar\omega_\mathrm{r}\frac{|\bra{v_1}\partial_w\ket{v_2}|^2}{E_1-E_2}=\frac{q_{N\beta}^2\big[N^2-e_{N\beta}(w)+w\partial_w e_{N\beta}(w)\big]^2}{\left\{\left[N^2-e_{N\beta}(w)\right]^2+4q_{N\beta}^2w^2\right\}^{5/2}}.
        \end{align}
        The second ratio on the right-hand side tends to $q^2/64(N-1)^3$ for $w\rightarrow 0$, and vanishes (not necessarily monotonically) for $w\rightarrow\infty$. Using this general result in the expression for the LZ phase in Eq.~\eqref{eq:LZphase} one finds
        \begin{align}
            \SASphase_{N\beta}^\mathrm{LZ}&=\frac{1}{\omega_\mathrm{r}}\int_{-\infty}^\infty\mathrm{d}t \left(\frac{\mathrm{d}w(t)}{\mathrm{d}t}\right)^2 \frac{|\bra{v_1}\partial_w\ket{v_2}|^2}{E_1-E_2}\\
            &=\frac{y_{N\beta}(\Omega_0)}{256(N-1)^3}\frac{\Omega_0^2}{\omega_\mathrm{r}^3\tspread},
        \end{align}
        where
        \begin{multline}\label{eq:yparams}
            y_{N\beta}(\Omega_0)=\int_{-\infty}^{\infty}\mathrm{d}\zeta \left(\frac{\partial_\zeta\Omega(\zeta)}{\Omega_0}\right)^2\\
                    \times\frac{64q_{N\beta}^2(N-1)^3\big[N^2-e_{N\beta}(w)+w\partial_we_{N\beta}(w)\big]^2}{\left\{\left[(N^2-e_{N\beta}(w)\right]^2+4q_{N\beta}^2w^2\right\}^{5/2}}.
        \end{multline}
        Here $w(\zeta)=\Omega(\zeta)/2\omega_\mathrm{r}$ and $\zeta=t/\tspread$ is a dimensionless time scaled to the (effective) pulse duration $\tspread$. The parameter $y_{N\beta}$ is dimensionless and constructed such as to be of order unity.
        
        \subsection{Formulas for LZ losses from Ref.~\cite{Vasilev2004}}\label{app:LandauZenerLosses}
        
        We reproduce here the functions entering formula~\eqref{eq:gamma2+} for the LZ loss parameter $\gamma_{2+}$:
        \begin{subequations}\label{eq:abcoeffs}
         \begin{align}
            \begin{split}
                 a_{\dPhi}(\Omega_0,\tspread)=
                 &\sqrt{2}\tspread \left\{ \vphantom{\sqrt{\sqrt{\left[\ln{\frac{\lambda^2}{\mu_{\dPhi}(2-\mu_{\dPhi})}}\right]^2+\pi^2 + \ln{\frac{\lambda^2}{\mu_{\dPhi}(2-\mu_{\dPhi})}}}}} (\sqrt{\lambda^2 +1} -1) \right. \\
                 &\times\sqrt{\frac{1}{2}\ln{\frac{\lambda^2}{\left[1+\nu_{\dPhi}(\sqrt{\lambda^2+1} -1)\right]^2 -1}}}\\
                 &\left. {} +\frac{1}{2}\sqrt{\sqrt{\left[\ln{\frac{\lambda^2}{\mu_{\dPhi}(2-\mu_{\dPhi})}}\right]^2+\pi^2 + \ln{\frac{\lambda^2}{\mu_{\dPhi}(2-\mu_{\dPhi})}}}} \right\}
               \end{split}   
         \end{align}
        and 
         \begin{align}
            \begin{split}
                 b_\dPhi(\Omega_0,\tspread)=
                 &\frac{4 \omega_\mathrm{r} }{2} \sqrt{2} \tspread \sqrt{\sqrt{4\ln{(m_{\dPhi}\lambda)}^2 + \pi^2} -2 \ln{(m_{\dPhi} \lambda)}},
               \end{split}   
         \end{align}
         \end{subequations}
         with
         \begin{align*}
            \lambda &\equiv \sqrt{2} \frac{\Omega_0}{4\omega_\mathrm{r}}.
         \end{align*}
         These equations correspond to formulas (53) and (44), respectively, in the work of Vasilev and Vitanov \cite{Vasilev2004}, while we inserted the asymptotic energy difference between the states $\ket{2,+}$ and $\ket{0,+}$:
         \begin{align}
              \lim_{t\rightarrow\pm\infty} \frac{E_{2,+}(t) -E_{0,+}(t)}{\hbar} = 4\omega_\mathrm{r}.
         \end{align}
         Eq.~\eqref{eq:gamma2+} from the main text follows from Eq.~(59) in \cite{Vasilev2004}. All formulas have been adapted to the notation used here. This requires in particular to identify the basic Hamiltonian of Vasilev and Vitanov in Eq.~(2) of their paper with Eq.~\eqref{eq:H2e+} of our derivation. Note that the latter features an increased coupling strength $\sqrt{2}\Omega_0$ in comparison to the former. Following the logic of Vasilev and Vitanov, we can find values for the set of free parameters $\mu_{\dPhi},\nu_{\dPhi},m_{\dPhi}$ in Eqs.~\eqref{eq:abcoeffs} to match the exact numerical results as presented in Fig.~\ref{fig:BraggBeamSplitter} and in Fig.~\ref{fig:BraggMirror}:
         \begin{align}\label{eq:abfreeparameters}
            \begin{split}
              m_{\frac{\pi}{2}} & = 0.918028 ; \quad     \nu_{\frac{\pi}{2}} = 0.693525 ; \quad     \mu_{\frac{\pi}{2}} = 0.790483  \\
               m_{\pi} & = 0.983601 ; \quad     \nu_{\pi} \hspace{1.4pt}= 0.596432; \quad     \mu_{\pi}\hspace{1.4pt} = 0.822102.
            \end{split}
         \end{align}
         The $\dPhi$ dependence results from the fact that the basic Hamiltonian in Eq.~(2) of Ref.~\cite{Vasilev2004} applies to a constant energy offset between the two levels. The inclusion of more states than  $\ket{2,\pm}$ and $\ket{0,+}$ required for our analysis, however, leads to ac Stark shifts such that the energy offset becomes $\Omega^2(t)$ dependent. We do not include this as the adaptation of the LZ-formula proposed by Vasilev and Vitanov in~\cite{Vasilev2004} is beyond the scope of the paper. To account for the different ac Stark shifts in case of a beam splitter and mirror pulse, we have optimized the parameters in Eq.~\eqref{eq:abfreeparameters} separately.
    
    \section{Doppler detuning}\label{app:Doppler}
        The first-order correction is described by
     \begin{subequations}
        \begin{align}
            \bra{+,N,p}Z_\alpha\ket{-,N,p}&=\!\!\!\int_{-\infty}^\infty\!\!\!\!\mathrm{d}t\;\frac{e^{i \dPhi(t)} \bra{+,N,p;t}{V}_\alpha(t)\ket{-,N,p;t}}{\hbar}
            \label{eq:Zlim},
       \end{align}
        where
        \begin{align}
            \dPhi(t)&=\int_{-\infty}^t\mathrm{d}t_1 \Big[E_{N+}(t_1)-E_{N-}(t_1)\Big].
        \end{align}
        \end{subequations}
        Note that the phase here is related to the differential phase in Eq.~\eqref{eq:phases} by $\lim_{t\rightarrow + \infty}\dPhi(t)=\dPhi$.
        The matrix element on the right-hand side of Eq.~\eqref{eq:Zlim} can be further simplified by noting that we can rewrite ${V}_\alpha(t)$ in Eq.~\eqref{eq:H1} as
        \begin{align}
            {V}_\alpha(t)&=i2\omega_\mathrm{r} t \hbar\Omega(t)\sum_{n\in \mathds{Z}_\alpha} \left(e^{2i\Lphase}\psigma{n+2,n}-\mathrm{H.c.}\right)\nonumber\\
            &=i2\omega_\mathrm{r}t\left[D_\alpha,\sum_{n\in \mathds{Z}_\alpha} \frac{\hbar\Omega(t)}{2}\left(e^{2i\Lphase}\psigma{n+2,n}+\mathrm{H.c.}\right)\right]\nonumber\\
            &=i2\omega_\mathrm{r}t\left[D_\alpha,H_\alpha(t)-\mathcal{L}_\alpha\right]\nonumber\\
            &=i2\omega_\mathrm{r}t\left[D_\alpha,H_\alpha(t)\right]\label{eq:Vsimple}
        \end{align}
        where $H_\alpha(t)$ and $\mathcal{L}_\alpha$ are given in Eqs.~\eqref{eq:Halpha0} and \eqref{eq:kineticL}, respectively, and we introduced the operator $D_\alpha=\sum_{n\in\mathds{N}_\alpha}n\hat{\sigma}_{n,n}$. This operator acts on the (anti)symmetric states (for $n>0$) as $D_\alpha\ket{\pm,n,p}=n\ket{\mp,n,p}$, and thus changes their parity. It also commutes with $\mathcal{L}_\alpha$, which we used in the last equality in \eqref{eq:Vsimple}. Taking into account the eigenvalue equation \eqref{eq:eigenvalueproblem} one finds
        \begin{multline}\label{eq:Vmatelem}
            \bra{+,N,p;t}{V}_\alpha(t)\ket{-,N,p;t}\\
            =-i 2\omega_\mathrm{r}t\Big[E_{N+}(t)-E_{N-}(t)\Big] \bra{+,N,p;t}{D}_\alpha\ket{-,N,p;t}
        \end{multline}
        This expression also shows that the diagonal matrix elements $\bra{\pm,N,p;t}{V}_\alpha(t)\ket{\pm,N,p;t}$ are proportional to ${(E_{N\pm}(t)-E_{N\pm}(t))=0}$. In order to interpret the matrix element $\bra{+,N,p;t}{D}_\alpha\ket{-,N,p;t}$ one can consider an expansion of the instantaneous energy eigenstates in terms of the asymptotic eigenstates, $\ket{\pm,N,p;t}=\sum_n c_{n\pm}(t)\ket{\pm,n,p}$. Using the fact that $D_\alpha$ flips the parity of the asymptotic eigenstates, one finds $\bra{+,N,p;t}{D}_\alpha\ket{-,N,p;t}=\sum_n n c^*_{n+}(t)c_{n-}(t)$. Due to the asymptotics of the energy eigenstates Eq.~\eqref{eq:energystateslimit} we have $\lim_{t\rightarrow\pm\infty}\bra{+,N,p;t}{D}_\alpha\ket{-,N,p;t}=N$. Inserting \eqref{eq:Vmatelem} into Eq.~\eqref{eq:Zlim} we arrive at 
        \begin{align}
            \bra{+,N,p}Z_\alpha\ket{-,N,p}&= 2N \tspread^2 \omega^2_r z_{N,\dPhi}(\Omega_0) e^{i\dPhi/2}.
        \end{align}
        which is Eq.~\eqref{eq:approxmatelemZ} from the main text. Here, $\dPhi$ is the differential phase from Eq.~\eqref{eq:phases} and $z_{N,\dPhi}$ is 
        \begin{align}\label{eq:zint}
            \begin{split}
            z_{N,\dPhi}(\Omega_0)=&-i\int_{-\infty}^\infty\mathrm{d}\zeta\, \zeta \frac{E_{N+}(\zeta\tspread)-E_{N-}(\zeta\tspread)}{\hbar\omega_\mathrm{r}}\\
            &\times\frac{\bra{+,N,p;\zeta\tspread}{D}_\alpha\ket{-,N,p;\zeta\tspread}}{N}e^{i [\dPhi(\zeta\tspread)-\dPhi/2]}\\
            \approx& -i\int_{-\infty}^\infty\mathrm{d}\zeta\, \zeta \frac{E_{N+}(\zeta\tspread)-E_{N-}(\zeta\tspread)}{\hbar\omega_\mathrm{r}}
            e^{i [\dPhi(\zeta\tspread)-\dPhi/2]}.
            \end{split}
        \end{align}
       In the last line of Eq.~\eqref{eq:zint}, we have approximated the rescaled matrix element to be unity. With this, our theory relies only on the simple calculation of instantaneous eigenenergies instead of the more involved computation of instantaneous energy eigenstates and their overlaps.
       
        The time integral and the integrand have been scaled to dimensionless units such that the value of $z_N$ is positive and on the order of unity as can be seen in Fig.~\ref{fig:xyz_Parameters}. The pulse length is $\tspread$ and $\zeta$ denotes a dimensionless time. The phase in Eq.~\eqref{eq:zint} has been adapted such as to assure that $z_N$ is real. In order to see this, we note that the argument of the exponential, $\dPhi(\zeta\tspread)-\dPhi/2$, is an odd function in $\zeta$ since, for a Gaussian pulse, $\dPhi(t)$ is essentially an error function. Because the rest of the integrand is an odd function in time, only the imaginary part of the exponential contributes to the integral in \eqref{eq:zint}, which makes $z_{N,\dPhi}$ real.
    
    \section{Fidelities}\label{app:Fidelities}
    The scattering matrices $B_\dPhi$ and  $B(p,\Omega_0,\tspread)$ are given in Eqs.~\eqref{eq:idealScattM} and \eqref{eq:BraggScatteringMatrixFinal}, respectively, such that $|B^\dagger_\dPhi B(p,\Omega_0,\tspread)|^2$  evaluates to ($\dPhi=\pi/2$)
     \begin{multline}\label{eq:FidpClassBeamsplitter}
    \left|\left[B^\dagger_\frac{\pi}{2} B(p,\Omega_0,\tspread)\right]_{11}\right|^2= \\ \frac{e^{-\Gamma}}{2[1+\eta^2(p)]} \Big\{1+[1+\eta^2(p)]\cosh\left(\gamma\right)
    +\sqrt{2} \eta(p) \sinh{(\gamma)} \Big\}\\ \simeq  \frac{e^{-\Gamma}}{2}\left[1+\cosh{(\gamma)}-\eta^2_{0,\frac{\pi}{2}} \left(\frac{p}{\hbar k}\right)^2 + \sqrt{2} \eta_{0,\frac{\pi}{2}} \frac{p}{\hbar k} \sinh{(\gamma)} \right],
    \end{multline}
    as well as ($\dPhi=\pi$)
    \begin{multline}\label{eq:FidpClassMirror}
        \left|\left[B^\dagger_\pi B(p,\Omega_0,\tspread)\right]_{11}\right|^2=\\
        \frac{e^{-\Gamma}}{2[1+\eta^2(p)]} \Big\{1-\eta^2(p)+[1+\eta^2(p)]\cosh\left(\gamma\right) +2 \eta(p) \sinh{(\gamma)} \Big\}\\
        \simeq  \frac{e^{-\Gamma}}{2}\left[1+\cosh{(\gamma)}-2\eta^2_{0,\pi} \left(\frac{p}{\hbar k}\right)^2 + 2 \eta_{0,\pi}  \frac{p}{\hbar k} \sinh{(\gamma)} \right].
    \end{multline}

    To obtain the last lines in these two equations we introduced the dimensionless parameter
    \begin{align}
        \eta_{0,\dPhi} =- 2 N \tspread^2 \omega_\mathrm{r}^2 z_{N,\dPhi}(\Omega_0),
    \end{align}
    and performed an expansion up to order $\mathcal{O}[p]^3$. 
\begin{widetext}
Assuming an atomic wave packet with a Gaussian momentum distribution with finite width $\pspread$,
    \begin{align}
        g(p,\pspread)&=(2\pi\pspread^2)^{-1/4}e^{-\frac{p^2}{4\pspread^2}}
    \end{align}

    we can immediately execute the integration in Eq. \eqref{eq:FidAverageGeneral} giving us the averaged beam splitter
    \begin{align}\label{eq:FidAverageBeamsplitter}
        \Fid_{\frac{\pi}{2},\pspread}(\Omega_0,\tspread) &\simeq \frac{e^{-\Gamma}}{2}\left\{\frac{\eta^2_{0,\frac{\pi}{2}}  \pspread}{\sqrt{2\pi }\hbar k} e^{-\frac{1}{8} \left(\frac{\pspread}{\hbar k }\right)^{-2}} + \left[1+ \cosh{(\gamma)}-\eta^2_{0,\frac{\pi}{2}}  \left(\frac{\pspread}{\hbar k }\right)^2\right] \mathrm{erf}\left(\frac{1}{2\sqrt{2}} \left(\frac{\pspread}{\hbar k}\right)^{-1}\right)\right\}
    \end{align}
  and mirror fidelity
    \begin{align}\label{eq:FidAverageMirror}
         \Fid_{\pi,\pspread}(\Omega_0,\tspread) &\simeq e^{-\Gamma}\left\{\frac{\eta^2_{0,\pi}  \pspread}{\sqrt{2\pi }\hbar k} e^{-\frac{1}{8} \left(\frac{\pspread}{\hbar k }\right)^{-2}} + \left[\frac{1}{2}\left[1+ \cosh{(\gamma)}\right]-\eta^2_{0,\pi}  \left(\frac{\pspread}{\hbar k }\right)^2 \right] \mathrm{erf}\left(\frac{1}{2\sqrt{2}} \left(\frac{\pspread}{\hbar k}\right)^{-1}\right)\right\}.
    \end{align}
   
\end{widetext}
    \section{Hilbert space dimensionality and numerical  integration}\label{AppendixNumerics}
    The results presented in this paper are the product of calculations in truncated finite dimensional Hilbert spaces. This applies to both, our analytics which requires us to diagonalize finite dimensional Hamiltonians to calculate their spectra (see Eqs.~\eqref{eq:eigenvalueproblem}) and to the full numerical integration of the Schr\"odinger equation. In each case, we truncate the momentum state basis like
    \begin{multline}
        \{\ket{-n_\mathrm{max} \hbar k +p },\ket{(-n_\mathrm{max}+2) \hbar k +p },...\\
        ...,\ket{(n_\mathrm{max}-2) \hbar k +p },\ket{n_\mathrm{max} \hbar k +p }\},
    \end{multline}
    where $n_\mathrm{max}$ is even (odd) if the diffraction order $N$ is even (odd) and perform the same truncation of the Hamiltonians in Eqs.~\eqref{eq:Hblocks} in the (anti)symmetric basis. 
    The truncations applied for our calculations are $n_\mathrm{max} = 6,7,8,11$ for the different Bragg orders $N=2,3,4,5$ respectively. The codes that generate the results presented in Figs. \ref{fig:Bragg_Fid_NoNorm_Num}-\ref{fig:xyz_Parameters} and Figs. \ref{fig:BraggCondition}-\ref{fig:BlochSpectra}  of this paper using these truncations are available online \cite{Siemss2020}.
    
    We note that these truncations are adequate for the analysis performed in the context of this article. Calculation of atom interferometer phases may require increased accuracy and therefore higher truncations as noted in Ref.~\cite{Gochnauer2019}. At the same time, we point out that our analytic model only relies on the computation of the spectra of these Hamiltonians. Such a step will therefore not add significantly to the complexity of the model.
    
    Throughout this study we compute time integrals and numerically solve the Schrödinger equations. To ensure that these calculation reflect the asymptotic nature of scattering theory on which our model is based on, we choose time intervals (expressed here in in units of $\omega^{-1}_\mathrm{r}$) ${\zeta \in [-22,22]}$ accordingly.
 
    \bibliographystyle{apsrev4-1_costum}
    \bibliography{bibliography}

    \end{document}